\documentclass[twocolumn,tighten,times]{aastex631}
\usepackage{enumitem}

\shorttitle{GATOS  VIII: Origin extended MIR emission in AGN}
\shortauthors{Lopez-Rodriguez, E. et al.}
\graphicspath{{./}{figures/}}

\begin{document}

\title{GATOS. VIII. \\ 
On the physical origin of the extended MIR emission in AGN}

\correspondingauthor{Enrique Lopez-Rodriguez}
\email{elopezrodriguez@stanford.edu}

\author[0000-0001-5357-6538]{Enrique Lopez-Rodriguez}
\affiliation{Kavli Institute for Particle Astrophysics \& Cosmology (KIPAC), Stanford University, Stanford, CA 94305, USA}

\author[0000-0001-8353-649X]{Cristina Ramos Almeida}
\affiliation{Instituto de Astrof\'sica de Canarias, Calle V\'ia L\'actea, s/n, E-38205, La Laguna, Tenerife, Spain}
\affiliation{Departamento de Astrof\'isica, Universidad de La Laguna, E-38206 La Laguna, Tenerife, Spain}

\author[0000-0002-4005-9619]{Miguel Pereira-Santaella}
\affiliation{Instituto de F\'isica Fundamental, CSIC, Calle Serrano 123, 28006, Madrid, Spain}

\author[0000-0002-9627-5281]{Ismael Garc\'ia-Bernete}
\affiliation{Department of Physics, University of Oxford, Keble Road, Oxford, OX1 3RH, UK}

\author[0000-0002-7052-6900]{Robert Nikutta}
\affiliation{NSF's NOIRLab, 950 North Cherry Avenue, Tucson, AZ 85719, USA}

\author[0000-0001-6794-2519]{Almudena Alonso-Herrero}
\affiliation{Centro de Astrobiolog\'ia (CSIC-INTA), Camino Bajo del Castillo s/n, 28692 Villanueva de la Can\~nada, Madrid, Spain}

\author[0000-0003-3589-3294]{Anelise Audibert}
\affiliation{Instituto de Astrof\'sica de Canarias, Calle V\'ia L\'actea, s/n, E-38205, La Laguna, Tenerife, Spain}
\affiliation{Departamento de Astrof\'isica, Universidad de La Laguna, E-38206 La Laguna, Tenerife, Spain}

\author[0000-0001-9791-4228]{Enrica Bellocchi}
\affiliation{Departmento de F\'isica de la Tierra y Astrof\'isica, Fac. de CC F\'isicas, 1066 Universidad Complutense de Madrid, 28040 Madrid, Spain}
\affiliation{Instituto de F\'isica de Part\'iculas y del Cosmos IPARCOS, Fac. CC 1068 F\'isicas, Universidad Complutense de Madrid, 28040 Madrid, Spain}

\author[0000-0002-8651-9879]{Andrew Bunker}
\affiliation{Department of Physics, University of Oxford, Keble Road, Oxford, OX1 3RH, UK}

\author[0000-0001-9520-7765]{Steph Campbell}
\affiliation{School of Mathematics, Statistics and Physics, Newcastle University, Newcastle upon Tyne, NE1 7RU, UK}

\author[0000-0003-2658-7893]{Fran\c{c}oise Combes}
\affiliation{LERMA, Observatoire de Paris, Colle\'ege de France, PSL University, 1072 CNRS, Sorbonne University, Paris}

\author[0000-0003-4949-7217]{Richard Davies}
\affiliation{Max Planck Institute for Extraterrestrial Physics (MPE), Giessenbachstr.1, 85748 Garching, Germany}

\author[0000-0003-0699-6083]{Tanio Diaz-Santos}
\affiliation{Institute of Astrophysics, Foundation for Research and Technology 
 Hellas (FORTH), Heraklion, GR-70013, Greece}

\author[0000-0003-4809-6147]{Lindsay Fuller}
\affiliation{Department of Physics and Astronomy, The University of Texas at San Antonio, 1 UTSA Circle, San Antonio, TX 78249, USA}

\author[0000-0003-3105-2615]{Poshak Gandhi}
\affiliation{Department of Physics \& Astronomy, University of Southampton, Highfield, Southampton SO171BJ, UK}

\author[0000-0003-0444-6897]{Santiago Garc\'ia-Burillo}
\affiliation{Observatorio Astron\'omico Nacional (OAN-IGN)-Observatorio de Madrid, Alfonso XII, 3, 28014 Madrid, Spain}

\author[0000-0002-2356-8358]{Omaira Gonz\'alez-Mart\'in}
\affiliation{Instituto de Radioastrononom\'ia y Astrof\'isica (IRyA), Universidad Nacional Aut\'onoma de M\'exico, Antigua Carretera a P\'atzcuaro \#8701, 1087 ExHda. San Jos\'e de la Huerta, Morelia, Michoac\'an, Mexico C.P. 1088 58089}

\author[0000-0002-4457-5733]{Erin K. S. Hicks}
\affiliation{Department of Physics \& Astronomy, University of Alaska Anchorage, Anchorage, AK 99508-4664, USA}
\affiliation{Department of Physics and Astronomy, The University of Texas at San Antonio, 1 UTSA Circle, San Antonio, TX 78249, USA}
\affiliation{Department of Physics, University of Alaska, Fairbanks, Alaska 99775-5920, USA}

\author[0000-0002-6353-1111]{Sebastian H\"onig}
\affiliation{Department of Physics \& Astronomy, University of Southampton, Highfield, Southampton SO171BJ, UK}

\author[0000-0002-4377-903X]{Kohei Ichikawa}
\affiliation{Global Center for Science and Engineering, Faculty of Science and Engineering, Waseda University}
\affiliation{Department of Physics, School of Advanced Science and Engineering, Faculty of Science and Engineering, Waseda University, 3-4-1,
Okubo, Shinjuku, Tokyo 169-8555, Japan}

\author[0000-0001-6186-8792]{Masatoshi Imanishi}
\affiliation{National Astronomical Observatory of Japan, National Institutes of Natural Sciences (NINS), 2-21-1 Osawa, Mitaka, Tokyo 181-8588, Japan}

\author[0000-0001-9452-0813]{Takuma Izumi}
\affiliation{National Astronomical Observatory of Japan, National Institutes of Natural Sciences, 2-21-1 Osawa, Mitaka, Tokyo 181-8588, Japan}
\affiliation{Department of Astronomy, School of Science, Graduate University for Advanced Studies (SOKENDAI), Mitaka, Tokyo 181-8588, 1095 Japan}

\author[0000-0002-0690-8824]{Alvaro Labiano}
\affiliation{Telespazio UK for the European Space Agency, ESAC, Camino 1097 Bajo del Castillo s/n, 28692 Villanueva de la Ca\~nada, Spain}

\author[0000-0003-4209-639X]{Nancy A. Levenson}
\affiliation{Space Telescope Science Institute, 3700 San Martin Drive, Baltimore, MD 21218, USA}

\author[0000-0001-7827-5758]{Christopher Packham}
\affiliation{Department of Physics and Astronomy, The University of Texas at San Antonio, 1 UTSA Circle, San Antonio, TX 78249, USA}
\affiliation{National Astronomical Observatory of Japan, National Institutes of Natural Sciences, 2-21-1 Osawa, Mitaka, Tokyo 181-8588, Japan}

\author[0000-0002-0001-3587]{David Rosario}
\affiliation{School of Mathematics, Statistics and Physics, Newcastle University, Newcastle upon Tyne, NE1 7RU, UK}

\author[0000-0001-6854-7545]{Dimitra Rigopoulou}
\affiliation{Department of Physics, University of Oxford, Keble Road, Oxford OX1 3RH, UK }
\affiliation{School of Sciences, European University Cyprus, Diogenes street, Engomi, 1516 Nicosia, Cyprus}

\author[0000-0002-2352-1736]{Daniel Rouan}
\affiliation{LESIA, Observatoire de Paris, Universit\'e PSL, CNRS, Sorbonne 1107 Universit\'e , Sorbonne Paris Cite\'e, 5 place Jules Janssen, 92195 1108 Meudon, France}

\author[0000-0002-2125-4670]{Taro Shimizu}
\affiliation{Max Planck Institute for Extraterrestrial Physics (MPE), Giessenbachstr.1, 85748 Garching, Germany}

\author[0000-0001-5146-8330]{Marko Stalevski}
\affiliation{Astronomical Observatory, Volgina 7, 11060 Belgrade, Serbia}
\affiliation{Sterrenkundig Observatorium, Universiteit Gent, Krijgslaan 281-S9, Gent B-9000, Belgium}

\author{Martin Ward}
\affiliation{Centre for Extragalactic Astronomy, Department of Physics, Durham University, South Road, Durham, DH1 3LE, UK}

\author[0000-0003-4937-9077]{Lulu Zhang}
\affiliation{Department of Physics and Astronomy, The University of Texas at San Antonio, 1 UTSA Circle, San Antonio, TX 78249, USA}

\author[0000-0001-5231-2645]{Claudio Ricci}
\affiliation{Instituto de Estudios Astrof\'isicos, Facultad de Ingenier\'ia y Ciencias, Universidad Diego Portales, Av. Ej\'ercito Libertador 441, Santiago, Chile}
\affiliation{Kavli Institute for Astronomy and Astrophysics, Peking University, Beijing 100871, China}

\author[0000-0001-8042-9867]{Donaji Esparza-Arredondo}
\affiliation{ Instituto de Radioastronom\'ia y Astrof\'isica (IRyA-UNAM), 3-72 (Xangari), 8701, Morelia, Mexico}

\author[0000-0002-7228-7173]{Bego\~na Garc\'ia-Lorenzo}
\affiliation{Instituto de Astrof\'sica de Canarias, Calle V\'ia L\'actea, s/n, E-38205, La Laguna, Tenerife, Spain}
\affiliation{Departamento de Astrof\'isica, Universidad de La Laguna, E-38206 La Laguna, Tenerife, Spain}



\begin{abstract}
The polar mid-infrared (MIR) emission detected within 10-100s parsecs in some active galactic nuclei (AGN) has been associated with dusty winds driven away by radiation pressure. The physical characterization of this extended polar emission remains uncertain. Here we combine $10-21\,\mu$m JWST/MIRI imaging observations with $7-25\,\mu$m JWST/MIRI\,MRS integral field spectroscopic observations of 6 nearby, $\bar{D}=35.4\pm4.6$\,Mpc, AGN from the GATOS Survey to quantify the nature of the extended MIR emission at $\sim75$ pc resolution at $21\,\mu$m. These AGN have similar bolometric luminosities, $\log_{10}(\bar{L}_{\rm{bol}}\,[\rm{erg\,s}^{-1}])=44.0\pm0.3$, span a wide range of optical outflow rates, $\dot{M}=0.003-0.21$ M$_{\odot}$ yr$^{-1}$, column densities, $\log_{10}(N_{\rm{H}}^{\rm{X-ray}}[\rm{cm^{-2}}])=22.2-24.3$, and Eddington ratios, $\lambda_{\rm{Edd}}=0.005-0.06$. We cross-correlate the line-only and continuum-only images and find a poor correlation, which indicates that the extended MIR continuum emission is spatially uncorrelated with the warm outflows associated with narrow emission lines within $10-15\,\mu$m. Line emission is resolved along the jet axis, while dust emission is perpendicular to it. The $75-450$\,pc continuum emission has a fairly constant dust temperature, $T_{\rm{d}}=132^{+7}_{-7}$\,K, and mass, $M_{\rm{d}}=728^{+29}_{-27}$\,M$_{\odot}$. Using the conditions of energy balance between radiation-pressure and gravity ($\lambda_{\rm{Edd}}$\,vs.\,N$_{\rm{H}}$), we find that our AGN sample is in the gravitationally bounded regime consistent with no detection of dusty winds. At $10\,\mu$m, the level of extended line emission contribution is correlated with the outflow kinetic energy and mass outflow rates. We find no correlation with the AGN properties. These results indicate that the radio jet may be triggering the gas outflow and line emission, while the extended dust emission is distributed in molecular clouds and/or shocked regions.
\end{abstract}



\section{Introduction}
\label{sec:intro}

Feedback from active galactic nuclei (AGN) is a key ingredient for regulating gas cooling and subsequent galaxy growth in galaxies. However, because of the intermittent and short-lived nature of nuclear activity, studying its impact on galaxies that are currently active is challenging from an observational point of view \citep{HRA2024}. A way to characterize the localized, in-situ effect of AGN feedback is by studying how the dusty and gas torus properties vary with AGN luminosity and Eddington ratio through the use of high angular resolution observations of nearby AGN (e.g., \citealt{Ramos2017,Ricci2017,GB2019,AH2021,GATOSI,Ricci23}). 

The torus has been pictured as an optically and geometrically thick structure capable of producing obscuration of the broad-line region (BLR) along certain line-of-sights \citep[LOS;][]{Antonucci1985}.  In the last ten years, the Atacama Large Millimeter/submillimeter Array (ALMA) has permitted the study of the molecular gas and cold dust in the torus with angular resolutions of a few parsecs ($\sim5-15$ pc). These dusty and molecular tori have a mean radius of $\sim42$ pc and a molecular gas mass of $\sim6\times10^{5}$ M$_{\odot}$, based on spatially resolved sub-mm observations at $\le0.1\arcsec$ resolution of $10-20$ nearby Seyfert galaxies \citep{GB2016b,AH2018,Imanishi2018,Combes2019,GATOSI,GB2024}. A gravitationally unstable dense molecular disk \citep{Izumi2023} and/or a strong and ordered magnetic field parallel to the equatorial axis of the dusty torus \citep{ELR2015,ELR2020} can further support the accretion flow and the outflowing material within the central parsecs. Thus, outflows are launched at sub-pc scales \citep[e.g.,][]{Begelman1983,Emmering1992,Proga1998,WD2001,Chelouche2001} with the bulk of the dust and molecular mass being cospatial with the equatorial axis of an optically and geometrically thick disk feeding gas from tens to pc scales onto the AGN \citep{ELR2018,ELR2020,AH2021,Nikutta2021a,Nikutta2021b,GATOSI}. The torus is now understood as an episodic and dynamic multi-phase structure that connects the central supermassive black hole (SMBH) and the circumnuclear region of the host galaxy via inflows and outflows \citep{Ramos2017,GB2019,GATOSI}.

Mid-infrared (MIR; N-band: $7-13~\mu$m) interferometric observations have detected extended emission on parsec scales mostly along the direction of the radio jet axis and accounting for a significant portion of the MIR emission \citep{Jaffe2004,Raban2009,Tristram2009,Tristram2014,Hoenig2012,Hoenig2013,Burtscher2013,LG2014,LG2016,LG2016b,Leftley2018,Isbell2022,Isbell2023,GamezRosas2022}. This extended MIR emission component is commonly referred to as ``polar dust'' and it has been interpreted as dusty winds driven by radiation pressure from the AGN \citep{Hoenig2017,Hoenig2019}, therefore becoming part of the feeding and feedback cycle in AGN. According to radiation hydrodynamical models, these dusty winds can extend up to scales of $\sim$1–100 pc and account for up to 90\% of the IR emission within the range of $30-100$ pc, while the central $\sim10$ pc accounts for 50\% of the total IR emission \citep{Williamson2020}. Because the current IR interferometric observations are restricted to a few bright and nearby AGN and field-of-views (FOVs) of a few pc-scales, MIR (N and Q bands: $\sim7-20~\mu$m) ground-based observations from single-dish telescopes have been used to search for polar MIR emission measured on scales of 10-100s parsecs \citep{Asmus2014,Asmus2015,Asmus2016}. These scales are much larger than the expected torus sizes at MIR  \citep[$\lesssim$10 pc;][]{Radomski2003,Mason2006,AH2011,ELR2018,Nikutta2021a} and in the sub-mm wavelengths \citep[$\sim40$ pc;][]{GATOSI,AH2021} where the bulk of the mass and molecular gas resides \citep{ELR2018,Nikutta2021a,GATOSI,AH2021}. 

Based on a sample of $149$ Seyfert galaxies observed with 8-m class telescopes, \citet{Asmus2016} reported that 18 of the objects (12\% of the sample) showed extended MIR emission along the radio jet axis cospatial with the [\ion{O}{3}] emission, i.e., the narrow-line-region (NLR). They estimated that the relative contribution of this polar MIR emission is at least 40\%, and it scales with the [O IV] fluxes. Furthermore, the IR-X-ray luminosities are found to be correlated with nuclear X-ray emission irrespective of AGN type and orientation, within smaller factors of order 2 \citep{Gandhi2009, Levenson2009, Asmus2015}. Thus, the polar MIR emission was interpreted as a continuation of the pc-scale dusty winds  \citep{Asmus2014,Asmus2015}. This interpretation was based on the assumption that the extended MIR emission arises from continuum dust emission cospatial with the outflowing material in the NLR. However, the potential line emission contribution included in the imaging filters was not taken into account. 

Previous work reported the presence of ambient dust in the NLR based on MIR observations \citep[i.e.,][]{Cameron1993,Bock2000,Radomski2003,ELR2016,GB2016,ELR2018,AH2021}. This dust was suggested to account for the suppression of line emission \citep[e.g.,][]{Heckman1981,DeRobertis1984,Whittle1985,MacAlpine1985,Osterbrock1989,Netzer1993} and variations in the spatial distribution of the emission across the NLR \citep{Quillen1999}. For example, MIR observations of NGC\,4151 show a dust contribution of $\sim27$\% at $10~\mu$m with a characteristic dust temperature of $165\pm15$ K at $\sim100$ pc from the AGN, consistent with ambient dust heated by the AGN \citep{Radomski2003}. NGC\,1068 has a warmer dust component of $220-260$ K across the northern region of the NLR, which may be due to the shock front generated by the radio jet and to the higher AGN luminosity of this Seyfert galaxy \citep{Bock2000}. Photoionization and shock models of the NLR estimated that the IR emission in the NLR accounts for $\sim10$\% of the total emission at $25~\mu$m \citep{Groves2006}. These models show the large number of emission lines within the MIR wavelength range that may be contributing to the total emission observed using broadband imaging filters \citep[see Fig. 5 by][]{Groves2006}. Thus, for years it has remained unclear whether the measured extended MIR emission is dominated by dust or if emission lines accounts for a substantial contribution. 

\vspace{-0.8cm}

\begin{deluxetable*}{lcccccccccccc}[ht!]
\centering
\tablecaption{Physical properties of our galaxy sample.
\label{table:tab1} 
}
\tablewidth{0pt}
\tablehead{\colhead{} &
\colhead{} &
\colhead{ESO137-G034}  & 
\colhead{MCG-05-23-016} & 
\colhead{NGC\,3081} & 
\colhead{NGC\,5506 } & 
\colhead{NGC\,5728}  & 
\colhead{NGC\,7172} &
\colhead{}}
\startdata
Redshift   & & 0.009 & 0.008 & 0.008 & 0.006 & 0.009 & 0.009 & (a) \\
Distance                        & [Mpc] &  38.9 & 36.2 & 34.0 & 26.0 & 39.7 & 37.6 & (b)\\
Scale                           & [pc "$^{-1}$] & 189 & 175 & 165 & 126 & 192 & 179 & (c)\\
$\log_{10}$ L$_{\rm{AGN}}$      & [erg s$^{-1}$] & 43.4 & 44.4 & 44.1 & 44.1 & 44.1 & 44.1 & (d)\\
$\dot{M}_{\rm{out}}^{\rm{Opt}}$                       & [M$_{\odot}$ yr$^{-1}$] & 0.52 & 0.003 & 0.04 & 0.21 & 0.09 & 0.005 & (e)\\
$\dot{M}_{\rm{out}}^{\rm{MIR}}$                       & [M$_{\odot}$ yr$^{-1}$] & 0.33 & 0.04 & 0.03 & 0.28 & 0.08 & 0.03 & (f)\\
$\log_{10} \dot{E}_{\rm{kin}}$  & [erg s$^{-1}$] & 40.7 & 37.8 & 39.0 & 40.6 & 40.3 & 38.4 & (g)\\
$\lambda_{\rm{Edd}}$            &  & 0.01 & 0.06 & 0.02 & 0.04 & 0.05 & 0.02 & (h)\\
$\log_{10}$ $N_{\rm{H}}^{\rm{X-ray}}$      & [cm $^{-2}$] & 24.3 & 22.2 & 23.9 & 22.4 & 24.2 & 22.9 & (i)\\
$\log_{10}$ $N_{\rm{H}}^{\rm{CO}}$         & [cm $^{-2}$] & 22.61 & - & 22.16 & 22.90 & 23.42 & 23.60 & (j)\\
PA$_{\rm{jet}}$                 & [$^{\circ}$] & 140 & 169 & 158 & 70 & 127 & 0 & (k)\\
PA$\rm{_{DustLane}}$             & [$^{\circ}$] & 165 & 58 & 122 & 90 & 33, 86 & 90 & (l)\\
Refs.   & & D1, K1, L1 & D2, K2, L1 & D1, K3, L2 & D1, K4, I1, L3 & D3, J4, L4 & D1, K4, L5 & (m)\\
\enddata
\tablenotetext{}{\textbf{Notes:} 
(a) Redshift taken from NED, 
(b) Distance in units of Mpc, 
(c) Scale in units of pc arcsec$^{-1}$, 
(d) AGN luminosity in units of erg s$^{-1}$, 
(e) Outflow rate in units of M$_{\odot}$ yr$^{-1}$ using the optical [\ion{O}{3}] emission, 
(f) Outflow rate in units of M$_{\odot}$ yr$^{-1}$ using the MIR [\ion{Ne}{5}] emission, 
(g) Kinetic energy in units of erg s$^{-1}$,
(h) Eddington ratio $\lambda_{\rm{Edd}}=L_{\rm{AGN}}/L_{\rm{Edd}}$ using $2-10$ keV observations,
(i) column density in units of cm$^{-2}$ based on modeling $0.3–150$ keV spectrum,
(j) column density in units of cm$^{-2}$ based on CO observations,
(k) Radio jet PA axis in units of $^{\circ}$ East of North,
(l) Dust lane PA axis in units of $^{\circ}$ East of North,
(m) References labeled as per column. }
\tablenotetext{}{\textbf{References:} 
(d)
    D1: \citet{Davies2020},
    D2: \citet{AH2011}.
    D3: \citet{Shimizu2019}.
(e) \citet{Davies2020}.
(f) \citet{GATOSIV}.
(g) \citet{Davies2020}.
(i) I1: \citet{Koss2017},
(j) \citet{GB2024},
(K) 
    K1: \citet{Morganti1999}, 
    K2: \citet{Mundell2009}, 
    K3: \citet{Nagar1999},
    K4: \citet{Kinney2000,Sebastian2020},
(l) 
    L1: \citet{Ferruit2000}, 
    L2: Nuclear bar from \citet{Erwin2004},
    L3: Edge-on disk from \citet{Malkan1998,Theios2016},
    L4: Large scale bar and nuclear stellar bar from \citet{Prada1999},
    L5: \citet{Smajic2012}.}
\vspace{-0.9cm}
\end{deluxetable*}

In the first part of the Galactic Activity, Torus and Outflow Survey (GATOS) manuscript series, aimed at characterizing and understanding MIR extended emission in nearby AGN, Campbell et al. (submitted) showed that line emission contributions are of $\sim10-60$\% across the $\sim 40-500\,$ pc NLR within the MIRI F1000W filter using data taken by JWST/MRS MIRI integral field spectroscopic observations of a sample of 11 nearby AGN (Note that these objects are the same as those presented in this work, Table \ref{table:tab1}). The observed polar MIR emission at scales of $\geq 40$ pc has a significant contribution from line emission and varies from object to object. Thus, this polar MIR emission must be taken into account when interpreting feeding and feedback mechanisms in AGN. 

In this second part of the manuscript series, we aim to quantify the physical properties of the MIR emission after removing the line emission contribution in the NLR of nearby Seyfert galaxies (same AGN sample as in Campbell et al. submitted). Here, we provide dust masses and temperatures that can be used to estimate energy budgets of AGN feedback mechanisms. This analysis can now be performed thanks to the combination of sensitivity, angular resolution, and spectral coverage provided by the integral field unit (IFU) of the Mid-InfRared Instrument \citep[MIRI;]{Rieke2015,Wright2015} onboard the JWST. Specifically, we make use of MIRI imaging and integral field spectroscopy with the Mid-Resolution Spectroscopy \citep[MRS;][]{Wells2015,Wright2023,Argyriou2023} IFU obtained through two different JWST Cycle 1 General Observer (GO) programs with the purpose of characterizing the physical properties of the extended MIR emission using the IFU data. We describe the observations and data reduction in Section \ref{sec:OBS_RED}.  Then, we present the analysis in Section \ref{sec:dis} and discuss it in Section \ref{sec:ObsEv}. Our main conclusions are summarized in Section \ref{sec:concusions}. We estimate distances and scales assuming H$_{0} = 69.6$ km s$^{-1}$ Mpc$^{-1}$, $\Omega_{\rm{M}} = 0.286$ and $\Omega_{\rm{v}} = 0.714$ \citep{Bennett2014}.

\section{Observations and AGN sample} \label{sec:OBS_RED}

\begin{figure*}
\centering
\includegraphics[width=\textwidth]{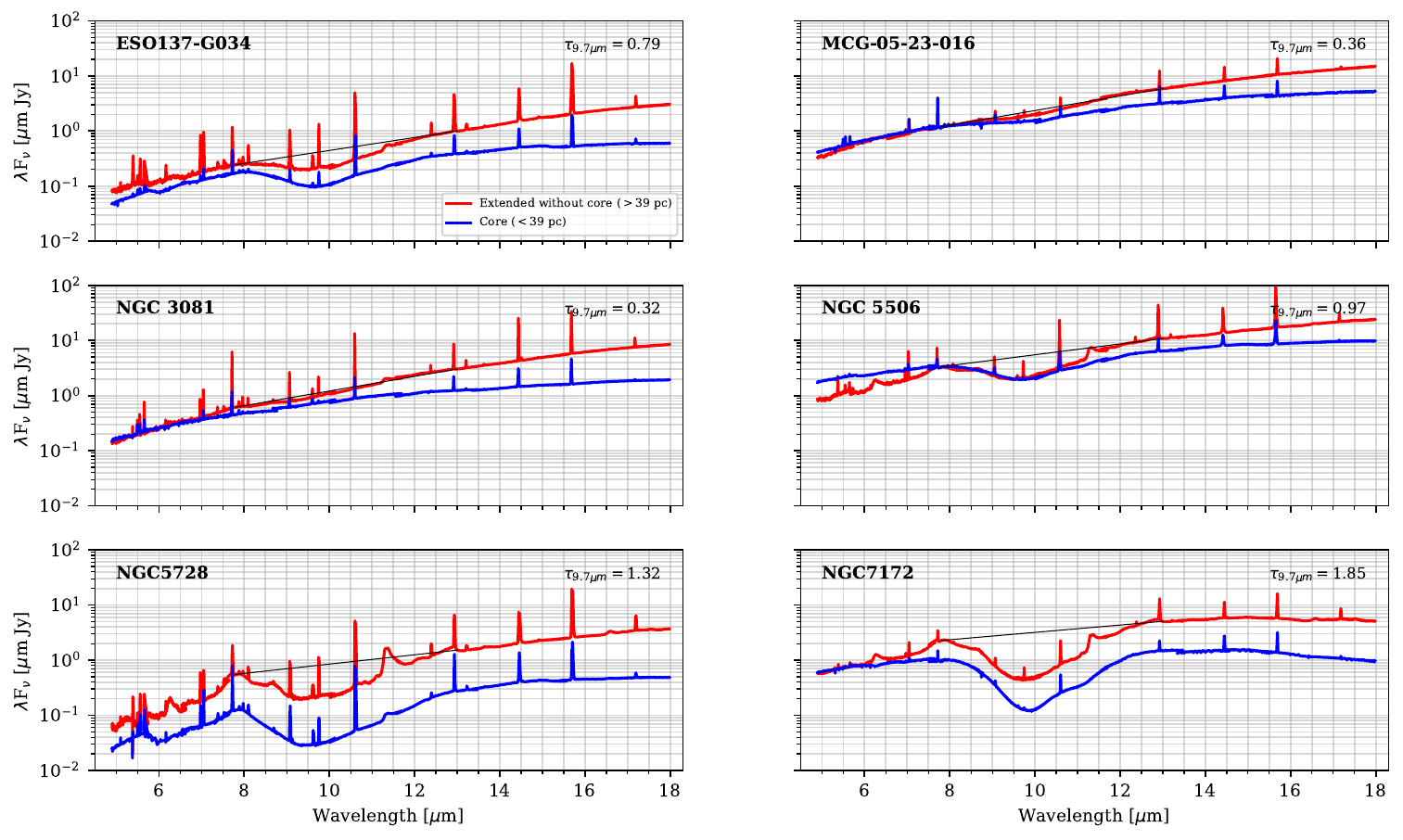}
\caption{MIR spectra of the AGN in our sample. The spectra of the extended emission without the core ($>39$ pc; red) and only the core ($\le39$ pc; blue) are shown. The black solid line shows the extrapolated continuum used to determine the silicate feature, $\tau_{\rm{9.7\mu m}}=-\ln(F_{9.7\mu \rm{m}}/F_{\rm{c}})$,  from the extended MIR emission without the core.}
 \label{fig:figX}
\end{figure*}

Our AGN sample is taken from the JWST Cycle 1 GO programs \#1670 (PI: Shimizu, T.) and \#2064 (PI: Rosario, D., H\"onig, S., Burtscher, L.) as part of the GATOS\footnote{GATOS website: \url{https://gatos.myportfolio.com/}} collaboration \citep{GATOSI,AH2021,GATOSIII}. GATOS aims to understand the properties of the dusty and molecular gas torus, feedback and feeding mechanisms, and their connection to their host galaxies in local AGN. The sample in this manuscript is part of the larger sample of AGN in GATOS drawn from the 70 Months Swift-BAT All-sky Hard X-ray Survey \citep{Baumgartner2013}. This ensures a nearly complete selection of AGN with luminosities $L_{14-150\rm{keV}} > 10^{42}$ erg s$^{-1}$ at distances of $10-40$ Mpc. The selection band also means that the sample is largely unbiased to obscuration/absorption even up to column densities of $N_{H}\sim10^{24}$ cm$^{-2}$. Both the AGN luminosity and absorbing column are available from the analysis of the X-ray data \citep{Ricci2017}. 

Specifically, JWST Program \#1670 aims to characterize and spatially resolve the warm phase of AGN outflows and map the dust continuum and Polycyclic Aromatic Hydrocarbons (PAHs) within the central few hundred pcs using MIRI/MRS observations. The sample was selected to include six AGN with similar bolometric luminosities and distances (mean $\log_{10} (L_{\rm{bol}}~[\rm{erg\,s}^{-1}]) = 44.0\pm0.3$ and D $=35.4\pm4.6$ Mpc), while spanning a wide range of ionized outflow rates ($\dot{M}_{\rm{out}}^{\rm{Opt}} =0.003-0.52$ M$_{\odot}$ yr$^{-1}$; $\dot{M}_{\rm{out}}^{\rm{MIR}} =0.03-0.33$ M$_{\odot}$ yr$^{-1}$), where the warm outflow rate at MIR wavelengths is less affected by dust obscuration. A first analysis of the Cycle 1 JWST/MRS data investigating the nuclear silicate features and water ices was presented in \citet{GATOSIII}. On the other hand, JWST Program \#2064 aims to characterize the extended MIR continuum emission (i.e., the polar dust) in local AGN using MIRI imaging observations. The sample was curated with AGN that showed extended MIR emission using sub-arcsecond imaging observations from single-dish telescopes \citep{Asmus2016,Asmus2019} and milli-arcsecond interferometric observations \citep{Burtscher2013,LG2016}. The data reduction is described in Appendix \ref{sec:RED}. A first analysis of the imaging data is presented in Campbell et al. (submitted).

Both JWST programs have several AGN in common, which makes it ideal for our goal of quantifying the line and continuum contributions across the MIR extended emission in AGN observed with broadband filters. We use all AGN observed with MRS (\#1670): ESO137-G034, MCG-05-23-016, NGC\,3081, NGC\,5506, NGC\,5728, NGC\,7172. The common objects observed (\#2064) with the MIRI imaging mode are NGC\,3081, NGC\,5728, and NGC\,7172. The AGN sample and physical properties are shown in Table \ref{table:tab1}.

\subsection{Line and continnum extraction}\label{subsec:ExtrContLine}

\begin{figure*}[ht!]
\centering
\includegraphics[scale=0.55]{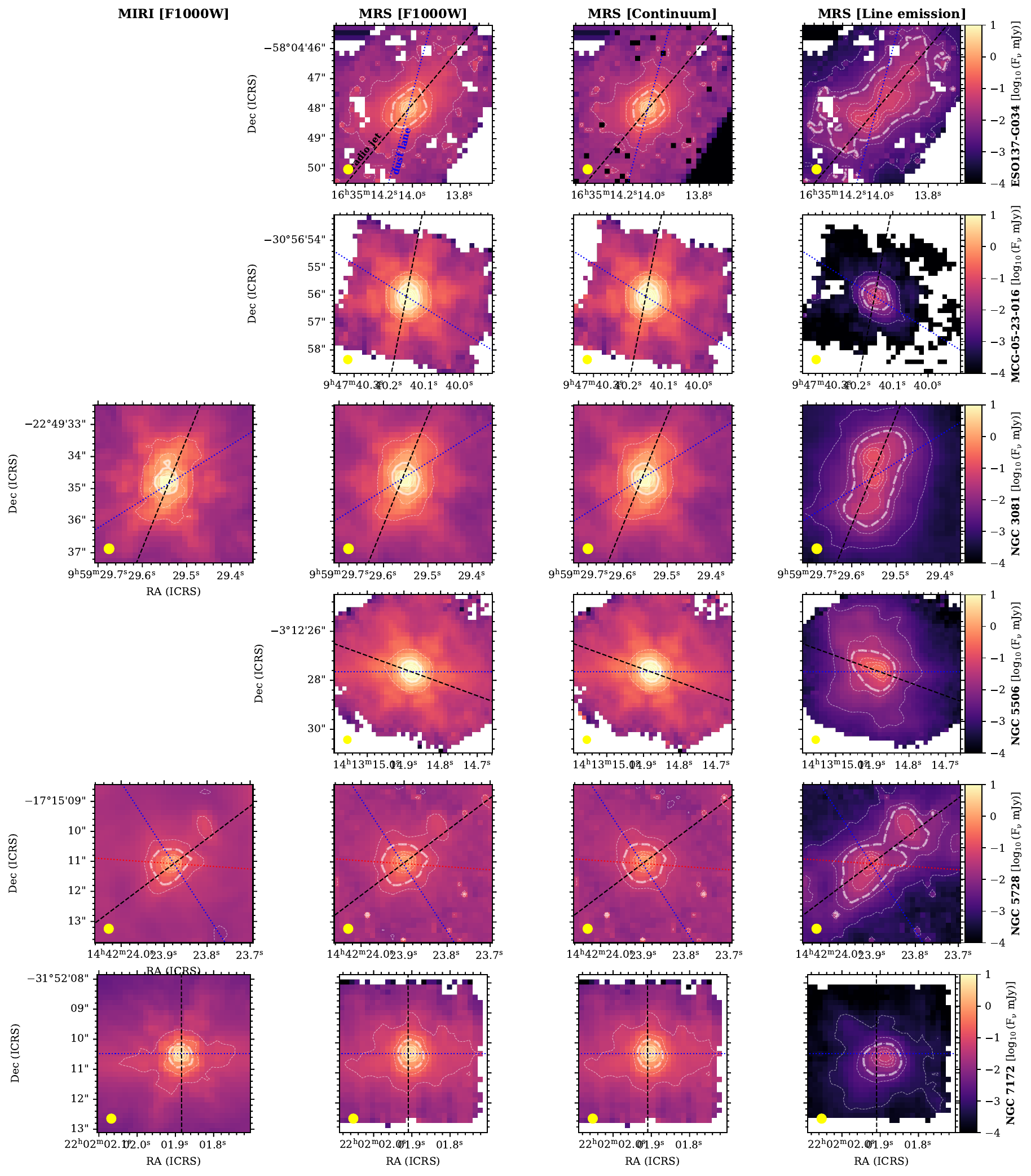}
\caption{Continuum and line emission images matching the wavelength range of the F1000W MIRI filter constructed from MRS observations. MIRI (first column), `MIRI-like' (second column), continuum (third column), and line (fourth column) images of the central $1$ kpc of ESO137-G0134 (first row), MCG-05-23-016 (second row), NGC\,3081 (third row), NGC\,5506 (fourth row), NGC\,5728 (fifth row), and NGC\,7172 (sixth row). Contours (white lines) increase in steps of $[1,5,10,30,50,70]$\% from the peak flux. The contour at a $10$\% level (thick white line) from the peak flux is shown for reference. The orientations of the radio jet (black dashed line), dust lane (blue dotted line), and stellar bar (red line; only for NGC\,5728) axes are shown. The beam (yellow circle) of the observations is shown.}
 \label{fig:fig2}
\end{figure*}

To characterize the morphology and physical properties of the extended MIR emission, we study the continuum and line emission of the central $1$ kpc of the AGN in our sample. Note that all observations have a larger FOV with MIR emission associated with star-forming rings and/or host galaxies. These features are studied in detail in several GATOS papers \citep{Davies2024,HM2024}. Campbell et al. (submitted) and Appendix \ref{sec:ExtLine} include a detailed explanation of the techniques used to extract the continuum and line emission using the MRS observations. Figure \ref{fig:figX} shows the $5.5-18~\mu$m spectra of the core ($\leq39$ pc) and the extended MIR emission without the core in an annulus of $39-450$ pc. For all objects, we use the same physical scale of $39$ pc in radius given by the lowest angular scale of the furthest object, i.e., NGC\,5728 at a distance of $39.7$ Mpc ($r = 192$ pc \arcsec$^{-1} \times 0.41\arcsec/2 \sim 39$ pc). The spectra contain multiple lines and PAH features that may contribute to the MIR emission (Campbell et al., submitted), and we point to \citet{Groves2006} for the similarities of the synthetic spectra containing both continuum and emission lines in the NLR using photoionization and shock models. For the integrated spectra of these objects, see \citet{GATOSIV}. Using the procedure described in Appendix \ref{sec:ExtLine}, we estimate the integrated continuum and line emission images in the central $1$ kpc of all galaxies in the F1000W filter (Figure \ref{fig:fig2}). 

Note the prominent and ubiquitous PAH features (e.g., 6.3, 11.3 $\mu$m) in the extended regions for most of the objects in our sample (Fig. \ref{fig:figX}). As we are interested in the dust contribution in the extended region, the PAHs are not analyzed in detail here. However, we discuss some interesting results already published on these objects for reference. We note that the strong PAH emission observed in the extended spectra of our targets, compared to the nuclear spectra, is consistent with previous works showing lower equivalent widths of all the PAH bands in AGN compared to those observed in star-forming galaxies \citep[e.g.,][]{AH2014,GB2022b}. These differences have been attributed to dilution of PAH features by the strong AGN continuum \citep[e.g.,][]{AH2014}, PAH destruction by the hard AGN radiation field \citep[e.g.,][]{Roche1991,Voit1992,Siebenmorgen2004,GB2015, RA2023}, and/or a relative lack of star formation activity in the central regions of AGN \citep[e.g.,][]{EA2018}.

The clear detection of PAH features in the extended regions of several of our targets indicates that ongoing star formation might be indeed present at circumnuclear scales. Recent GATOS JWST work has shown that PAHs can survive not only in the extended regions of AGN, but also along the outflow direction where AGN–host coupling is strong and even within the nuclear regions, although their properties are significantly altered by the harsh AGN radiation field \citep[e.g.,][]{GB2022c, GB2024b} compared to those in star-forming regions (e.g. \citealt{GB2022c, Rigopoulou2024}). In addition to the radiation field, shocks may also play an important role in modifying PAH properties in AGN (e.g. \citealt{GB2024b, Zhang2024}). Thus, GATOS plan to investigate the kinematics of PAHs in future work to better quantify this role, making use of Principal Component Analysis (PCA)-based techniques such as those presented by \citet{Donnan2024}.

\section{Physical properties of the extended MIR continuum emission} \label{sec:dis}

\subsection{MIR continuum and line emission morphologies}\label{sec:linecontribution}

\begin{figure*}[ht!]
\centering
\includegraphics[width=\textwidth]{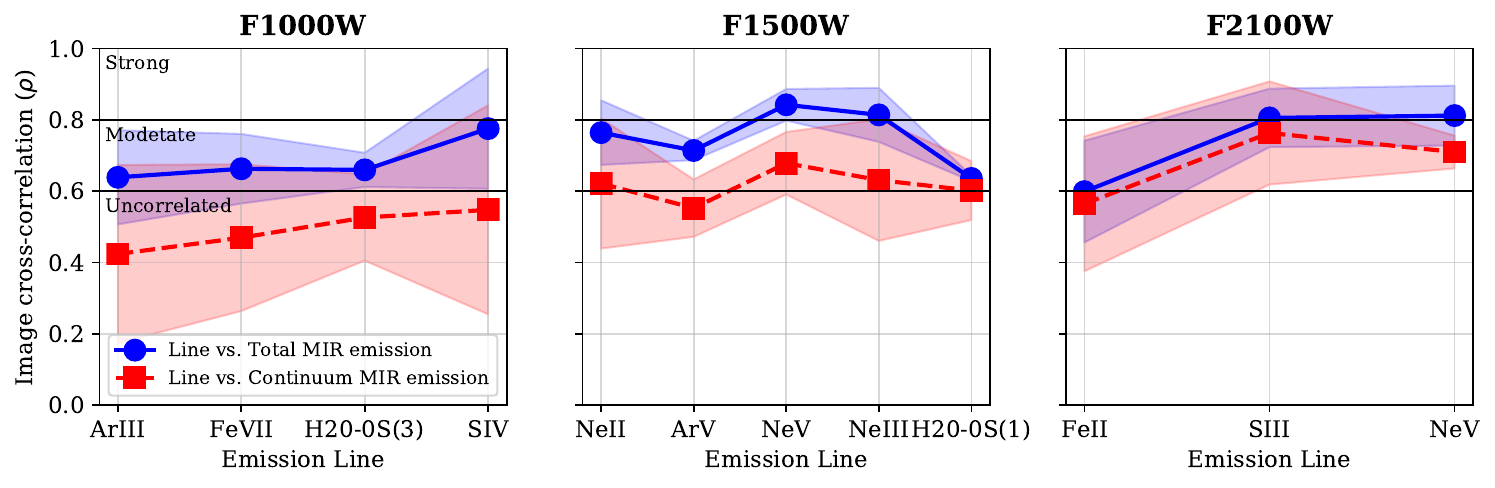}
\caption{Normalized correlation coefficients between line emission and the total MIR and continuum emission images. For all galaxies with detected extended MIR emission, we show the mean and standard deviation (shadowed regions) of the normalized correlation coefficients, $\rho$, between the line and total MIR emission (solid line), and line and continuum (dashed line) emission in the F1000W (left), F1500W (middle), and F2100W (right) filters. The ranges for strong, moderate, and uncorrelated coefficients are displayed. The correlations for each galaxy and filter are shown in Appendix \ref{app:AppC} (Fig. \ref{fig:fig1_appC}).}
 \label{fig:fig8}
\end{figure*}

If outflowing material is filled with dust, then one would expect that the spatial distribution of molecular outflows would be highly spatially correlated with the spatial distribution of continuum MIR emission. We study the correlation between morphological structures in the continuum and line emission within the MIRI filters. Figure \ref{fig:fig2} shows an example of the comparison between the integrated emission line maps and the total (i.e., continuum+line) and continuum-only MIR emission for individual galaxies in the F1000W filter. In Appendix \ref{app:AppB} (Figures \ref{fig:fig6}-\ref{fig:fig7_F2100W}), we show the visual comparison between the emission line maps and the total and continuum-only MIR emission for our galaxies and all filters. It is visually apparent that the total MIR extended emission of the images has similar morphological structures as the line emission images for ESO137-G014, NGC\,3081, NGC\,5728, and NGC\,7172. Specifically, we find the largest similarities with  [\ion{S}{4}] (Ionization potential (IP): 34.8 eV), [\ion{Ne}{5}] (IP: 97.1 eV), [\ion{Ne}{3}] (IP: 41.0 eV), and [\ion{S}{3}] (IP: 23.3 eV). These morphological similarities are less obvious when compared with the continuum-only emission images (Appendix \ref{app:AppB}: Figures  \ref{fig:fig6}-\ref{fig:fig7_F2100W}). 

To quantify the spatial correspondence between line, continuum and total emission maps, we compute the normalized correlation coefficient, $\rho$, between two images as (i.e., Pearson correlation coefficient)
\begin{equation}
    \rho = \frac{\sum_{i}^{n}\sum_{j}^{n}(I_{\rm{1},ij} - \overline{I_{\rm{1}}})(I_{\rm{2},ij} - \overline{I_{\rm{2}}})}{\sqrt{\sum_{i}^{n}\sum_{j}^{n}(I_{\rm{1,ij}} - \overline{I_{\rm{1}}})^{2}} \sqrt{\sum_{i}^{n}\sum_{j}^{n}(I_{\rm{2,ij}} - \overline{I_{\rm{2}}})^{2}}}
\end{equation}
\noindent
where $I_{\rm{1},ij}$ and $I_{\rm{2},ij}$ are the images to be compared at the $i,j$ pixel and $\overline{I_{\rm{1}}}$ and $\overline{I_{\rm{2}}}$ are the mean of the images.  The value of $\rho$ ranges from $1$ to $-1$ with $\rho = 1$ if two images are identical, $\rho =0$ if the images are uncorrelated, and $\rho=-1$ if the images are anti-correlated. In this analysis, we consider the ranges of $\rho<0.6$, $\rho = [0.6,0.8)$, and $\rho\ge0.8$ as poor (i.e., uncorrelated), moderate, and strong correlations, respectively.

Figure \ref{fig:fig8} shows the normalized correlation coefficients for the line and total MIR emission (solid lines), and line and continuum emission (dashed lines). We show the mean and standard deviation of the image cross correlation for the AGN with detected extended emission, i.e., ESO137-G014, NGC\,3081, NGC\,5728, and NGC\,7172. To avoid an overestimation of the normalized correlation coefficients, we masked the central PSF (i.e., the core) at a constant spatial scale, i.e., a radius of $39$ pc at F1000W, and $75$ pc at F1500W and F2100W. This spatial distance corresponds to a $\sim15-20$\% flux level from the peak pixel of each galaxy. In addition, to avoid an underestimation of the normalized correlation coefficients, we masked the fluxes below $1$\% of the peak flux of each image. These two conditions ensure that the correlation is performed using only the bona fide extended emission for each image. Appendix \ref{app:AppC} shows the correlation coefficients for each galaxy and emission line within each filter (Fig. \ref{fig:fig1_appC}).

Overall, we find that the total MIR emission images have a moderate correlation ($\rho\simeq0.6-0.8$) with the morphological structures of the line emission images in the $10-21~\mu$m wavelength range. This correlation becomes poor ($\rho\le0.6$) when the line emission and continuum emission images are compared in the F1000W and F1500W filters. This result indicates that the $39-450$ pc extended MIR continuum emission is uncorrelated with the extended emission lines in the $10-18~\mu$m wavelength window. Moderate correlations are found in the F2100W filter.  This correlation may be due to the low angular resolution of the F2100W observations and a diffuse and extended cold dust component from the host galaxy, making the extended emission in both line and continuum appear diffuse and smooth across the NLR (see images in Appendix \ref{app:AppB}). We describe the morphological comparison of each galaxy in detail in Appendix \ref{app:AppC}.

\subsection{Extended MIR continuum and line emission and the radio jet}\label{subsec:MIRJet}

Figure \ref{fig:fig2} shows that the line emission morphology is elongated along the radio jet axis for all galaxies. Except for MCG-05-23-016, its line emission morphology is partially resolved along the jet axis. This result may indicate that the radio jet may be triggering the gas outflow and line emission. In terms of the dust emission morphology, for most of the objects, i.e., ESO137-G034, MCG-05-23-016, NGC\,5728, and NGC\,7172, the continuum emission is perpendicular to the radio jet axis. This result may indicate that the dust is gravitationally bound in a disk/torus structure (Section \ref{subsec:windvsdisk}). The relationship between dust properties and survival with jet properties will be explored in a companion manuscript (Haidar et al., in prep.).

\subsection{Dust mass and temperature}\label{subsec:TdMd}

\begin{figure*}
\centering
\includegraphics[width=\textwidth]{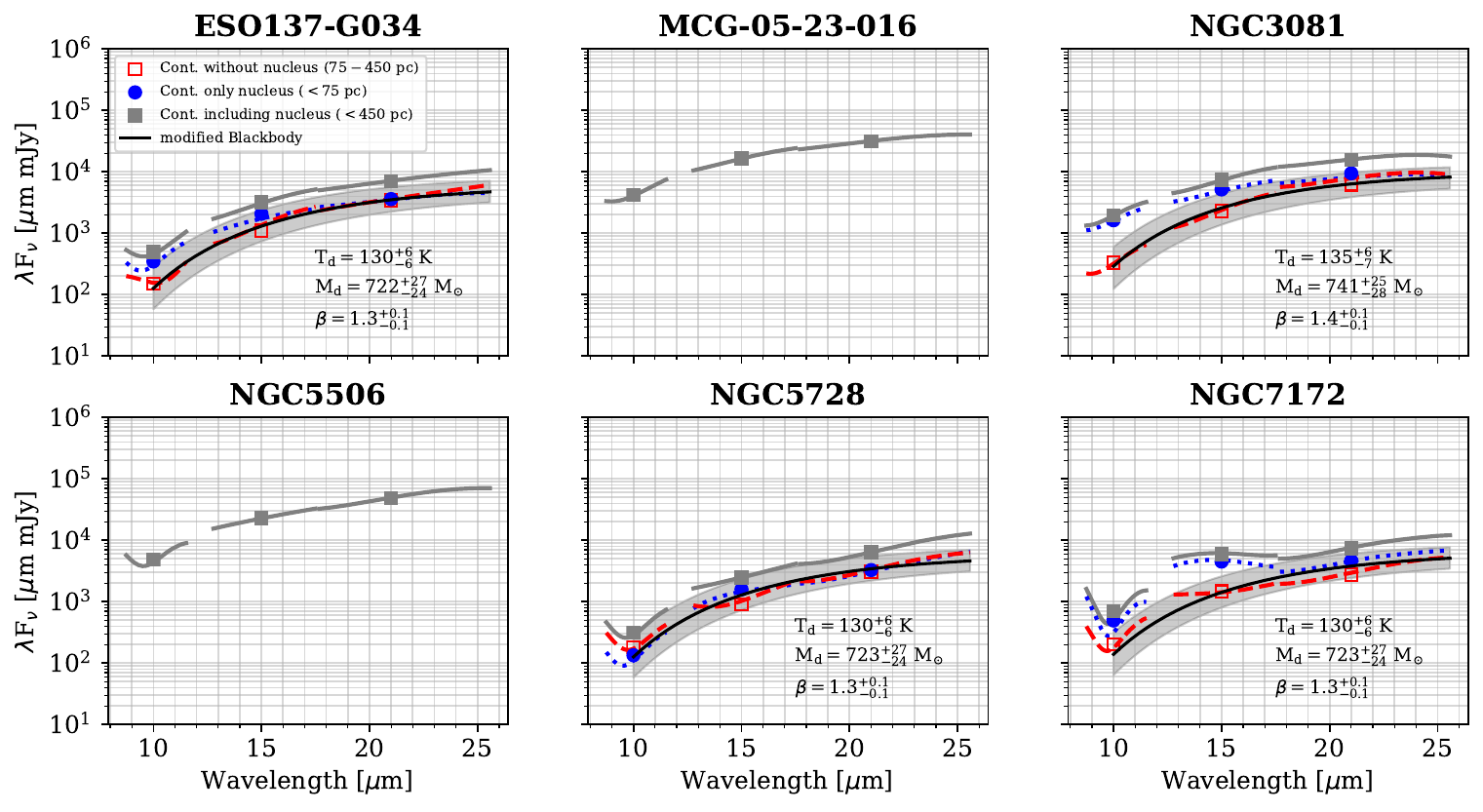}
\caption{SEDs of the continuum emission within the MIRI imaging filters of the AGN in our sample. The SEDs of the nucleus ($<75$ pc; blue dotted line with circles), the full FOV including the nucleus ($<450$pc; grey solid line with squares) and without including the nucleus ($75-450$ pc; red dashed line with open squares) are shown. The best fit of the blackbody function (black line) to the continuum extended emission in the $10-25~\mu$m wavelength range excluding the nucleus ($r<75$ pc; red dashed line) and the first interquartile ($25-75$\%; gray shadowed area) are displayed. In each panel, we show the characteristic dust temperature, $T_{\rm{d}}$, dust mass, $M_{\rm{d}}$, and emissivity index, $\beta$.}
 \label{fig:fig10}
\end{figure*}

We compute the SEDs within the three simulated MIRI imaging filters for the entire $1$ kpc FOV (including the core), for the $75-450$ pc extended continuum emission (without the core), and only for the $\le75$ pc nuclear continuum emission. To perform an analysis with a consistent spatial scale across the AGN sample, we define the core as the central $75$ pc in radius. Figure \ref{fig:fig10} shows the $9-25~\mu$m continuum SEDs.

We characterize the $9-25~\mu$m continuum SEDs using a single-temperature modified blackbody function expressed as

\begin{equation}\label{eq:mBB}
F_{\nu} (M_{\rm d},\beta,T_{\rm d}) = \frac{M_{\rm d}}{D^{2}}k_{\lambda_{0}}\left(\frac{\nu}{\nu_{0}}\right)^{\beta}B_{\nu}(T_{\rm d})
\end{equation}
\noindent
where M$_{\rm d}$ is the dust mass, $D$ is the distance to the source, $k_{\lambda_{0}}$ is the dust mass absorption coefficient of $0.29$ m$^{2}$ kg$^{-1}$ at a wavelength of $\lambda_{0} = 250~\mu$m  \citep{Wiebe2009}, $\beta$ is the dust emissive index, and $B_{\nu}(T_{\rm d})$ is the blackbody function at a characteristic dust temperature $T_{\rm d}$.

We have three free model parameters: $M_{\rm d}$, $\beta$, and $T_{\rm d}$ that we fit within the $9-25~\mu$m wavelength range using the continuum SEDs of the MIRI imaging filters F1000W, F1500W, and F2100W (red dashed lines in Fig. \ref{fig:fig10}). The free parameters are sampled within the range of $M_{\rm{d}} = [100,1000]$ M$_{\odot}$, $\beta = [1,2]$, and $T_{\rm d} = [50,200]$ K using 1,000 steps. We minimize the normalized $\chi^{2}$ to find the best-fit parameters and use the first quartile to quantify the uncertainties of each best-fit parameter. 

Figure \ref{fig:fig10} shows the best fit modified blackbody function for the galaxies (i.e., ESO137-G034, NG\,3081, NGC\,5728, and NGC\,7172) with extended continuum emission. Using these galaxies, we estimate a mean $T_{\rm{d}} = 132^{+7}_{-7}$ K, $M_{\rm{d}} = 728^{+29}_{-27}$ M$_{\odot}$, and $\beta = 1.3^{+0.1}_{-0.1}$ for the MIR extended continuum emission within an annulus of $75-450$ pc around the AGN.

\section{Observational and theoretical evidences}\label{sec:ObsEv}

We discuss several observational measurements of the extended MIR emission and put them in the context of AGN and outflow properties.

\subsection{Line emission contribution and AGN properties}\label{subsec:correlations}

\begin{figure*}
\centering
\includegraphics[width=\textwidth]{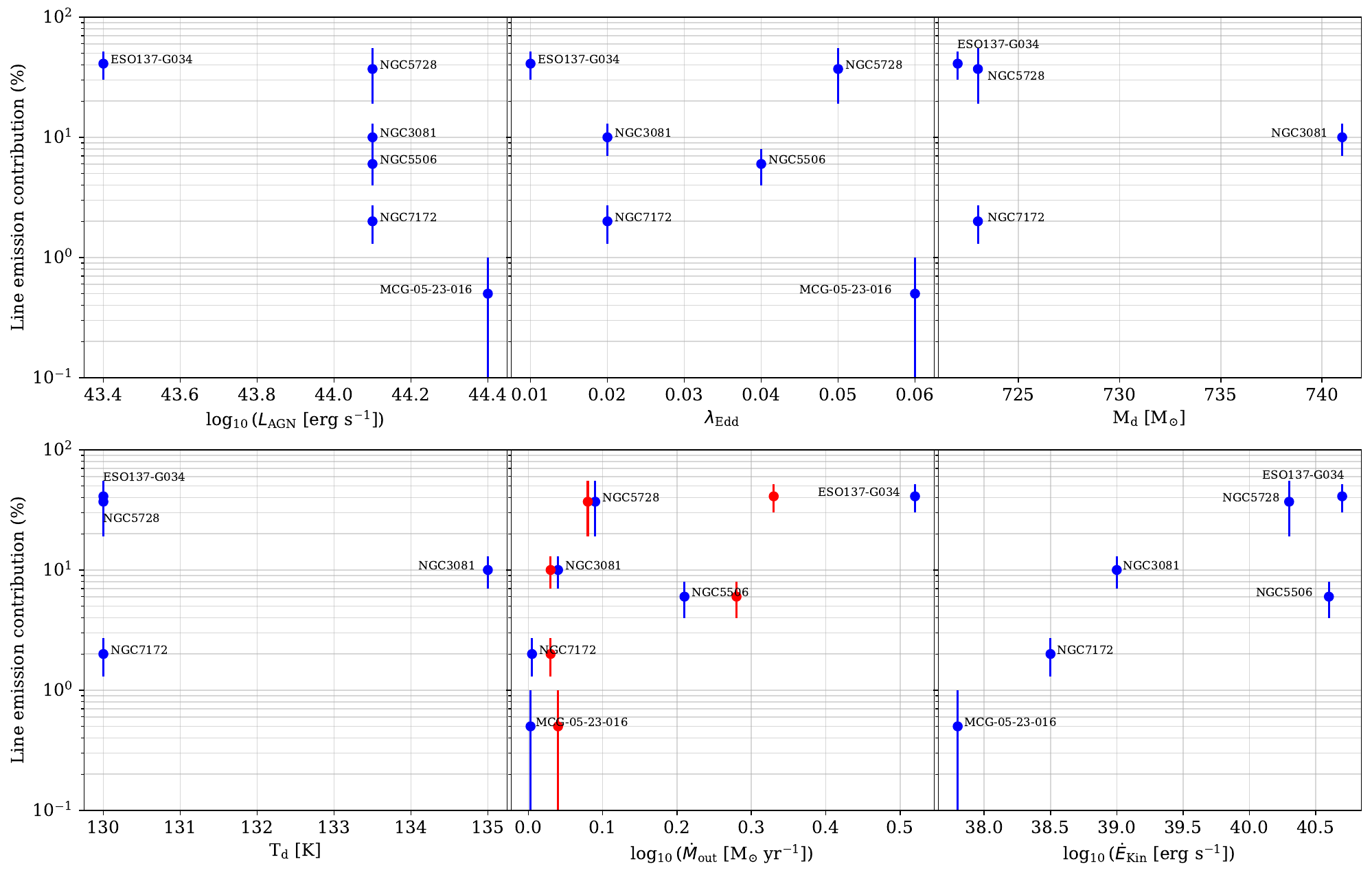}
\caption{Correlations of the line emission contribution with AGN and outflows properties. The line emission contribution from Campbell et al. (submitted) is plotted as a function of the AGN luminosity (top left), Eddington ratio (top middle), dust mass (top right), dust temperature (bottom left), mass outflow rate (bottom middle), and outflow kinetic energy (bottom right). The outflow mass rates using optical  [\ion{O}{3}]  emission (blue) and MIR [\ion{Ne}{5}] emission (red) are displayed. These values are shown in Table \ref{table:tab1}.}
 \label{fig:fig5}
\end{figure*}

One of the key results of Campbell et al. (submitted) is that the level of emission line contamination varies significantly from object to object.  Specifically, the  emission line contributions at at $10\,\mu$m within the $39-450$ pc radius across the NLR are $41\pm11$\%, $0.5\pm0.5$\%, $10\pm3$\%, $6\pm2$, $37\pm18$\%, and $2\pm0.7$\% for ESO137-G034, MCG-05-23-016 , NGC\,3081, NGC\,5506, NGC\,5728, and NGC\,7172, respectively. Here, we investigate whether the amount of emission line contribution is somehow correlated with the AGN and outflow properties listed in Table \ref{table:tab1}. Figure \ref{fig:fig5} shows the studied correlations. Given the small number of statistics, we provide tentative trends until a larger AGN sample is used.

Among the possible correlations, we find one between the level of line emission contribution and the ionized outflow kinetic energy, $\dot{E}_{\rm{kin}}$.  A weak correlation is shown between the line emission contribution and the ionized and warm mass outflow rates. We find that the AGN properties given by the bolometric luminosity, $L_{\rm{AGN}}$, and Eddington ratio, $\lambda_{\rm{Edd}}$, do not correlate with the level of line emission contribution across the NLR. Additionally, we find that the line emission contribution is not correlated with dust mass or dust temperature across the NLR. This agrees with the result that the line emission morphology is spatially correlated with the jet axis (Fig. \ref{fig:fig2}, Section \ref{subsec:MIRJet}), which also suggests that line emission contribution is not correlated with the Eddington ratio or AGN properties.

These results suggest that the ionized and warm outflow properties may be related to the level of line emission contribution across the NLR, rather than the AGN properties. As the MIR spectra have a higher contribution of emission lines in the $10~\mu$m window (Fig. \ref{fig:figX}), this results in a higher level of line emission contribution at $10\,\mu$m. This may explain the overestimation of dust emission using a single image from the interferometric observations. Additionally, the low level of dust across the NLR may also be explained by the outflow kinetic energy, which can produce the destruction of dust grains across the NLR, for example, via kinetic sputtering \citep[e.g.,][]{Jones2004}.

\subsection{The dust mass}

We estimate a fairly constant dust mass content of $728^{+29}_{-27}$ M$_{\odot}$ at a constant dust temperature of $132^{+7}_{-7}$ K, within the $75-450$ pc across the NLR for the AGN with extended MIR emission after line emission removal (Section \ref{subsec:TdMd}). In addition, we found that the line emission contribution depends on the outflow properties (Section \ref{subsec:correlations}). These results may indicate that the dust emission producing the $5.5-25\,\mu$m emission is uncorrelated with the outflow properties.

Assuming a homogeneous distribution of dust within an annulus of $75-450$ pc and a width of $75$ pc with a quarter of the volume filled with dust (i.e., NLR), we estimate a dust surface density of $\sim10^{-3}$ M$_{\odot}$ pc$^{-2}$ and a dust density of $\sim10^{-12}$ cm$^{-3}$, assuming a typical dust grain mass $m_{\rm{d}}=1.26\times10^{-24}$ g (with radius of $0.1$ $\mu$m and density $3.5$ g cm$^{-3}$ for silicates with chemical composition MgFeSiO$_{4}$, \citealt[e.g.,][]{WD2001,Zubko2004}). These results are similar to the physical conditions in the solar neighborhood ISM, $n_{\rm{d,\odot}} \sim 1.1 \times 10^{-12}$ cm$^{-3}$ and several orders of magnitude smaller than the typical particle dust density of $10$ cm$^{-3}$ for the cold ISM with T$_{\rm{d}}\sim100$ K \citep{Whittet1992,Whittet2022}. The low density brightness of the dust within the $75-450$ pc in comparison with the typical cold ISM in the Galaxy is compatible with the low $A_{\rm{V}}/N_{\rm{H}}$ ratio in AGN measured by \citet{Maiolino2001a,Maiolino2001b}. 

Our results indicate that most of the estimated dust mass may be located in several molecular clouds or shocked regions of $\le10$s pc-scales (i.e., much smaller than the beam of the observations) covering the NLR for NGC\,3081 and NGC\,5728. For these two galaxies, the MIR extended emission is very compact around the nucleus, extending along the dust lane for NGC\,5728, and along the S-shape spiral arms parallel to the jet axis for NGC\,3081. The MIR emission is more extended in ESO137-G034, which may arise from a diffuse dust screen in the NLR galaxy. The result that the dust may be located in molecular clouds, shocked regions, and/or diffuse components in the NLR indicates that the $75-450$ pc MIR emission is not a dust-filled component along the entire NLR, but rather a small-volume filling factor located in molecular clouds and shocked regions within the NLR. This result is consistent with the poor correlation between the spatial morphologies of the dust continuum and line emission tracing outflowing material in these objects (Section \ref{sec:linecontribution}).

\subsection{Dusty wind vs. Disk}\label{subsec:windvsdisk}

\begin{figure*}
\centering
\includegraphics[width=\textwidth]{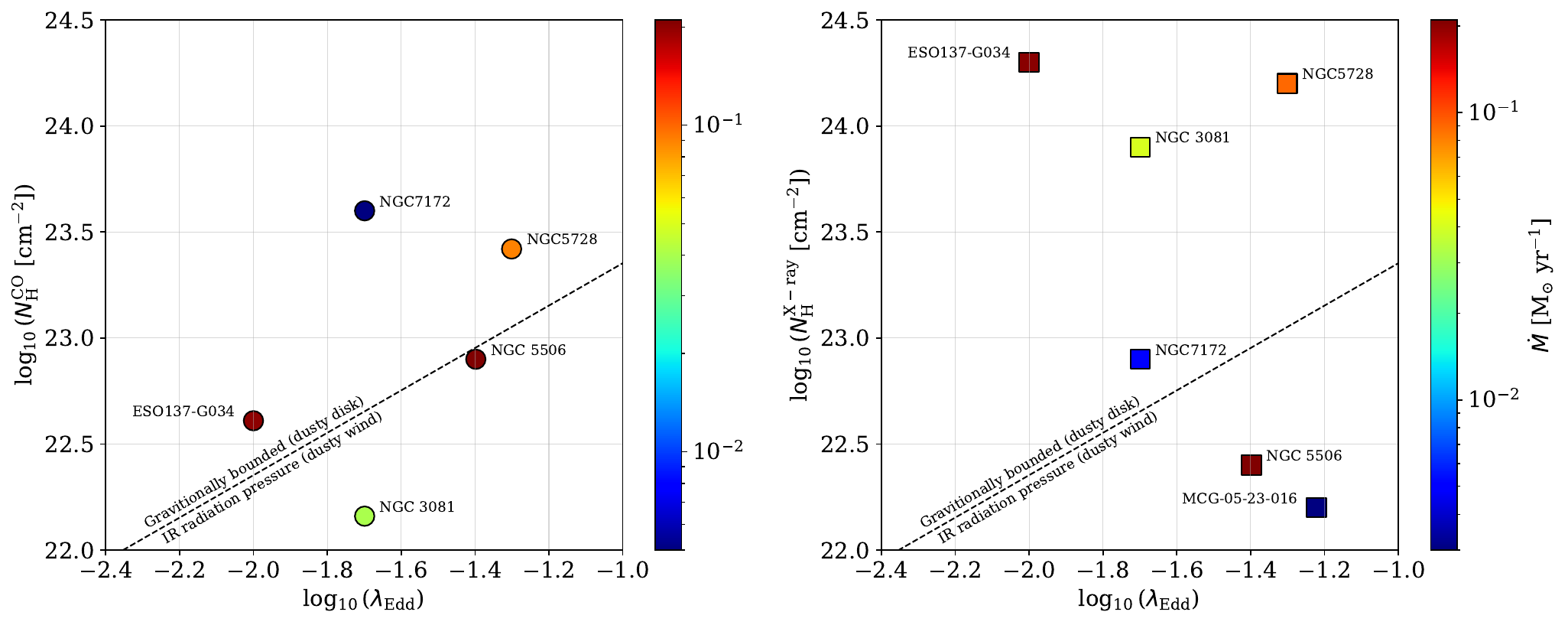}
\caption{Column density as a function of the Eddington ratio as a proxy of the AGN radiation pressure vs. gravity balance to generate dusty winds. The CO nuclear column density (left) and X-ray nuclear column density (right), the Eddington ratios, and the outflow mass ratio (colorscale) are shown in Table \ref{table:tab1}. The theoretical limit (Eq. \ref{eq:winds}) between gravitationally bounded and IR radiation pressure bounded (dashed line) is taken from \citet{Venanzi2020}.}
 \label{fig:fig11}
\end{figure*}

Several of our objects have low-surface brightness extended MIR emission within $75-450$ pc-scales (Fig. \ref{fig:fig2}) with a $T_{\rm{d}}=132^{+7}_{-7}$ K and $M_{\rm{d}} = 728^{+29}_{-27}$ M$_{\odot}$ (Section \ref{subsec:TdMd}). The dusty tori of nearby Seyfert galaxies have been measured to have a median molecular mass of $\sim6\times10^{5}$ M$_{\odot}$ and a diameter of $\sim42$ pc \citep{GATOSI}. These results may suggest that AGN radiation pressure may not be efficient to lift up dust up to $75-450$s pc-scales in these relatively low Eddington ratio objects.  This result has physical consequences to explain the required momentum to carry that mass out by only taking into account radiative pressure-driven mechanisms and other physical mechanisms, i.e., dragging effects, dust destruction mechanisms, anisotropic radiation, photoevaporation, and efficiency of the radiative-driven mechanism, may need to be included.

From a theoretical framework, three-dimensional radiation-dynamical simulations of dusty winds in AGN estimated that to observe dusty winds with clouds of N$_{\rm{H}} = 10^{23}$ and $10^{24}$ cm$^{-2}$, $\lambda_{\rm{Edd}} \leq 0.09$ and $0.15$ is required, respectively \citep{Venanzi2020}. $N_{H}$ is the column density of the cloud, and $\lambda_{\rm{Edd}}$ is the Eddington ratio of the AGN for dust-free gas. The energy balance between AGN radiation and gravity is given by 

\begin{equation}\label{eq:winds}
    \lambda_{\rm{Edd}} = \frac{2}{3}\sigma_{\rm{T}}N_{\rm{H}} = 0.044 N_{\rm{H,23}}
\end{equation}
\noindent
\citep[eq. 18 by][]{Venanzi2020}, where $\sigma_{T} = 6.6525 \times 10^{-25}$ cm$^{-2}$ is the Thomson scattering cross-section, and $N_{\rm{H,23}}$ is the normalized column density to $N_{\rm{H}}/(10^{23}$ cm$^{-2}$). 

Figure \ref{fig:fig11} shows the nuclear hydrogen column density as a function of the Eddington ratio (cf. \citealt{Fabian2008} where such diagrams were first discussed in the context of dusty gas). The nuclear hydrogen column density was estimated using CO and X-ray observations, and the Eddington ratios are shown in Table \ref{table:tab1}. The X-ray observations measure the column density along the LOS for a given inclination, which may be different to the column density of the molecular gas.  A comparison of both X-ray and CO column densities using $45$ Seyfert galaxies from the GATOS sample \citep{AH2021,GATOSI,GB2024} showed that galaxies with high X-ray obscuration, $\log_{10} (N_{\rm{H}}^{\rm{X-ray}} [\rm{cm}^{-2}]) >22$, lie below the 1:1 relation. Thus, our galaxy sample with $\log_{10} (N_{\rm{H}}^{\rm{X-ray}} [\rm{cm}^{-2}]) >22.2$ may be considered as lower-limits of the CO column density. However, this is not always the case. The CO column densities may be $1-2$ orders lower than the X-ray column density (Fig. \ref{fig:fig11}). The CO column densities were estimated using ALMA CO(2-1) within apertures of $\sim22-93$ pc. The differences between X-ray and CO column densities may be caused by several factors. For example, a) the difference of the gas and dust traced along the LOS, b) the size, absorption and number of clumps traced by the X-rays, and c) a deficit of molecular gas in the core of the galaxy. The deficit of nuclear molecular gas may underestimate the column density of the molecular gas lifted by the AGN but it also may imply that this molecular gas has already been lifted up by the AGN or cleared up by some ejection event from the AGN. 

We find that most objects in our sample lie within the gravitationally bounded region. This result implies that the molecular gas and dust are mainly located on an optically and geometrically thick disk in which the AGN radiation pressure may not be enough to lift up the material to generate dusty winds. This result is unclear for NGC\,3081, in which X-ray to CO column densities vary by $1.79$dex, potentially by a deficit of CO in the core \citep{Ramakrishnan2019}. For MCG-05-23-016, it was not possible to estimate the CO column density due to low-quality molecular gas data.

Recent results using $>40$ Seyfert galaxies show that most of the objects with $\log_{10} (N_{\rm{H}}^{\rm{X-ray}} [\rm{cm}^{-2}]) >22$ lie on the gravitationally bounded region (i.e., dusty disk) \citep{GB2022, GB2024}. For these objects, the IR SEDs were better explained by dusty torus models torus models \citep{Nenkova2008a,Nenkova2008b,GM2023} rather than dusty+wind models \citep{Hoenig2017}. In addition, within the sample of $12$ Seyfert galaxies observed by GATOS, $>60$\% of the objects lie on the gravitational bounded region \citep{AH2021, GB2024}. Many objects show upper limits on their column density derived from CO data \citep{GB2024}, indicating a deficit of molecular gas within the central $<50$ pc, which locates them at the boundary of the gravitationally bounded vs. IR polar outflows with a $1\sigma$ statistical significance. From these objects, NGC\,3227, NGC\,1365, and NGC\,4388 are in the dusty wind region within $1\sigma$ statistical significance due to the large uncertainties of the Eddington ratio. Even though these studies favor the gravitational bounded regime, the large uncertainties of these measurements make it unclear if the $N_{\rm{H}}-\lambda_{\rm{Edd}}$ plot can reliably distinguish the dusty wind scenario. For the column density, the molecular gas may have already been cleared from the nuclear region due to a previous ejection event or inefficient accretion flows, which may produce a false positive, favoring the dusty wind scenario. For the Eddington ratio, the black hole mass can have an uncertainty of several orders of magnitude, resulting in many objects falling within the $1\sigma$ uncertainty of the boundary between the two regimes. Overall, the presence of a polar dusty outflow in these sources can be ruled out at distances of $>40$ pc from the AGN; however, our findings indicate that if this component is present, it is very compact, with a diameter of $<40$ pc (as sampled here) in Seyfert galaxies.

\subsection{The 9.7\,$\mu$m silicate feature}

The dusty winds are thought to be optically thin and have a large covering factor along the polar direction at radius $\sim100$ pc \citep[e.g.,][]{Asmus2016,Asmus2019}. Under this condition, the dusty winds should not show a $9.7\,\mu$m silicate feature in absorption. According to dusty wind models \citep{Hoenig2017}, these dusty winds must be silicate-poor, as this dust is launched from the inner region of the torus, where silicates should be sublimated. We extract the spectra of the extended MIR emission without the core ($39-450$ pc) for the AGN in our sample (Figure \ref{fig:figX}) and estimate the silicate feature $\tau_{\rm{9.7\mu m}}=-\ln(F_{9.7\mu m}/F_{\rm{c}})$, where $F_{9.7\mu m}$ and $F_{c}$ are the observed and interpolated continuum fluxes at $9.7~\mu$m. We interpolated the continuum over the silicate feature within the  $7.8-13~\mu$m wavelength range. We find that all objects have a silicate feature in absorption in the range of $\tau_{9.7\mu m} = 0.32-1.85$. The $7.7\,\mu$m PAH feature affects the determination of the $9.7\,\mu$m silicate feature very challenging. Specifically, this feature in the extended spectra causes an overestimation of the silicate features. Here, we are interested in the $9.7\,\mu$m silicate feature itself rather than in an accurate estimation of it. The result of measuring the $9.7\,\mu$m silicate feature is already a result to discuss in the context of the polar winds. We note that NGC\,3081 and MCG-05-23-016 have the shallowest feature, and in combination with the $7.7\,\mu$m PAH feature, makes it a tentative detection. The MIR spectra of NGC\,3081 using CanariCam on the 10.4-m Gran Telescopio Canarias \citep{GM2013} and of MCG-05-23-016 using MRS/JWST \citep{EA2025} show a shallow $9.7\,\mu$m silicate feature confirming the detection. This result indicates that the observed extended continuum emission arises from a dust component in the host galaxy for ESO137-G034 and NGC\,5728, and by a dust lane for NGC\,3081 and NGC\,7172. This silicate feature is present in the extended emission for the point-like dominated sources, MCG-05-23-016 and NGC\,5506, due to the extended emission of the central PSF. For MCG-05-23-016 and NGC\,5506, the core and extended SEDs show different shapes at the longest wavelength, which indicates that there may be an extended cold dust component. This component may be the extended emission along the dust lane direction observed at $11.2~\mu$m narrow-band image, which is also observed by \citet{GB2016}. The detection of silicate absorption rules out that the dust emission observed in several of our objects arises from dusty winds in the NLR.

\subsection{IR polarization}

If the dusty winds are optically thin, this is an excellent condition to obtain the polarization signature of aligned dust grains at IR and sub-mm wavelengths. If dust grains are driven away by the AGN radiation pressure, the long axis of the grains will align along the direction of the AGN radiation. This dust alignment mechanism is known as the `k-RAT' alignment \citep{LH2007}. In case k-RAT is important in AGN,  the observational signature by polarization absorption will produce an azimuthal pattern (E-vector) with the position angle of polarization perpendicular to the radial direction to the AGN along the dusty wind with polarization levels of few \% in the $\sim1-10~\mu$m wavelength range. The observational signature due to polarized thermal emission will be a radial pattern (E-vector) pointing to the AGN along the dusty wind at levels of few \% at wavelengths $\sim10-1000~\mu$m. However, there is no observational evidence of the k-RAT alignment mechanism in any astrophysical object (including SF regions, AGB stars) \citep[e.g.,][]{Tazaki2017,Hull2022,LG2023}, favoring that dust grains are aligned by magnetic fields (B-fields) in the ISM (i.e., `B-RAT' alignment mechanism). In case B-RAT is important, the long axis of the dust grains will align perpendicular to the local B-field orientation. Thus, the polarization angle due to B-RAT is perpendicular to that from k-RAT.

The $1-5~\mu$m  wavelength regime is dominated by dust scattering in the NLR or magnetically aligned dust grains parallel to the dust lane of the galaxy \citep{Brindle1990a,Brindle1990b,Brindle1990c,Packham1997,Simpson2002,Watanabe2003,ELR2015,ELR2017}. MIR ($7-12~\mu$m ) polarimetric observations show a) unpolarized cores in all radio-quiet AGN, and b) scattering or magnetically aligned dust grains in the NLR as dominant polarization mechanisms \citep{ELR2018}. These observations were performed using CanariCam on the 10.4-m Gran Telescopio CANARIAS (GTC) with an angular resolution of $0.3\arcsec$. These observations showed extended IR polarization at scales of $\sim25-90$ pc along the NLR in NGC\,1068 and NGC\,4151 \citep{ELR2016,ELR2018}. For NGC\,4151, the PA of polarization is roughly perpendicular to the radio jet axis with polarization levels of $\sim1$\%. For NGC\,1068, the PA of polarization angle was measured to be uniform at $\sim44^{\circ}$ and with high polarization levels, $\sim7$\%, at $\sim10-40$ pc north from the AGN. For both objects, the polarization extends beyond the NLR, indicating that dichroic absorption by magnetically aligned dust grains in the host galaxy is the main polarization mechanism. These results suggested that an extended dust component irradiated by the AGN re-radiating at $7-15~\mu$m and magnetically aligned by the local B-field in the galaxy is present. IR ($1-12~\mu$m) polarization rules out the presence of dust associated with dusty winds in the NLR at $10-100$ pc-scales.

At sub-mm wavelengths, $860~\mu$m ALMA polarization observations with a resolution of $0.07\arcsec$ ($4.2$ pc) resolved the polarized thermal emission of the dusty torus of NGC\,1068 \citep{ELR2020}. The polarization is dominated by magnetically aligned dust grains with a B-field orientation parallel, $109\pm2^{\circ}$, to the equatorial axis of the torus and with a polarization level of $3.7\pm0.5$\%. These observations also found a highly polarized, $\ge5$\%, component with a position angle of $\sim0^{\circ}$ East of North (E-vector) extending up to $\sim30$ pc North above the dusty torus. The high polarization and the spectral index of this component \citep{GB2019} indicated that the polarization is due to synchrotron polarized emission with a toroidal B-field perpendicular to the radio jet axis. Thus, resolved thermal polarized emission from dust grains at pc-scales has not found the signature of dusty winds, neither by magnetically nor radiatively aligned dust grains. Further, emission line polarization associated with molecular outflows is required to characterize the correlation between velocity fields and magnetic fields in outflows.

\section{Conclusions} \label{sec:concusions}

We have presented an analysis of the physical properties of the MIR extended emission in a sample of 6 Seyfert galaxies using JWST/MIR MRS IFU observations from the GATOS survey. These galaxies are nearby, $35.4\pm4.6$ Mpc, have similar luminosities, $\log_{10} (L_{\rm{bol}}~[\rm{erg\,s}^{-1}]) = 44.0\pm0.3$, and a wide range of outflow rates, $\dot{M}_{\rm{out}}^{\rm{Opt}} = 0.003-0.52$ M$_{\odot}$ yr$^{-1}$, column densities, $\log_{10} (N_{\rm{H}}^{\rm{X-ray}} [\rm{cm^{-2}}]) = 22.2-24.3$, and Eddington ratios, $\lambda_{\rm{Edd}} = 0.005-0.06$. These galaxies were selected from a Cycle 1 JWST GO Program curated with objects that showed prior extended MIR emission using IR interferometric and single-dish telescope observations. In our companion manuscript, Campbell et al. (submitted) estimated that the extended MIR emission is highly contaminated by line emission contribution ranging from a few percent up to $60$\% across the NLR. Building on these results, we quantified the physical properties of the MIR extended emission. Our main results are:

\begin{itemize}
\item Our morphological comparison, using the Pearson correlation coefficient, showed that the continuum images (i.e., line emission subtracted) are weakly correlated ($\rho<0.6$) with the emission line images within the $39-450$ pc radius. These results suggest that the dust continuum emission is not spatially correlated with the distribution of line emission from the warm gas associated with outflows in the NLR. 

\item We compute the SEDs of the extended MIR continuum emission within the $75-450$ pc in radius and fitted a modified blackbody function. We estimate that the extended MIR continuum component is characterized with $T_{\rm{d}} = 132^{+7}_{-7}$ K, $M_{\rm{d}} = 728^{+29}_{-27}$ M$_{\odot}$, and $\beta=1.3^{+0.1}_{-0.1}$. We also show that all SEDs show silicate features.

\item We find that the line emission morphology is elongated along the radio jet axis, while the $10\,\mu$m continuum dust emission is preferentially oriented perpendicular to the radio axis.

\item We find a correlation between the level of line emission contribution and the outflow kinetic energy. A weak correlation is shown between the line emission contribution and the mass outflow rate, $\dot{M}_{\rm{out}}$. We find that the AGN properties given by the bolometric luminosity, $L_{\rm{AGN}}$, and Eddington ratio, $\lambda_{\rm{Edd}}$, do not correlate with the level of line emission contribution across the NLR. These results indicate that the outflow properties define the level of line emission contribution across the NLR rather than the AGN properties.

\end{itemize}

We argued that the weakly correlation between line and continuum emission morphologies, the correlation between line emission and outflow kinetic energy, and the additional observed silicate features, dust polarization of the MIR continuum component are incompatible with dust to be directly associated with dusty winds in the $39-450$ pc radius across the NLR. In addition, our objects favour the gravitational bound regime (i.e., a dusty disk) proposed by three-dimensional radiation-dynamical simulations of dusty winds in AGN.

Our results put constraints on the extension of the MIR emission associated with dusty winds to be at scales $\le20$ pc, which coincides with the median diameter, $\sim42$ pc, of the molecular and dusty torus \citep{GATOSI}. Note that the bulk of the dust mass is better traced at sub-mm wavelengths \citep{ELR2018b,AH2021,Nikutta2021a,Nikutta2021b}. This locates the potential dusty wind in the funnel of the torus width \citep[e.g.,][]{GamezRosas2022}.  To physically characterize a dusty component and associate it with a wind, high-angular resolution ($<0.1"$) and $1-1000~\mu$m IFU observations are required, which is beyond the capabilities of the JWST except for a few nearby, $<10$ Mpc, AGN. Alternatively, the use of the Aperture Masking Interferometer (AMI) on the JWST is a very promising observing mode, which can provide $1-10$ pc scale resolution in nearby AGN \citep{LR2025}.

\textbf{Our observations suggest that} the extended $39-450$ MIR emission arises from dust in molecular clouds or shocked regions across the NLR. In addition, \textbf{we estimated that} the extended MIR emission component has a fairly constant dust temperature and mass, and it is uncorrelated with AGN and outflow properties. This result favors the idea that the dust component is not spatially correlated with the warm outflows in the NLR. However, a statistical sample spanning a wide range of AGN and outflow properties is required to confirm these results. This work calls for a reassessment of many observational and theoretical studies of the extended IR emission in AGN.


\begin{acknowledgments}
This work is based on observations made with the NASA/ESA/CSA James Webb Space Telescope. The data were obtained from the Mikulski Archive for Space Telescopes at the Space Telescope Science Institute, which is operated by the Association of Universities for Research in Astronomy, Inc., under NASA contract NAS 5-03127 for JWST. These observations are associated with programs \#1670 and \#2064.  
\textbf{The specific observations analyzed can be accessed via \dataset[doi: 10.17909/wks7-4j90]{https://doi.org/10.17909/wks7-4j90}.}
Support for programs \#1670 and \#2064 were provided by NASA through a grant from the Space Telescope Science Institute, which is operated by the Association of Universities for Research in Astronomy, Inc., under NASA contract NAS 5-03127.
E.L.-R. acknowledges support from the NASA Astrophysics Decadal Survey Precursor Science (ADSPS) Program (NNH22ZDA001N-ADSPS) with ID 22-ADSPS22-0009 and agreement number 80NSSC23K1585.
E.L.-R. acknowledges support for Program numbers \#4611and \#5017 was provided through a grant from the STScI under NASA contract NAS5-03127.
This research has made use of the Spanish Virtual Observatory (https://svo.cab.inta-csic.es) project funded by MCIN/AEI/10.13039/501100011033/ through grant PID2020-112949GB-I00.
S.G.B.  acknowledges support from the Spanish grant PID2022-138560NB-I00, funded by
MCIN/AEI/10.13039/501100011033/FEDER, EU
IGB is supported by the Programa Atracción de Talento Investigador ``C\'esar Nombela'' via grant 2023-T1/TEC-29030 funded by
the Community of Madrid.
A.A.H. acknowledges support from grant PID2021-124665NB-I00 funded by the Spanish Ministry of Science and Innovation and the State Agency of Research MCIN/AEI/10.13039/501100011033 and ERDF A way of making Europe.
D.E.A. is supported by the ``Becas Estancia Postdoctorales por M\'exico'' EPM(1) 2024 (CVU:592884) program of SECIHTI.
MPS acknowledges support under grants RYC2021-033094-I, CNS2023-145506 and PID2023-146667NB-I00 funded by MCIN/AEI/10.13039/501100011033 and the European Union NextGenerationEU/PRTR/
\end{acknowledgments}

%

\vspace{5mm}
\facilities{
    JWST (MIRI, MRS),
    SVO \citep{SVO2012,SVO2020}}


\software{
    astropy \citep{astropy:2022,astropy:2018,astropy:2013},
    aplpy \citep{aplpy2012,aplpy2019},
    matplotlib \citep{matplotlib2007},
    numpy \citep{numpy2020},
    pandas \citep{mckinney-proc-scipy-2010,reback2020pandas},
     scipy \citep{2020SciPy-NMeth}.}



\appendix

\section{Data reduction} \label{sec:RED}

JWST Program \#1670 uses MRS IFU spectroscopy in four wavelength channels: ch1 ($4.9-7.65$ $\mu$m), ch2 ($7.51-11.71$ $\mu$m), ch3 ($11.55-18.02$ $\mu$m), and ch4 ($17.71-28.1$ $\mu$m) with a spectral resolution of $R\sim3700-1300$ \citep{Labiano2021}. The pixel sizes are $0.196\arcsec$, $0.196\arcsec$, $0.245\arcsec$, and $0.273\arcsec$ for ch1, ch2, ch3, and ch4, respectively. The Full-Width-at-Half-Maximums (FWHMs) are $0.35\arcsec$, $0.41\arcsec$, $0.6\arcsec$, and $0.9\arcsec$ in ch1, ch2, ch3, and ch4, thus the Point-Spread-Functions (PSFs) of MRS are undersampled from a factor of $\sim2$ at $\sim5$ $\mu$m to a factor of $\sim1.1$ at $25$ $\mu$m \citep[see fig. 12 by][]{Argyriou2023}. The FOVs are $3.2\arcsec \times3.2\arcsec$, $4.0\arcsec\times4.7\arcsec$, $5.2\arcsec\times6.1\arcsec$, and $6.6\arcsec\times7.6\arcsec$ for ch1, ch2, ch3, and ch4, respectively. We take a flux calibration uncertainty of $5.6$\% for the MRS observations \citep{Argyriou2023}. Further details about this mode can be found at \citet{Argyriou2023} and \citet{Rigby2023}. 

We processed our MIRI MRS observations using the JWST Calibration pipeline (version 1.11.4) and the calibration context 1130. We followed the standard MRS pipeline procedure by \citet{Labiano2021} and the same configuration of the pipeline stages described in \cite{GB2022b,PereiraSantaella2022,GATOSIII} to reduce the data. Some hot and cold pixels are not identified by the current pipeline, so we added some extra steps as described in \citet{GATOSIII}. The data reduction is described in detail in \citet{GB2022b,GATOSIII,PereiraSantaella2022}. For each MRS channel, the total exposure (i.e., on-source) times are $2293$s, $1121$s, $4587$s, $2198$s, $4205$s, and $1110$s for  ESO137-G034, MCG-05-23-016, NGC\,3081, NGC\,5506, NGC\,5728, NGC\,7172, respectively.

JWST Program \#2064 uses MIRI imaging in three broadband filters: F1000W ($\lambda_{\rm{c}} = 10.0$ $\mu$m, $\Delta\lambda = 1.80$ $\mu$m), F1500W ($\lambda_{\rm{c}} = 15.0$ $\mu$m, $\Delta\lambda = 2.92$ $\mu$m), and F2100W ($\lambda_{\rm{c}} = 21.0$ $\mu$m, $\Delta\lambda = 4.58$ $\mu$m). The pixel size is $0.11\arcsec$ and the FOV of the SUB256 array is $28.2\arcsec\times18.2\arcsec$. The  FWHMs are  $0.33\arcsec$, $0.49\arcsec$, $0.67\arcsec$ for F1000W, F1500W, F2100W, respectively. Further details about MIRI can be found at \citet{Rigby2023} and \citet{Libralato2023}. 

For all MIRI imaging observations, the SUB256 array and the FASTR1 readout pattern were used to avoid saturation with 50 integrations, 5 groups, and 1 frame per observation within a 4-points dither pattern. For the F1500W, and F2100W bands, background field observations were taken with the same instrumental configuration and 2-point dither to remove the thermal background. The data were reduced using the standard JWST pipeline python package v10.2 with CRDS reference files \textsc{jwst\_1097.pmap}. All objects have a total exposure (on-source) time of $358$s at each MIRI filter. We take a flux calibration uncertainty\footnote{JWST MIRI photometric calibration: \url{https://jwst-docs.stsci.edu/jwst-data-calibration-considerations/jwst-calibration-uncertainties\#JWSTCalibrationUncertainties-MIRI}} of $3$\% for the MIRI imaging observations.

\section{Extraction of the MIR extended emission using MRS IFU data} \label{sec:ExtLine}

To estimate the contribution of the continuum and line emission across the extended MIR emission, we obtain the emission lines using the MRS observations within the MIRI imaging filters. As an example of this procedure, Figure \ref{fig:fig0} illustrates the steps to compute the emission line and continuum images using NGC\,5728 in the F1000W filter. We perform the following steps:

\begin{figure*}
\centering
\includegraphics[width=\textwidth]{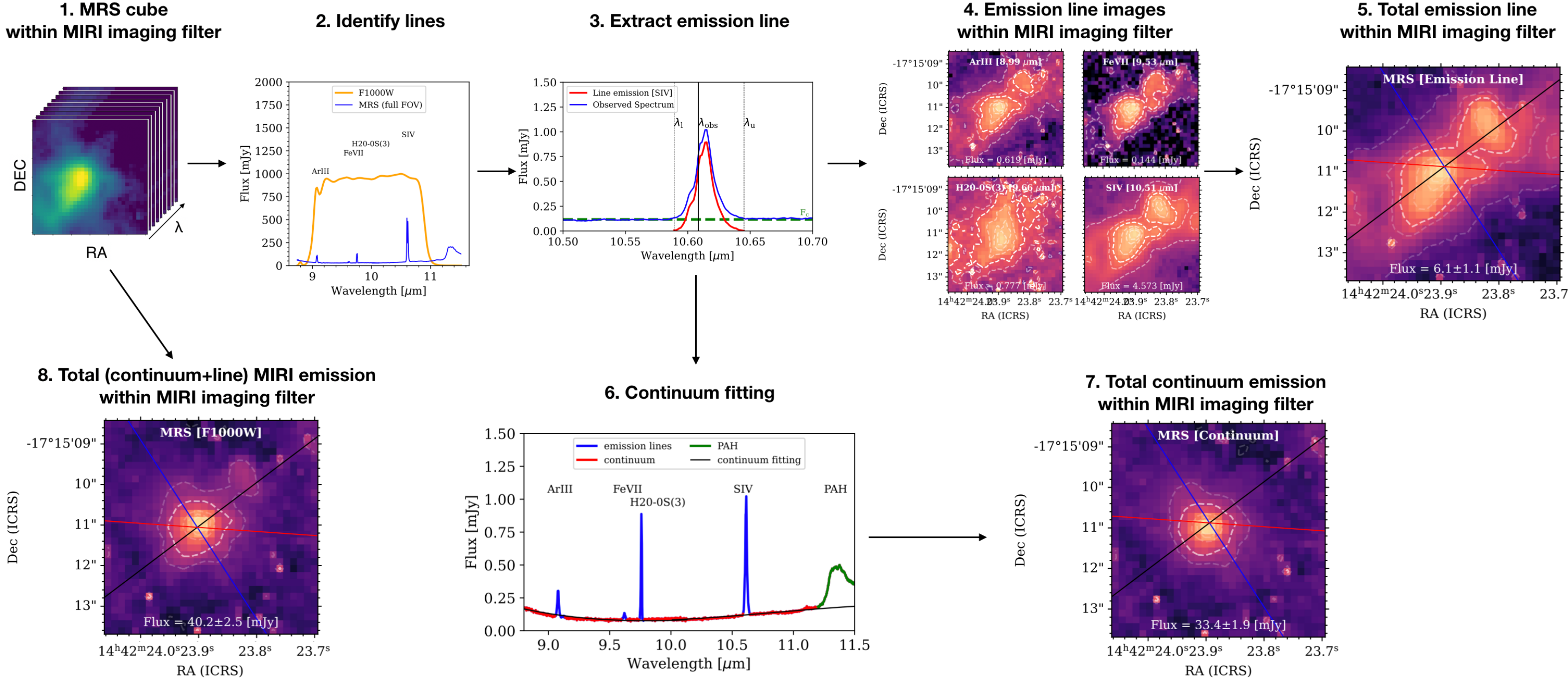}
\caption{Steps to extract the emission lines from the MRS observations and to compute the line emission, line-free, and `MIRI-like' (continuum+line) images within a MIRI imaging filter. As an example, we show the MRS observations of NGC\,5728 and the F1000W filter.
\textbf{Step 1:} Select MRS subchannels within the wavelength range of a given MIRI imaging filter. 
\textbf{Step 2:} Identify emission lines within the wavelength range of a MIRI imaging filter. The F1000W throughput (orange line), the integrated spectrum (blue line) within the full datacube, and the identified emission lines are shown.
\textbf{Step 3:} Extract the emission line per spaxel. The [\ion{S}{4}] emission line (blue line), the wavelength at the peak flux, $\lambda_{\rm{obs}}$, and the lower, $\lambda_{\rm{l}}$, and upper, $\lambda_{\rm{u}}$, wavelengths at which the emission line flux satisfies $ F_{\lambda_{\rm{l,u}}} \le F_{\rm{c}}+1\sigma_{\rm{c}}$ (green dashed line), are shown. The continuum-subtracted line (red solid line) is shown.
\textbf{Step 4:} Compute emission line images. The integrated emission lines (e.g., [\ion{Ar}{3}], [\ion{Fe}{7}], H$_{2}$0-0S(3), and [\ion{S}{4}]) identified in the F1000W filter of the central $1$ kpc of NGC\,5728 are shown.  
\textbf{Step 5:} Compute the total contribution of the emission lines within the MIRI filter. 
\textbf{Step 6:} Fit the continuum emission. The spectrum of a single spaxel with the third-order spline curve (black line) fitted to the continuum emission (red line) with masked emission lines (blue line) and broad lines (e.g., PAH at  $11.3$ $\mu$m; green line) is shown. 
\textbf{Step 7:} Compute the total contribution of the continuum emission within the MIRI imaging filter. 
\textbf{Step 8:} Compute the total (continuum+line) MIR emission within the MIRI imaging filter.
For all images, the filter throughput was multiplied by the MRS cube of the line and continuum emission. Contours (white lines) increase in steps of [1, 5, 10, 30, 50, 70]\% from the peak flux. The contour at a 10\% level (thick white line) from the peak flux is shown for reference. The radio axis (black line), dust lane (blue line), and stellar bar (red line) are shown.
}
 \label{fig:fig0}
\end{figure*}

\begin{enumerate}[leftmargin=*]
    \item We select the MRS subchannels within the wavelength range of a given MIRI imaging filter. The wavelength range of a MIRI imaging filter is given by its throughput that we take from the Spanish Virtual Observatory\footnote{JWST filter throughputs were obtained from the SVO at \url{http://svo2.cab.inta-csic.es/svo/theory/fps3/index.php?mode=browse&gname=JWST&asttype=}} \citep[SVO;][]{SVO2012,SVO2020}. We use the full wavelength range of the filter down to zero throughout response in this analysis. We found flux jumps between subchannels at spaxels with high intensity within the FOV. These flux jumps may be due to features in the flatfield and/or background subtraction. To correct the flux jumps, we scale adjacent subchannels such as the subchannel with higher flux is equal to the subchannel with lower flux. We select this approach because it minimizes morphological artifacts due to overestimated fluxes in regions with steep flux variations (e.g., spaxels surrounding the core). 
    
    \item We identify the emission lines in the MRS observations within the wavelength range of the MIRI imaging filter. To optimize the emission line identification, we maximize the signal-to-noise ratio (SNR) of the emission lines by computing the total spectra of the full FOV of the MRS cube. We identify the most common emission lines present in AGN reported by  \citet{Satyapal2021} and using the atomic line list for atoms and molecules from the Infrared Space Observatory (ISO) Spectrometer Data Center\footnote{The atomic line list v2.04 can be found at \url{https://www.mpe.mpg.de/ir/ISO/linelists/index.html}}. We identify the following most prominent emission lines in the MIRI imaging filters: 

    \begin{itemize}
        \item F1000W: [\ion{Ar}{3}] [$8.9914$ $\mu$m], [\ion{Fe}{7}] [$9.5267$ $\mu$m], H$_{2}$0-0S(3) [$9.6649$ $\mu$m], and [\ion{S}{4}] [$10.5105$ $\mu$m].
        
        \item F1500W: [\ion{Ne}{2}] [$12.8136$ $\mu$m], [\ion{Ar}{5}] [$13.1022$ $\mu$m], [\ion{Ne}{5}] [$14.3217$ $\mu$m], [\ion{Ne}{3}] [$15.551$ $\mu$m], and H$_{2}$0-0S(1) [$17.0348$ $\mu$m].
        
        \item F2100W: [\ion{Fe}{2}] [$17.9360$ $\mu$m], [\ion{S}{3}] [$18.7130$ $\mu$m], and [\ion{Ne}{5}] [$24.3175$ $\mu$m]
    \end{itemize}
    
    \item We extract the emission lines spaxel per spaxel. Note that emission lines can be composed of multiple components, however a multi component analysis of these lines will be the subject study of upcoming GATOS papers. We identify a line-free region using the total spectrum computed in step 2. Specifically, we use the line-free regions of $\Delta\lambda =$ $[9.3,9.5]$, $[14.0,14.2]$, and $[21.0,22.0]$ $\mu$m for F1000W, F1500W, and F2100W, respectively. For each spaxel, we estimate the standard deviation of the continuum, $\sigma_{\rm{c}}$, within the $\Delta\lambda$ wavelength range. Then, we find the wavelength at the location of the peak line emission, $\lambda_{\rm{obs}}$, and estimate the adjacent continuum flux, $F_{\rm{c}}$, as the mean of the spectrum at $\pm0.1\,\mu$m from the peak flux. This ensures we cover broad components, if present. We set the lower, $\lambda_{\rm{l}}$, and upper wavelength, $\lambda_{\rm{u}}$, ranges of the emission line when the line flux is equal or smaller than $1\sigma_{\rm{c}}$ of the adjacent continuum level, i.e., $F_{\lambda_{\rm{l,u}}} \le F_{\rm{c}}+1\sigma_{\rm{c}}$. We subtract the continuum flux from the emission line within the $[\lambda_{\rm{l}},\lambda_{\rm{u}}]$ wavelength range. 

    \item At this stage, we have subcubes of the identified emission lines within the wavelength range of a given MIRI imaging filter. We scale the fluxes of the subcubes across the wavelength range using the throughput curve of the MIRI imaging filter. Specifically, the area of the throughput curve is set to unity and resampled to the wavelength pixel scale of the MRS observations. Then, we multiply the scaled and resampled throughput curve to the emission line fluxes at each wavelength. We compute the image of each emission line by integrating the flux per spaxel. 

    \item The total contribution of emission lines within the MIRI imaging filter is estimated by integrating all the emission line images. We scale the fluxes of the subcubes across the wavelength range using the throughput curve of the MIRI imaging filter as mentioned in step 4. 

    \item The MRS observations contain many more fainter emission lines than those identified above, thus adding the flux after extracting the lines from step 3 will still contain residual emission line contamination. To obtain a line-free spectrum, we fit the continuum emission of the spectrum per spaxel. We mask the emission lines identified in step 3 as well as broad features present within the filter, e.g., the PAH at $11.3$ $\mu$m in the F1000W filter. Then, we fit a third-order spline curve to the continuum spectrum per spaxel.

    \item The total contribution of the continuum emission within the MIRI filter is estimated by integrating the spline curve per spaxel (black line in step 6 in Figure \ref{fig:fig0}). We scale the fluxes of the subcubes across the wavelength range using the throughput curve of the MIRI imaging filter as mentioned in step 4. For all filters, the MRS images have the same morphology as the MIRI images. We estimate a median flux difference of $3\pm1$\% within the central $1$ kpc of NGC\,3081, NGC\,5728, and NGC\,7172 between the direct images and those computed with the MRS observations. Our estimated median flux difference is within the flux calibration uncertainty of the MIRI ($3$\%) and MRS ($5.6$\%) observations (Section \ref{sec:OBS_RED}). 

    \item Finally, the total (continuum+line) MIR image is estimated by integrating each spaxel of the MRS cube within the MIRI imaging filter. We scale the fluxes of the subcubes across the wavelength range using the throughput curve of the MIRI imaging filter as mentioned in step 4.  
\end{enumerate}

\section{Comparison with line emission images}\label{app:AppB}

In this section, we show the emission line images with overlaid total MIR emission and continuum contours. 

\begin{figure*}
\centering
\includegraphics[scale=0.6]{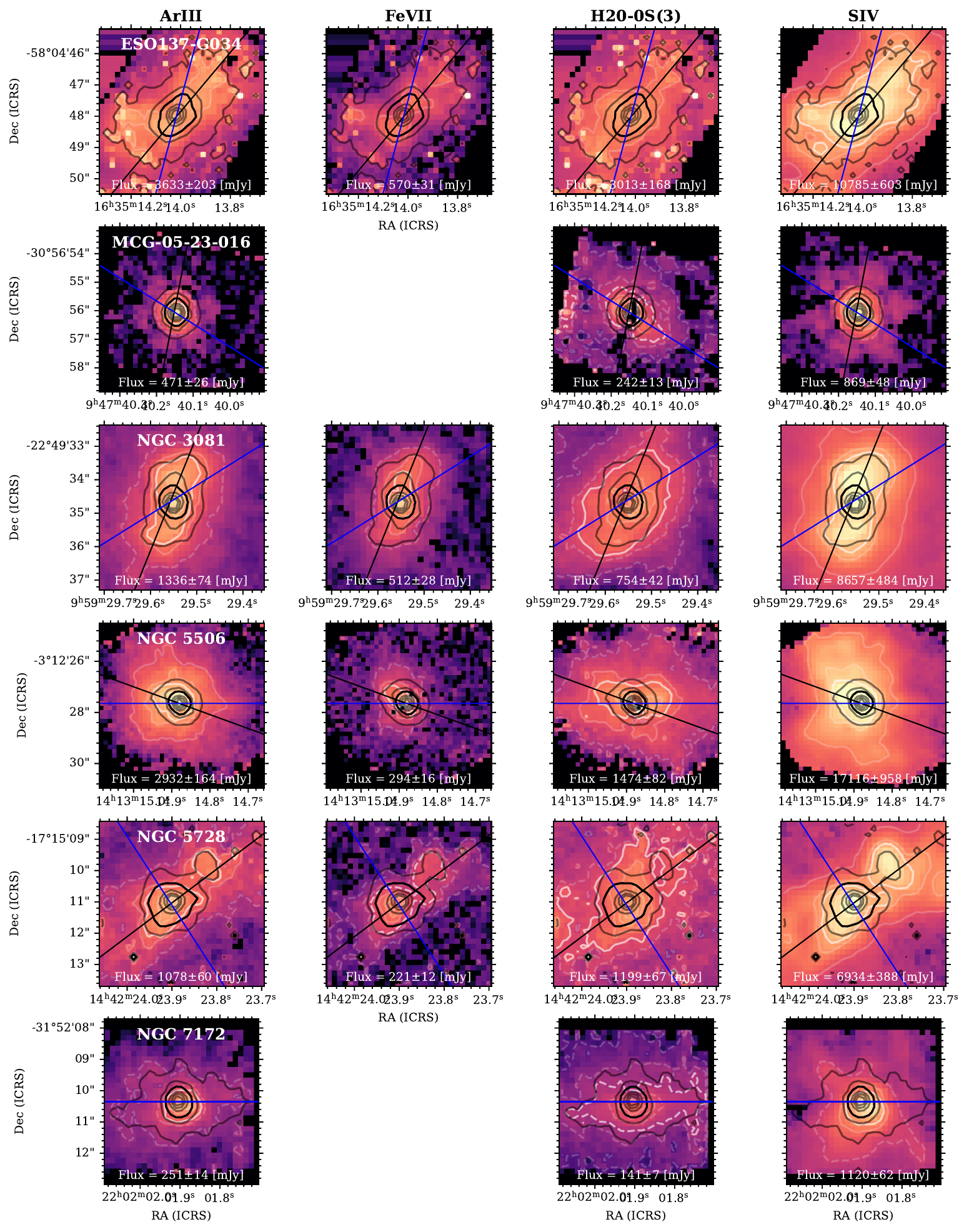}
\caption{Line emission vs. total MIR emission image in the F1000W filter. From left to right and top to bottom, we show the integrated line emission images (colorscale) of [\ion{Ar}{3}], [\ion{Fe}{7}], H$_{2}$0-0S(3), and [\ion{S}{4}] for ESO137-G034, MCG-05-23-016, NGC\,3081, NGC\,5506, NGC\,5728, and NGC\,7172. Line (white) and total MIR emission (black) contours increase in steps of $[1,5,10,30,50,70]$\% from the peak flux. The contour at a $10$\% level (thick line) from the peak flux is shown for reference. The radio jet (black line) and dust lane (blue line) are shown. The integrated flux of the line emission within the $1$ kpc FOV is shown at the bottom of each panel.}
 \label{fig:fig6}
\end{figure*}

\begin{figure*}[ht!]
\centering
\includegraphics[scale=0.6]{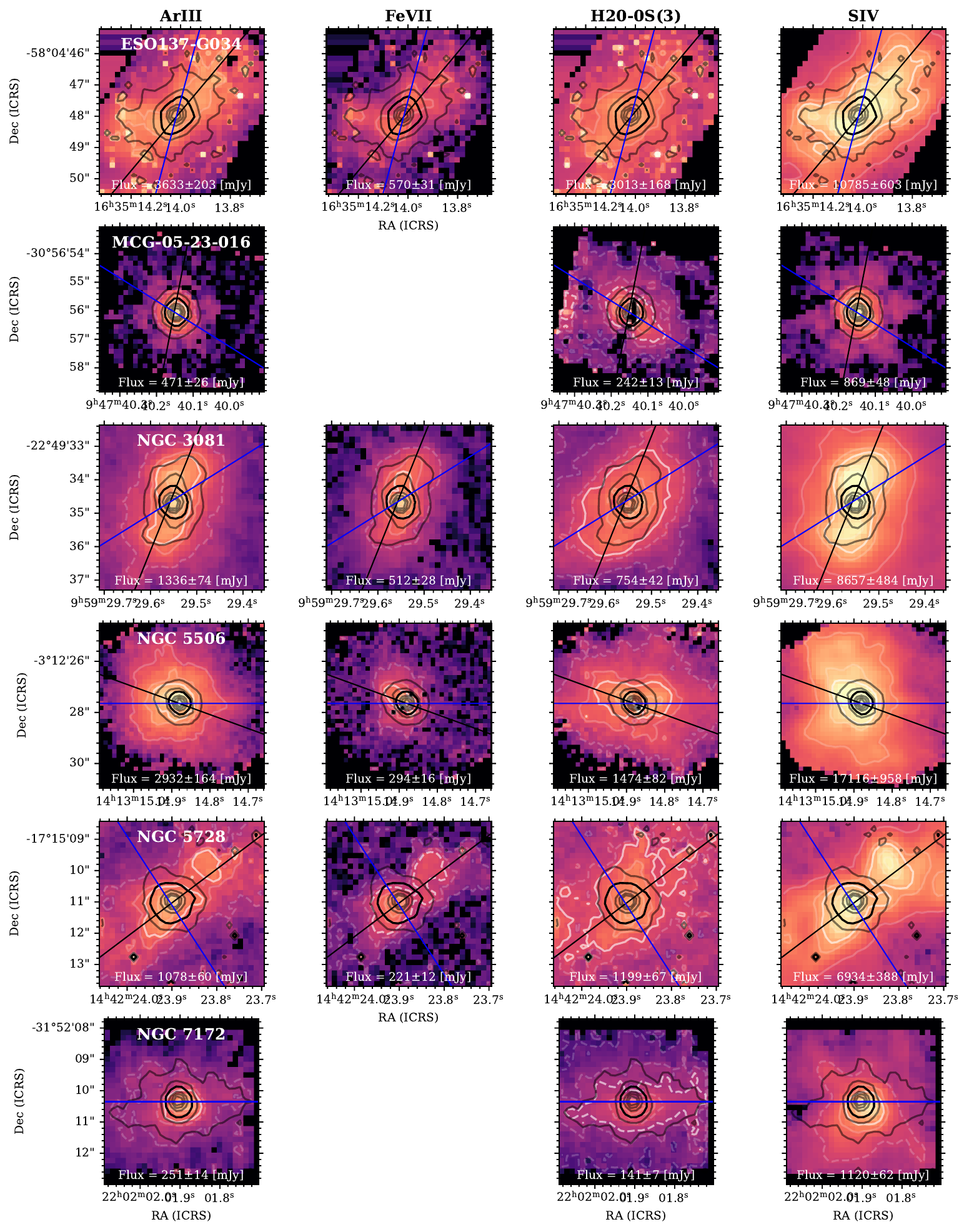}
\caption{Line emission vs. continuum emission image in the F1000W filter. From left to right and top to bottom, we show the integrated line emission images (colorscale) of [\ion{Ar}{3}], [\ion{Fe}{7}], H$_{2}$0-0S(3), and [\ion{S}{4}] for ESO137-G034, MCG-05-23-016, NGC\,3081, NGC\,5506, NGC\,5728, and NGC\,7172. Line (white) and continuum (black) contours increase in steps of $[1,5,10,30,50,70]$\% from the peak flux. The contour at a $10$\% level (thick line) from the peak flux is shown for reference. The radio jet (black line) and dust lane (blue line) are shown. The integrated flux of the line emission within the $1$ kpc FOV is shown at the bottom of each panel.}
 \label{fig:fig7}
\end{figure*}

\begin{figure*}
\centering
\includegraphics[width=\textwidth]{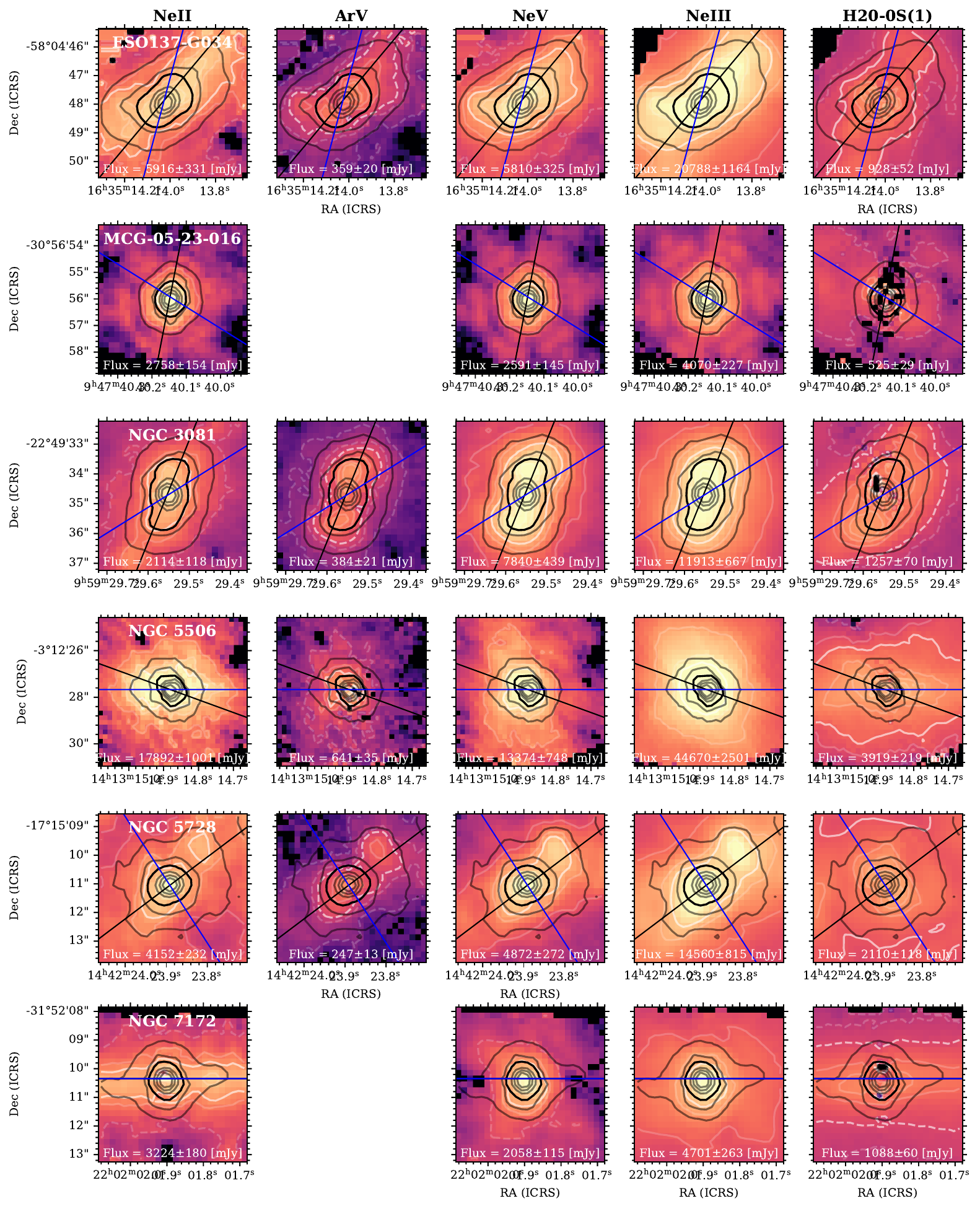}
\caption{Line emission vs. total MIR emission image in the F1500W filter. From left to right and top to bottom, we show the integrated line emission images (colorscale) of [\ion{Ne}{2}], [\ion{Ar}{5}], [\ion{Ne}{5}], [\ion{Ne}{3}], and H$_{2}$0-0S(1), for ESO137-G034, MCG-05-23-016, NGC\,3081, NGC\,5506, NGC\,5728, and NGC\,7172. Line (white) and total MIR emission (black) contours increase in steps of $[1,5,10,30,50,70]$\% from the peak flux. The contour at a $10$\% level (thick line) from the peak flux is shown for reference. The radio jet (black line) and dust lane (blue line) are shown. The integrated flux of the line emission within the $1$ kpc FOV is shown at the bottom of each panel.}
 \label{fig:fig6_F1500W}
\end{figure*}

\begin{figure*}
\centering
\includegraphics[width=\textwidth]{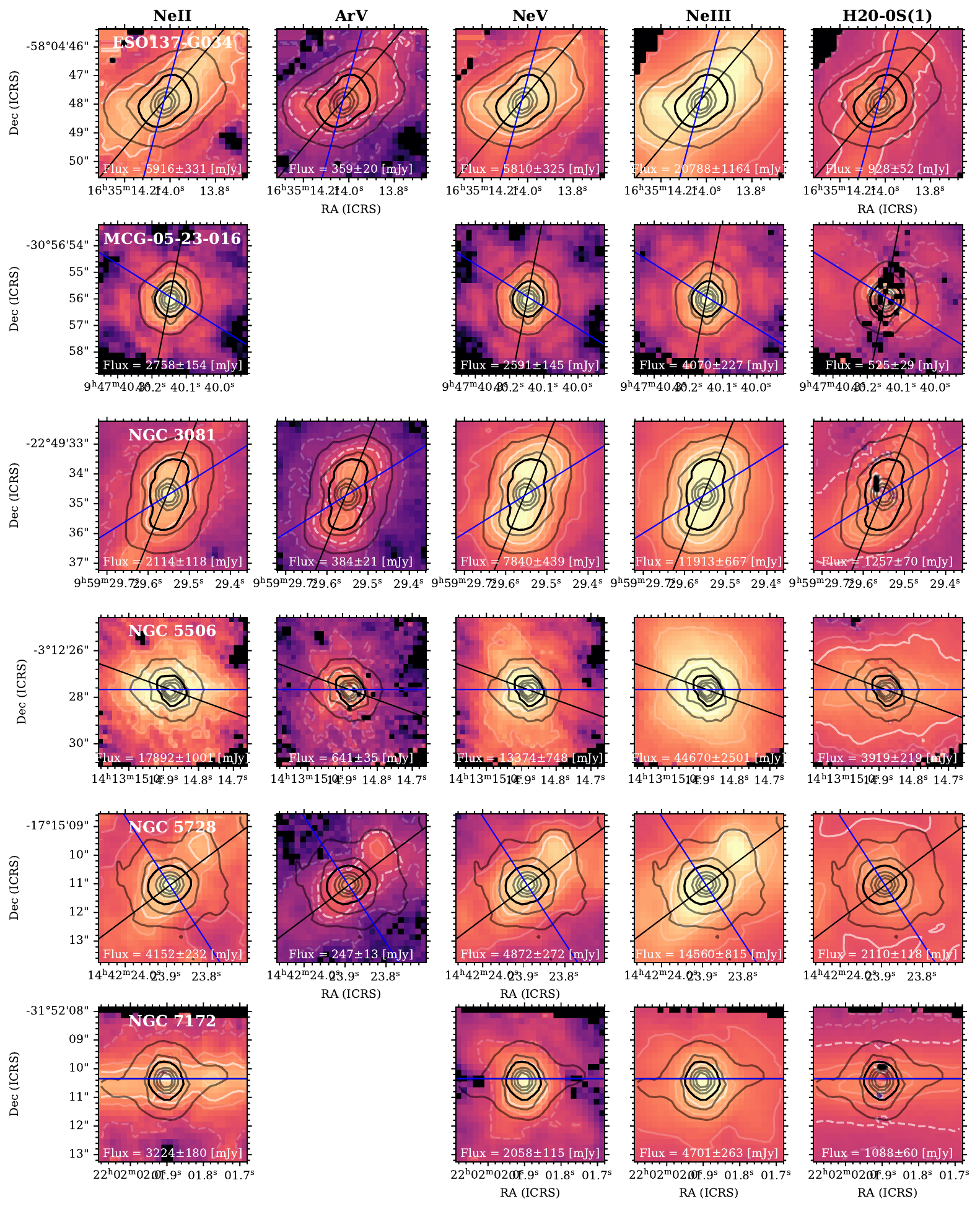}
\caption{Line emission vs. Continuum image emission in the F1500W filter. From left to right and top to bottom, we show the integrated line emission images (colorscale) of [\ion{Ne}{2}], [\ion{Ar}{5}], [\ion{Ne}{5}], [\ion{Ne}{3}], and H$_{2}$0-0S(1), for ESO137-G034, MCG-05-23-016, NGC\,3081, NGC\,5506, NGC\,5728, and NGC\,7172. Line (white) and total MIR emission (black) contours increase in steps of $[1,5,10,30,50,70]$\% from the peak flux. The contour at a $10$\% level (thick line) from the peak flux is shown for reference. The radio jet (black line) and dust lane (blue line) are shown. The integrated flux of the line emission within the $1$ kpc FOV is shown at the bottom of each panel.}
 \label{fig:fig7_F1500W}
\end{figure*}

\begin{figure*}
\centering
\includegraphics[scale=0.65]{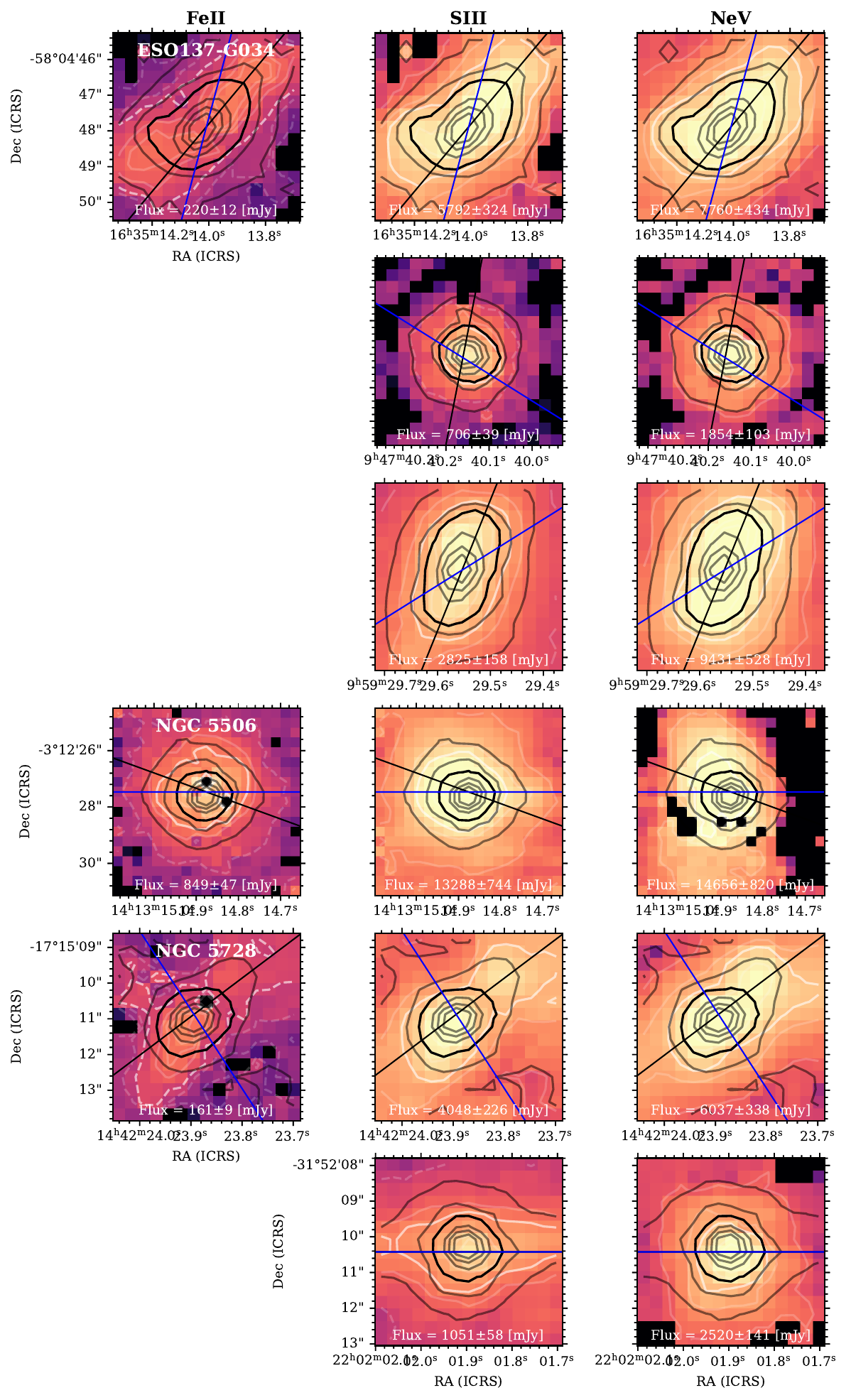}
\caption{Line emission vs. total MIR emission image in the F2100W filter. From left to right and top to bottom, we show the integrated line emission images (colorscale) of [\ion{Fe}{2}], [\ion{S}{3}], and [\ion{Ne}{5}] for ESO137-G034, MCG-05-23-016, NGC\,3081, NGC\,5506, NGC\,5728, and NGC\,7172. Line (white) and total MIR emission (black) contours increase in steps of $[1,5,10,30,50,70]$\% from the peak flux. The contour at a $10$\% level (thick line) from the peak flux is shown for reference. The radio jet (black line) and dust lane (blue line) are shown. The integrated flux of the line emission within the $1$ kpc FOV is shown at the bottom of each panel.}
 \label{fig:fig6_F2100W}
\end{figure*}

\begin{figure*}
\centering
\includegraphics[scale=0.65]{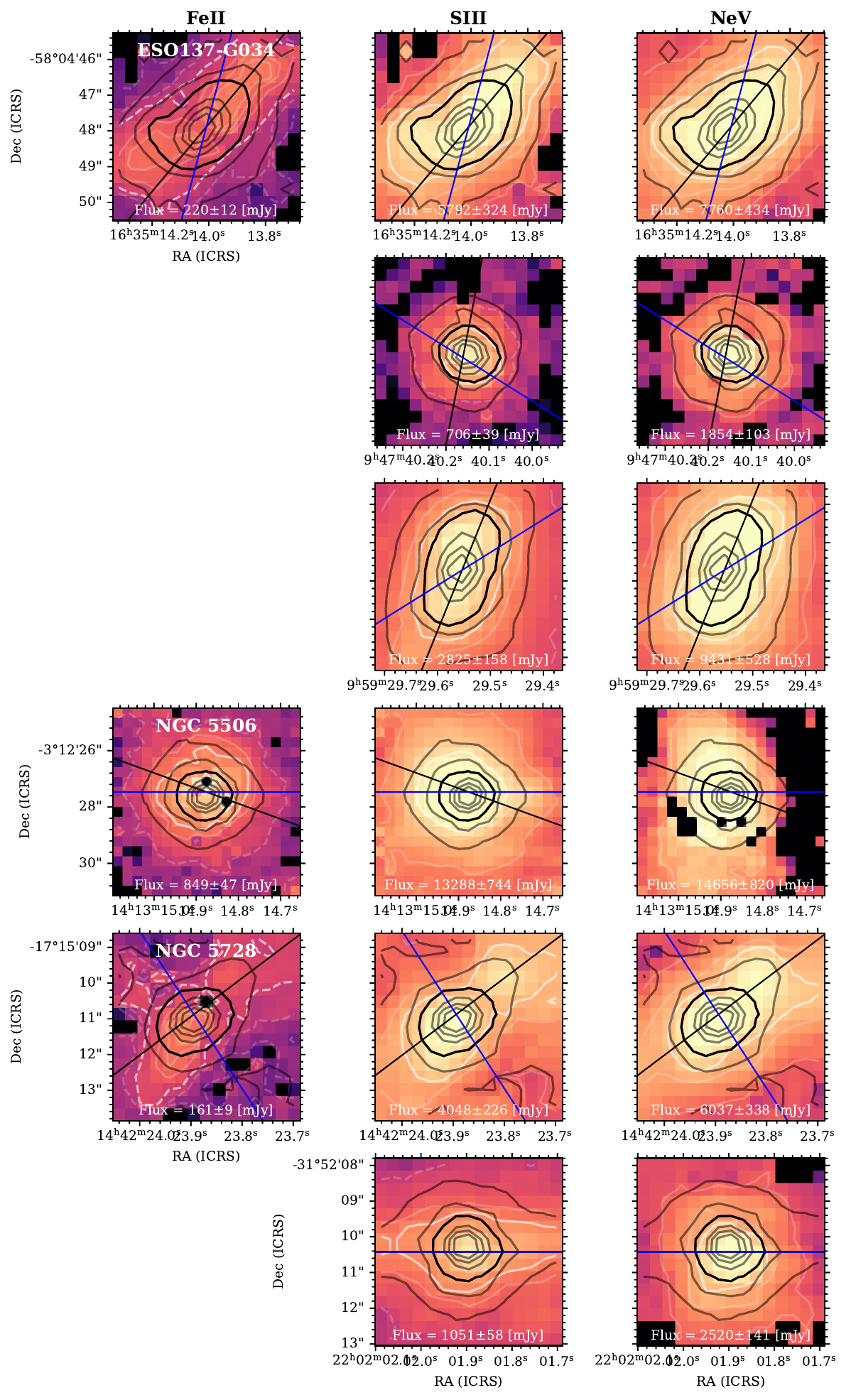}
\caption{Line emission vs. Continuum image emission in the F2100W filter. From left to right and top to bottom, we show the integrated line emission images (colorscale) of [\ion{Fe}{2}], [\ion{S}{3}], and [\ion{Ne}{5}] for ESO137-G034, MCG-05-23-016, NGC\,3081, NGC\,5506, NGC\,5728, and NGC\,7172. Line (white) and total  (black) contours increase in steps of $[1,5,10,30,50,70]$\% from the peak flux. The contour at a $10$\% level (thick line) from the peak flux is shown for reference. The radio jet (black line) and dust lane (blue line) are shown. The integrated flux of the line emission within the $1$ kpc FOV is shown at the bottom of each panel.}
 \label{fig:fig7_F2100W}
\end{figure*}

\section{Image cross-correlation analysis for individual galaxies}\label{app:AppC}

\begin{figure*}
\centering
\includegraphics[width=\textwidth]{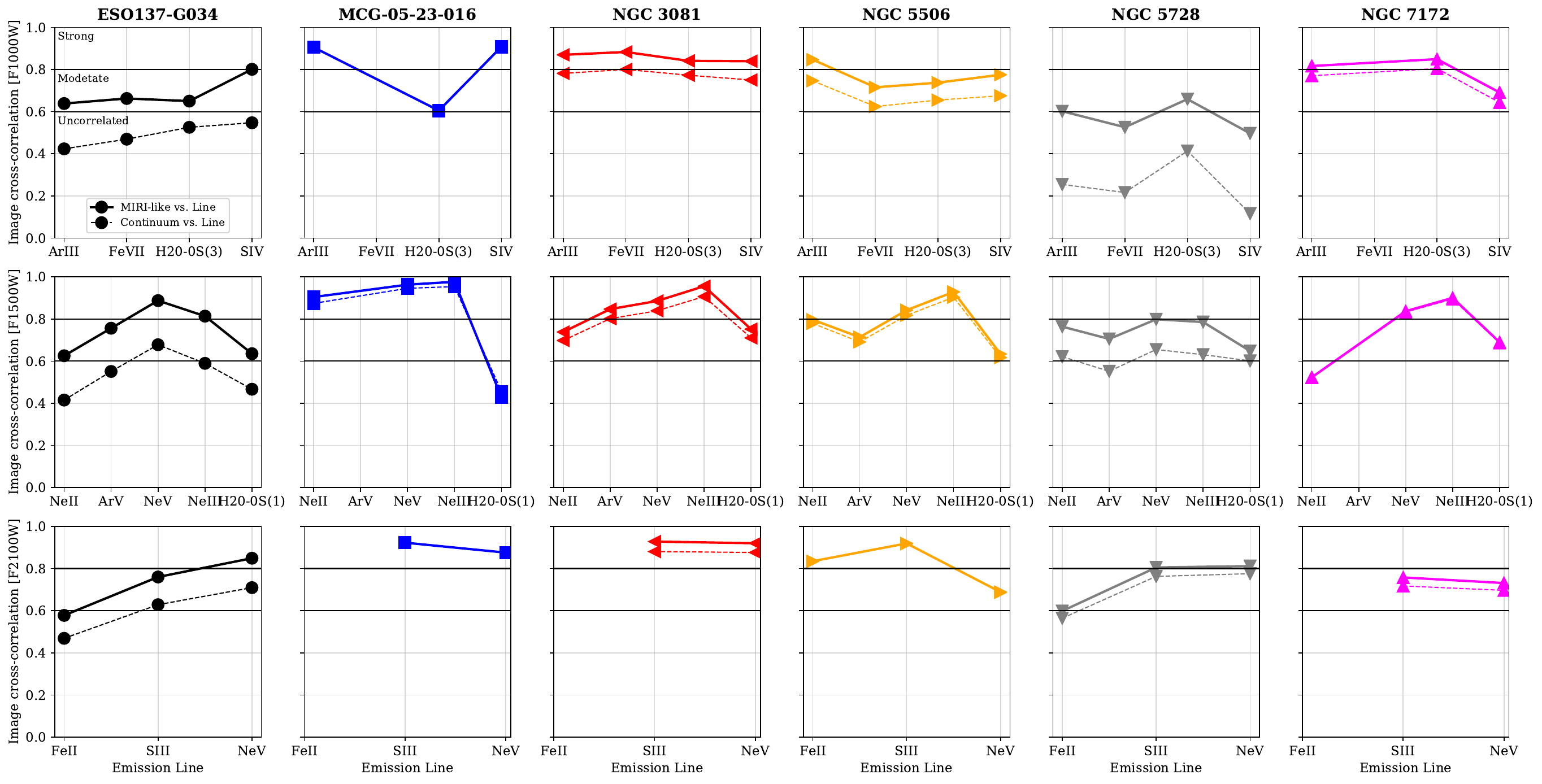}
\caption{Normalized correlation coefficients between line, total MIR emission, and continuum images for each object and emission line. The normalized correlation coefficients, $\rho$, between the line and dust-only (solid line), and line and continuum (dashed line) emission in the F1000W (top), F1500W (middle), and F2100W (bottom) filters for each galaxy as shown in the title. The ranges for strong, moderate, and uncorrelated coefficients are displayed.}
 \label{fig:fig1_appC}
\end{figure*}

The normalized correlation coefficients between line, total MIR emission, and continuum images for each object and emission line are shown in Figure \ref{fig:fig1_appC}. These coefficients are then discussed on an object-by-object basis.

\textbf{ESO137-G034} has a strong correlation ($\rho=0.80$) with [\ion{S}{4}] and lower-end moderate correlation ($\rho=0.64-0.66$) between the [\ion{Ar}{3}], [\ion{Fe}{7}] and H$_{2}$0-0S(3) lines and the total MIR emission images in the F1000W filter. This correlation is mainly due to the extended and low surface brightness emission in all images. In the F1500W filter, the strongest correlations are with [\ion{Ne}{5}] ($\rho=0.89$) and [\ion{Ne}{3}] ($\rho=0.81$). Both lines have broad extended emission along the radio jet axis with low level surface brightness also covering the full FOV, which makes the correlation coefficient to be high with the total MIR emission image. We measure moderated correlations ($\rho=0.63-0.76$) with the [\ion{Ne}{3}], H$_{2}$0-0S(1) and [\ion{Ar}{3}] lines (in increment order). Both [\ion{Ne}{3}] and [\ion{Ar}{3}] lines have narrow extended ($\ge450$ pc) emission mostly concentrated along the radio jet.  The H$_{2}$0-0S(1) line has extended ($\ge450$ pc) emission parallel to the dust lane with another extended ($\sim200$ pc) component at a PA$\sim50^{\circ}$ perpendicular to the radio jet axis (PA $=140^{\circ}$). In the F21000W filter, the total MIR emission image has a strong correlation with the [\ion{Ne}{5}] ($\rho=0.85$) and moderated to poor correlations with [\ion{S}{3}] ($\rho=0.76$) and [\ion{Fe}{2}] ($\rho=0.58$) lines. The [\ion{Ne}{5}] line has a broad and extended emission along the radio jet with low-level surface brightness also covering the full FOV. This emission is more narrow and less extended for the [\ion{S}{3}] and [\ion{Fe}{2}] lines. The extension and orientation of these emission lines are cospatial with the extended, $\sim400$ pc ($2.1\arcsec$), NLR as observed with [\ion{O}{3}] emission lines \citep{Ferruit2000}.

For all lines and filters, the continuum emission images are poorly correlated $\rho<0.65$ with the line emission images of ESO1347-G034. The extended continuum emission is wider and smoother than the extended line emission. This result indicates that the extended ($\sim39-450$ pc) MIR continuum emission of ESO1347-G034 cannot be spatially correlated with any of the multi-phase outflows in the $10-21~\mu$m wavelength range. This galaxy has a long dust lane passing across the nucleus at a PA$\sim165^{\circ}$ \citep{Ferruit2000}.

\textbf{MCG-05-23-016} has a very strong correlation ($\rho=0.91$) with [\ion{Ar}{3}] and [\ion{S}{4}], because both total MIR continuum and line emission are point-like sources.  As this source is unresolved, the continuum emission images have similar correlation coefficients to the total MIR emission images in all filters. Interestingly, the correlation is poorly correlated with H$_{2}$0-0S(3) ($\rho=0.61$) and H$_{2}$0-0S(1)  ($\rho=0.43$). The reason is that these line emissions are extended and resolved along the dust lane orientation (PA$\sim58^{\circ}$). Both H$_{2}$0-0S(3) and H$_{2}$0-0S(1) lines have an extended component at the $1$\% level from the peak flux mostly located in the eastern region of the galaxy with extensions $>350$ pc from the core. Extended, $\sim175$ pc ($\sim1\arcsec$), NLR has been observed using H$_{\alpha}$ line emission \citep{Prieto2014}, N-band imaging \citep{GB2016}, and \ion{O}{3} line emission at a PA$\sim40^{\circ}$ \citep{Ferruit2000}. This galaxy has a dust lane crossing the nucleus with an extension of $\sim360$ pc ($\sim2\arcsec$) at a PA$\sim58^{\circ}$ \citep{Prieto2014}. CO emission has not been detected within the nucleus, and the extended emission is not attributed to dust emission \citep{Rosario2018}. We find that the MIR extended emission arises from the molecular gas, H$_{2}$0-0S(3) \citep{EA2025}.

\textbf{NGC\,3081} has moderate to strong correlations ($\rho=0.74-0.96$) between all line emissions and the total MIR emission images in all filters. The strongest correlations are found to be for [\ion{Fe}{7}] ($\rho=0.88$), [\ion{Ne}{3}] ($\rho=0.96$), and [\ion{S}{3}] ($\rho=0.93$) in the F1000W, F1500W, and F2100W filters, respectively. In all cases, the line emission has an S-shape along the north-west to the south-east direction up to a distance of $\sim400$ pc from the core. Note that the S-shape extended emission in the north-south direction is coincident with the inner parts of two spiral arms as observed in [\ion{O}{3}] and [\ion{N}{2}] line emissions \citep{Ferruit2000}. N- and Q-band images showed extended emission \citep{GB2016}. In addition bipolar outflows have been observed in [\ion{O}{3}], H$_{\alpha}$, and [\ion{N}{2}] along the NLR with an S-shape at a PA $\sim120^{\circ}$ \citep{Ruiz2005, SM2016}.

The strong correlations with the total MIR emission images slightly drop toward the high-end of moderated correlations ($\rho=0.75-0.80$) for the continuum emission images in the F1000W filter. The smaller differences between the correlation coefficients are for the molecular lines H$_{2}$0-0S(3) and H$_{2}$0-0S(1) in the F1000W and F15000W, respectively. These lines seem to have a compact ($r\sim100$ pc) component parallel to the dust lane and another extended ($r\sim400$ pc) component at a PA$\sim70^{\circ}$. This galaxy has a nuclear bar at a PA$\sim122^{\circ}$ passing through the core with two spiral arms \citep{Erwin2004,Buta2004}, while the radio jet is at a PA$\sim158^{\circ}$ \citep{Nagar1999}. These results indicate that the extended MIR continuum emission may arise from a combination of dust emission from the dust lane and heated regions in the inner arms irradiated by the radio jet.

\textbf{NGC5506} has moderate to strong correlations ($\rho=0.70-0.92$) between all line emissions and the `MIRI-like' images in all filters, except for the molecular gas H$_{2}$0-0S(1) in the F1500W filter with a poor correlation of $\rho=0.63$. For all the coronal lines, the emission is extended mostly in the eastern region up to $r\sim400$ pc in the north-south direction with a compact, $r\sim150$ pc, elongated extension parallel to the radio jet axis. This galaxy has [\ion{O}{3}] extended, kpc-scale, NLR in the north-south direction \citep{Wilson1985}. Q-band extended emission was also observed \citep{GB2016}. The molecular gas, H$_{2}$0-0S(3) and H$_{2}$0-0S(1), show extended ($r>500$ pc) emission along the east-west direction parallel to the dust lane axis. The `MIRI-like' images seem to have a morphology composed of an extended component along the dust lane and another along the radio jet axis in the F1500W and F2100W filters. This morphology causes it to be strongly correlated with the emission lines. 

The line and continuum emission images in the F1000W filter have moderate correlation coefficients ($\rho=0.65-0.75$). The drop in the correlation indicates that the $\sim1-30$\% flux from the peak pixel is arising from line emission contribution in the total MIR emission images. The correlation is still moderate to strong in both the F1500W and the F2100W. Previous MIR observations at sub-arcsecond resolution have shown an unresolved core at $\sim10$ $\mu$m, while a north-east extension of $\sim126$ ($\sim1\arcsec$) was observed at $11.9$ $\mu$m \citep{Raban2008}. In addition, the $3.6$ cm radio emission shows an $\sim252$ pc ($\sim2\arcsec$) extended and diffuse structure surrounding the core \citep{Wehrle1987}. This diffuse emission is also observed in our line emission images. We suggest that the extended radio emission is heating the dust in the host galaxy producing a cold dust component to emit at the $1-30$\% flux level from the peak of the continuum emission in the F1500W and F2100W filters. 

\textbf{NGC5728} has poor to moderate correlations ($\rho=0.50-0.66$) in the F1000W filter and moderate correlations ($\rho=0.60-0.81$) in the F1500W and F2100W filters between the line emission and the total MIR emission images. The coronal lines show an extended emission component parallel to the radio jet axis with a bright spot at $r\sim250$ pc in the north-west region. The molecular lines show a compact ($r<200$ pc) emission component mostly perpendicular to the dust lane. The NLR extends $\sim1.5$ kpc ($\sim7.8\arcsec$) at a PA $\sim118^{\circ}$ \citep{Schommer1988,MA1995}. In addition, the H$_{2}$0-0S(3) line has an elongation in the north-west region highly offset from the radio jet, while the H$_{2}$0-0S(1) line has an extended and diffuse component filling the full FOV ($r>0.5$ kpc). The correlation between the total MIR emission and line emission images may be produced by the bright north-west spot at $r\sim250$ pc and some of the extended MIR emission in the south-east region along the radio jet axis. 

The correlation coefficients indicate that the continuum and line emission images are uncorrelated ($\rho=0.12-0.41$) in the F1000W filter. The continuum images show an unresolved core with an elongated component along the dust lane axis \citep[$\sim33^{\circ}$;][]{Prada1999} and another along the stellar bar \citep[$\sim86^{\circ}$;][]{Prada1999}. The extended MIR continuum emission is not spatially correlated with any of the coronal lines. The correlation coefficient is similar in the F2100W filter because the MIR continuum emission is also extended along the radio axis. We suggest that the increasing MIR continuum emission from F1500W to F21000W may arise from a cold dust component heated by the jet interacting with the ISM of the host galaxy.

\textbf{NGC7172} has moderate to strong correlations ($\rho=0.64-0.84$) between all line emissions and the `MIRI-like' images in all filters, except for [\ion{Ne}{2}] in the F1500W with an uncorrelated coefficient ($\rho=0.52$). The lines [\ion{Ar}{3}], H$_{2}$0-0S(3), [\ion{Ne}{2}], and H$_{2}$0-0S(1) show extended emission along the east-west direction, which is parallel to both the radio and dust lane axes. These lines are correlated with the extended MIR emission along the same direction measured in the total MIR emission images. Extended MIR emission was also detected by \citet{GB2016}. The lines [\ion{S}{4}], [\ion{Ne}{5}], [\ion{Ne}{3}], and [\ion{Ne}{5}] show extended emission along the north-south direction up to $\sim400$ pc above and below the disk of the galaxy. These lines are correlated with the slightly elongated emission at a level of $\sim10$\% from the peak emission in the north-south direction. The correlation coefficients only drop by $\sim0.05$ when compared with the continuum images in the F1000W filter. We suggest that the continuum emission arises from the dust lane of this edge-on galaxy.


\bibliography{references}{}

\begin{thebibliography}{}
\expandafter\ifx\csname natexlab\endcsname\relax\def\natexlab#1{#1}\fi
\providecommand{\url}[1]{\href{#1}{#1}}
\providecommand{\dodoi}[1]{doi:~\href{http://doi.org/#1}{\nolinkurl{#1}}}
\providecommand{\doeprint}[1]{\href{http://ascl.net/#1}{\nolinkurl{http://ascl.net/#1}}}
\providecommand{\doarXiv}[1]{\href{https://arxiv.org/abs/#1}{\nolinkurl{https://arxiv.org/abs/#1}}}

\bibitem[{{Alonso-Herrero} {et~al.}(2011){Alonso-Herrero}, {Ramos Almeida},
  {Mason}, {Asensio Ramos}, {Roche}, {Levenson}, {Elitzur}, {Packham},
  {Rodr{\'\i}guez Espinosa}, {Young}, {D{\'\i}az-Santos}, \&
  {P{\'e}rez-Garc{\'\i}a}}]{AH2011}
{Alonso-Herrero}, A., {Ramos Almeida}, C., {Mason}, R., {et~al.} 2011, \apj,
  736, 82, \dodoi{10.1088/0004-637X/736/2/82}

\bibitem[{{Alonso-Herrero} {et~al.}(2018){Alonso-Herrero}, {Pereira-Santaella},
  {Garc{\'\i}a-Burillo}, {Davies}, {Combes}, {Asmus}, {Bunker},
  {D{\'\i}az-Santos}, {Gandhi}, {Gonz{\'a}lez-Mart{\'\i}n},
  {Hern{\'a}n-Caballero}, {Hicks}, {H{\"o}nig}, {Labiano}, {Levenson},
  {Packham}, {Ramos Almeida}, {Ricci}, {Rigopoulou}, {Rosario}, {Sani}, \&
  {Ward}}]{AH2018}
{Alonso-Herrero}, A., {Pereira-Santaella}, M., {Garc{\'\i}a-Burillo}, S.,
  {et~al.} 2018, \apj, 859, 144, \dodoi{10.3847/1538-4357/aabe30}

\bibitem[{{Alonso-Herrero} {et~al.}(2021){Alonso-Herrero},
  {Garc{\'\i}a-Burillo}, {H{\"o}nig}, {Garc{\'\i}a-Bernete}, {Ramos Almeida},
  {Gonz{\'a}lez-Mart{\'\i}n}, {L{\'o}pez-Rodr{\'\i}guez}, {Boorman}, {Bunker},
  {Burtscher}, {Combes}, {Davies}, {D{\'\i}az-Santos}, {Gandhi},
  {Garc{\'\i}a-Lorenzo}, {Hicks}, {Hunt}, {Ichikawa}, {Imanishi}, {Izumi},
  {Labiano}, {Levenson}, {Packham}, {Pereira-Santaella}, {Ricci}, {Rigopoulou},
  {Roche}, {Rosario}, {Rouan}, {Shimizu}, {Stalevski}, {Wada}, \&
  {Williamson}}]{AH2021}
{Alonso-Herrero}, A., {Garc{\'\i}a-Burillo}, S., {H{\"o}nig}, S.~F., {et~al.}
  2021, \aap, 652, A99, \dodoi{10.1051/0004-6361/202141219}

\bibitem[{{Antonucci} \& {Miller}(1985)}]{Antonucci1985}
{Antonucci}, R.~R.~J., \& {Miller}, J.~S. 1985, \apj, 297, 621,
  \dodoi{10.1086/163559}

\bibitem[{{Argyriou} {et~al.}(2023){Argyriou}, {Glasse}, {Law}, {Labiano},
  {{\'A}lvarez-M{\'a}rquez}, {Patapis}, {Kavanagh}, {Gasman}, {Mueller},
  {Larson}, {Vandenbussche}, {Glauser}, {Royer}, {Dicken}, {Harkett},
  {Sargent}, {Engesser}, {Jones}, {Kendrew}, {Noriega-Crespo}, {Brandl},
  {Rieke}, {Wright}, {Lee}, \& {Wells}}]{Argyriou2023}
{Argyriou}, I., {Glasse}, A., {Law}, D.~R., {et~al.} 2023, \aap, 675, A111,
  \dodoi{10.1051/0004-6361/202346489}

\bibitem[{{Asmus}(2019)}]{Asmus2019}
{Asmus}, D. 2019, \mnras, 489, 2177, \dodoi{10.1093/mnras/stz2289}

\bibitem[{{Asmus} {et~al.}(2015){Asmus}, {Gandhi}, {H{\"o}nig}, {Smette}, \&
  {Duschl}}]{Asmus2015}
{Asmus}, D., {Gandhi}, P., {H{\"o}nig}, S.~F., {Smette}, A., \& {Duschl}, W.~J.
  2015, \mnras, 454, 766, \dodoi{10.1093/mnras/stv1950}

\bibitem[{{Asmus} {et~al.}(2016){Asmus}, {H{\"o}nig}, \& {Gandhi}}]{Asmus2016}
{Asmus}, D., {H{\"o}nig}, S.~F., \& {Gandhi}, P. 2016, \apj, 822, 109,
  \dodoi{10.3847/0004-637X/822/2/109}

\bibitem[{{Asmus} {et~al.}(2014){Asmus}, {H{\"o}nig}, {Gandhi}, {Smette}, \&
  {Duschl}}]{Asmus2014}
{Asmus}, D., {H{\"o}nig}, S.~F., {Gandhi}, P., {Smette}, A., \& {Duschl}, W.~J.
  2014, \mnras, 439, 1648, \dodoi{10.1093/mnras/stu041}

\bibitem[{{Astropy Collaboration} {et~al.}(2013){Astropy Collaboration},
  {Robitaille}, {Tollerud}, {Greenfield}, {Droettboom}, {Bray}, {Aldcroft},
  {Davis}, {Ginsburg}, {Price-Whelan}, {Kerzendorf}, {Conley}, {Crighton},
  {Barbary}, {Muna}, {Ferguson}, {Grollier}, {Parikh}, {Nair}, {Unther},
  {Deil}, {Woillez}, {Conseil}, {Kramer}, {Turner}, {Singer}, {Fox}, {Weaver},
  {Zabalza}, {Edwards}, {Azalee Bostroem}, {Burke}, {Casey}, {Crawford},
  {Dencheva}, {Ely}, {Jenness}, {Labrie}, {Lim}, {Pierfederici}, {Pontzen},
  {Ptak}, {Refsdal}, {Servillat}, \& {Streicher}}]{astropy:2013}
{Astropy Collaboration}, {Robitaille}, T.~P., {Tollerud}, E.~J., {et~al.} 2013,
  \aap, 558, A33, \dodoi{10.1051/0004-6361/201322068}

\bibitem[{{Astropy Collaboration} {et~al.}(2018){Astropy Collaboration},
  {Price-Whelan}, {Sip{\H{o}}cz}, {G{\"u}nther}, {Lim}, {Crawford}, {Conseil},
  {Shupe}, {Craig}, {Dencheva}, {Ginsburg}, {Vand erPlas}, {Bradley},
  {P{\'e}rez-Su{\'a}rez}, {de Val-Borro}, {Aldcroft}, {Cruz}, {Robitaille},
  {Tollerud}, {Ardelean}, {Babej}, {Bach}, {Bachetti}, {Bakanov}, {Bamford},
  {Barentsen}, {Barmby}, {Baumbach}, {Berry}, {Biscani}, {Boquien}, {Bostroem},
  {Bouma}, {Brammer}, {Bray}, {Breytenbach}, {Buddelmeijer}, {Burke},
  {Calderone}, {Cano Rodr{\'\i}guez}, {Cara}, {Cardoso}, {Cheedella}, {Copin},
  {Corrales}, {Crichton}, {D'Avella}, {Deil}, {Depagne}, {Dietrich}, {Donath},
  {Droettboom}, {Earl}, {Erben}, {Fabbro}, {Ferreira}, {Finethy}, {Fox},
  {Garrison}, {Gibbons}, {Goldstein}, {Gommers}, {Greco}, {Greenfield},
  {Groener}, {Grollier}, {Hagen}, {Hirst}, {Homeier}, {Horton}, {Hosseinzadeh},
  {Hu}, {Hunkeler}, {Ivezi{\'c}}, {Jain}, {Jenness}, {Kanarek}, {Kendrew},
  {Kern}, {Kerzendorf}, {Khvalko}, {King}, {Kirkby}, {Kulkarni}, {Kumar},
  {Lee}, {Lenz}, {Littlefair}, {Ma}, {Macleod}, {Mastropietro}, {McCully},
  {Montagnac}, {Morris}, {Mueller}, {Mumford}, {Muna}, {Murphy}, {Nelson},
  {Nguyen}, {Ninan}, {N{\"o}the}, {Ogaz}, {Oh}, {Parejko}, {Parley}, {Pascual},
  {Patil}, {Patil}, {Plunkett}, {Prochaska}, {Rastogi}, {Reddy Janga},
  {Sabater}, {Sakurikar}, {Seifert}, {Sherbert}, {Sherwood-Taylor}, {Shih},
  {Sick}, {Silbiger}, {Singanamalla}, {Singer}, {Sladen}, {Sooley},
  {Sornarajah}, {Streicher}, {Teuben}, {Thomas}, {Tremblay}, {Turner},
  {Terr{\'o}n}, {van Kerkwijk}, {de la Vega}, {Watkins}, {Weaver}, {Whitmore},
  {Woillez}, {Zabalza}, \& {Astropy Contributors}}]{astropy:2018}
{Astropy Collaboration}, {Price-Whelan}, A.~M., {Sip{\H{o}}cz}, B.~M., {et~al.}
  2018, \aj, 156, 123, \dodoi{10.3847/1538-3881/aabc4f}

\bibitem[{{Astropy Collaboration} {et~al.}(2022){Astropy Collaboration},
  {Price-Whelan}, {Lim}, {Earl}, {Starkman}, {Bradley}, {Shupe}, {Patil},
  {Corrales}, {Brasseur}, {N{"o}the}, {Donath}, {Tollerud}, {Morris},
  {Ginsburg}, {Vaher}, {Weaver}, {Tocknell}, {Jamieson}, {van Kerkwijk},
  {Robitaille}, {Merry}, {Bachetti}, {G{"u}nther}, {Aldcroft},
  {Alvarado-Montes}, {Archibald}, {B{'o}di}, {Bapat}, {Barentsen}, {Baz{'a}n},
  {Biswas}, {Boquien}, {Burke}, {Cara}, {Cara}, {Conroy}, {Conseil}, {Craig},
  {Cross}, {Cruz}, {D'Eugenio}, {Dencheva}, {Devillepoix}, {Dietrich},
  {Eigenbrot}, {Erben}, {Ferreira}, {Foreman-Mackey}, {Fox}, {Freij}, {Garg},
  {Geda}, {Glattly}, {Gondhalekar}, {Gordon}, {Grant}, {Greenfield}, {Groener},
  {Guest}, {Gurovich}, {Handberg}, {Hart}, {Hatfield-Dodds}, {Homeier},
  {Hosseinzadeh}, {Jenness}, {Jones}, {Joseph}, {Kalmbach}, {Karamehmetoglu},
  {Ka{l}uszy{'n}ski}, {Kelley}, {Kern}, {Kerzendorf}, {Koch}, {Kulumani},
  {Lee}, {Ly}, {Ma}, {MacBride}, {Maljaars}, {Muna}, {Murphy}, {Norman},
  {O'Steen}, {Oman}, {Pacifici}, {Pascual}, {Pascual-Granado}, {Patil},
  {Perren}, {Pickering}, {Rastogi}, {Roulston}, {Ryan}, {Rykoff}, {Sabater},
  {Sakurikar}, {Salgado}, {Sanghi}, {Saunders}, {Savchenko}, {Schwardt},
  {Seifert-Eckert}, {Shih}, {Jain}, {Shukla}, {Sick}, {Simpson},
  {Singanamalla}, {Singer}, {Singhal}, {Sinha}, {Sip{H{o}}cz}, {Spitler},
  {Stansby}, {Streicher}, {{{S}}umak}, {Swinbank}, {Taranu}, {Tewary},
  {Tremblay}, {Val-Borro}, {Van Kooten}, {Vasovi{'c}}, {Verma}, {de Miranda
  Cardoso}, {Williams}, {Wilson}, {Winkel}, {Wood-Vasey}, {Xue}, {Yoachim},
  {Zhang}, {Zonca}, \& {Astropy Project Contributors}}]{astropy:2022}
{Astropy Collaboration}, {Price-Whelan}, A.~M., {Lim}, P.~L., {et~al.} 2022,
  apj, 935, 167, \dodoi{10.3847/1538-4357/ac7c74}

\bibitem[{{Baumgartner} {et~al.}(2013){Baumgartner}, {Tueller}, {Markwardt},
  {Skinner}, {Barthelmy}, {Mushotzky}, {Evans}, \& {Gehrels}}]{Baumgartner2013}
{Baumgartner}, W.~H., {Tueller}, J., {Markwardt}, C.~B., {et~al.} 2013, \apjs,
  207, 19, \dodoi{10.1088/0067-0049/207/2/19}

\bibitem[{{Begelman} {et~al.}(1983){Begelman}, {McKee}, \&
  {Shields}}]{Begelman1983}
{Begelman}, M.~C., {McKee}, C.~F., \& {Shields}, G.~A. 1983, \apj, 271, 70,
  \dodoi{10.1086/161178}

\bibitem[{{Bennett} {et~al.}(2014){Bennett}, {Larson}, {Weiland}, \&
  {Hinshaw}}]{Bennett2014}
{Bennett}, C.~L., {Larson}, D., {Weiland}, J.~L., \& {Hinshaw}, G. 2014, \apj,
  794, 135, \dodoi{10.1088/0004-637X/794/2/135}

\bibitem[{{Bock} {et~al.}(2000){Bock}, {Neugebauer}, {Matthews}, {Soifer},
  {Becklin}, {Ressler}, {Marsh}, {Werner}, {Egami}, \& {Blandford}}]{Bock2000}
{Bock}, J.~J., {Neugebauer}, G., {Matthews}, K., {et~al.} 2000, \aj, 120, 2904,
  \dodoi{10.1086/316871}

\bibitem[{{Brindle} {et~al.}(1990{\natexlab{a}}){Brindle}, {Hough}, {Bailey},
  {Axon}, \& {Sparks}}]{Brindle1990c}
{Brindle}, C., {Hough}, J.~H., {Bailey}, J.~A., {Axon}, D.~J., \& {Sparks},
  W.~B. 1990{\natexlab{a}}, \mnras, 247, 327

\bibitem[{{Brindle} {et~al.}(1990{\natexlab{b}}){Brindle}, {Hough}, {Bailey},
  {Axon}, {Ward}, {Sparks}, \& {McLean}}]{Brindle1990a}
{Brindle}, C., {Hough}, J.~H., {Bailey}, J.~A., {et~al.} 1990{\natexlab{b}},
  \mnras, 244, 577

\bibitem[{{Brindle} {et~al.}(1990{\natexlab{c}}){Brindle}, {Hough}, {Bailey},
  {Axon}, {Ward}, {Sparks}, \& {McLean}}]{Brindle1990b}
---. 1990{\natexlab{c}}, \mnras, 244, 604

\bibitem[{{Burtscher} {et~al.}(2013){Burtscher}, {Meisenheimer}, {Tristram},
  {Jaffe}, {H{\"o}nig}, {Davies}, {Kishimoto}, {Pott}, {R{\"o}ttgering},
  {Schartmann}, {Weigelt}, \& {Wolf}}]{Burtscher2013}
{Burtscher}, L., {Meisenheimer}, K., {Tristram}, K.~R.~W., {et~al.} 2013, \aap,
  558, A149, \dodoi{10.1051/0004-6361/201321890}

\bibitem[{{Buta} {et~al.}(2004){Buta}, {Byrd}, \& {Freeman}}]{Buta2004}
{Buta}, R.~J., {Byrd}, G.~G., \& {Freeman}, T. 2004, \aj, 127, 1982,
  \dodoi{10.1086/382239}

\bibitem[{{Cameron} {et~al.}(1993){Cameron}, {Storey}, {Rotaciuc}, {Genzel},
  {Verstraete}, {Drapatz}, {Siebenmorgen}, \& {Lee}}]{Cameron1993}
{Cameron}, M., {Storey}, J. W.~V., {Rotaciuc}, V., {et~al.} 1993, \apj, 419,
  136, \dodoi{10.1086/173467}

\bibitem[{{Chelouche} \& {Netzer}(2001)}]{Chelouche2001}
{Chelouche}, D., \& {Netzer}, H. 2001, \mnras, 326, 916,
  \dodoi{10.1046/j.1365-8711.2001.04586.x}

\bibitem[{{Combes} {et~al.}(2019){Combes}, {Garc{\'\i}a-Burillo}, {Audibert},
  {Hunt}, {Eckart}, {Aalto}, {Casasola}, {Boone}, {Krips}, {Viti}, {Sakamoto},
  {Muller}, {Dasyra}, {van der Werf}, \& {Martin}}]{Combes2019}
{Combes}, F., {Garc{\'\i}a-Burillo}, S., {Audibert}, A., {et~al.} 2019, \aap,
  623, A79, \dodoi{10.1051/0004-6361/201834560}

\bibitem[{{Davies} {et~al.}(2020){Davies}, {Baron}, {Shimizu}, {Netzer},
  {Burtscher}, {de Zeeuw}, {Genzel}, {Hicks}, {Koss}, {Lin}, {Lutz},
  {Maciejewski}, {M{\"u}ller-S{\'a}nchez}, {Orban de Xivry}, {Ricci}, {Riffel},
  {Riffel}, {Rosario}, {Schartmann}, {Schnorr-M{\"u}ller}, {Shangguan},
  {Sternberg}, {Sturm}, {Storchi-Bergmann}, {Tacconi}, \&
  {Veilleux}}]{Davies2020}
{Davies}, R., {Baron}, D., {Shimizu}, T., {et~al.} 2020, \mnras, 498, 4150,
  \dodoi{10.1093/mnras/staa2413}

\bibitem[{{Davies} {et~al.}(2024){Davies}, {Shimizu}, {Pereira-Santaella},
  {Alonso-Herrero}, {Audibert}, {Bellocchi}, {Boorman}, {Campbell}, {Cao},
  {Combes}, {Delaney}, {D{\'\i}az-Santos}, {Eisenhauer}, {Esparza Arredondo},
  {Feuchtgruber}, {F{\"o}rster Schreiber}, {Fuller}, {Gandhi},
  {Garc{\'\i}a-Bernete}, {Garc{\'\i}a-Burillo}, {Garc{\'\i}a-Lorenzo},
  {Genzel}, {Gillessen}, {Gonz{\'a}lez Mart{\'\i}n}, {Haidar}, {Hermosa
  Mu{\~n}oz}, {Hicks}, {H{\"o}nig}, {Imanishi}, {Izumi}, {Labiano}, {Leist},
  {Levenson}, {Lopez-Rodriguez}, {Lutz}, {Ott}, {Packham}, {Rabien}, {Ramos
  Almeida}, {Ricci}, {Rigopoulou}, {Rosario}, {Rouan}, {Santos}, {Shangguan},
  {Stalevski}, {Sternberg}, {Sturm}, {Tacconi}, {Villar Mart{\'\i}n}, {Ward},
  \& {Zhang}}]{Davies2024}
{Davies}, R., {Shimizu}, T., {Pereira-Santaella}, M., {et~al.} 2024, \aap, 689,
  A263, \dodoi{10.1051/0004-6361/202449875}

\bibitem[{{De Robertis} \& {Osterbrock}(1984)}]{DeRobertis1984}
{De Robertis}, M.~M., \& {Osterbrock}, D.~E. 1984, \apj, 286, 171,
  \dodoi{10.1086/162585}

\bibitem[{{Emmering} {et~al.}(1992){Emmering}, {Blandford}, \&
  {Shlosman}}]{Emmering1992}
{Emmering}, R.~T., {Blandford}, R.~D., \& {Shlosman}, I. 1992, \apj, 385, 460,
  \dodoi{10.1086/170955}

\bibitem[{{Erwin}(2004)}]{Erwin2004}
{Erwin}, P. 2004, \aap, 415, 941, \dodoi{10.1051/0004-6361:20034408}

\bibitem[{{Esparza-Arredondo} {et~al.}(2025){Esparza-Arredondo}, {Ramos
  Almeida}, {Audibert}, {Pereira-Santaella}, {Garc{\'\i}a-Bernete},
  {Garc{\'\i}a-Burillo}, {Shimizu}, {Davies}, {Hermosa Mu{\~n}oz},
  {Alonso-Herrero}, {Combes}, {Speranza}, {Zhang}, {Campbell}, {Bellocchi},
  {Bunker}, {D{\'\i}az-Santos}, {Garc{\'\i}a-Lorenzo},
  {Gonz{\'a}lez-Mart{\'\i}n}, {Hicks}, {Labiano}, {Levenson}, {Ricci},
  {Rosario}, {Hoenig}, {Packham}, {Stalevski}, {Fuller}, {Izumi},
  {L{\'o}pez-Rodr{\'\i}guez}, {Rigopoulou}, {Rouan}, \& {Ward}}]{EA2025}
{Esparza-Arredondo}, D., {Ramos Almeida}, C., {Audibert}, A., {et~al.} 2025,
  \aap, 693, A174, \dodoi{10.1051/0004-6361/202452488}

\bibitem[{{Fabian} {et~al.}(2008){Fabian}, {Vasudevan}, \&
  {Gandhi}}]{Fabian2008}
{Fabian}, A.~C., {Vasudevan}, R.~V., \& {Gandhi}, P. 2008, \mnras, 385, L43,
  \dodoi{10.1111/j.1745-3933.2008.00430.x}

\bibitem[{{Ferruit} {et~al.}(2000){Ferruit}, {Wilson}, \&
  {Mulchaey}}]{Ferruit2000}
{Ferruit}, P., {Wilson}, A.~S., \& {Mulchaey}, J. 2000, \apjs, 128, 139,
  \dodoi{10.1086/313379}

\bibitem[{{G{\'a}mez Rosas} {et~al.}(2022){G{\'a}mez Rosas}, {Isbell}, {Jaffe},
  {Petrov}, {Leftley}, {Hofmann}, {Millour}, {Burtscher}, {Meisenheimer},
  {Meilland}, {Waters}, {Lopez}, {Lagarde}, {Weigelt}, {Berio}, {Allouche},
  {Robbe-Dubois}, {Cruzal{\`e}bes}, {Bettonvil}, {Henning}, {Augereau},
  {Antonelli}, {Beckmann}, {van Boekel}, {Bendjoya}, {Danchi}, {Dominik},
  {Drevon}, {Gallimore}, {Graser}, {Heininger}, {Hocd{\'e}}, {Hogerheijde},
  {Hron}, {Impellizzeri}, {Klarmann}, {Kokoulina}, {Labadie}, {Lehmitz},
  {Matter}, {Paladini}, {Pantin}, {Pott}, {Schertl}, {Soulain}, {Stee},
  {Tristram}, {Varga}, {Woillez}, {Wolf}, {Yoffe}, \& {Zins}}]{GamezRosas2022}
{G{\'a}mez Rosas}, V., {Isbell}, J.~W., {Jaffe}, W., {et~al.} 2022, \nat, 602,
  403, \dodoi{10.1038/s41586-021-04311-7}

\bibitem[{{Gandhi} {et~al.}(2009){Gandhi}, {Horst}, {Smette}, {H{\"o}nig},
  {Comastri}, {Gilli}, {Vignali}, \& {Duschl}}]{Gandhi2009}
{Gandhi}, P., {Horst}, H., {Smette}, A., {et~al.} 2009, \aap, 502, 457,
  \dodoi{10.1051/0004-6361/200811368}

\bibitem[{{Garc{\'\i}a-Bernete} {et~al.}(2016){Garc{\'\i}a-Bernete}, {Ramos
  Almeida}, {Acosta-Pulido}, {Alonso-Herrero}, {Gonz{\'a}lez-Mart{\'\i}n},
  {Hern{\'a}n-Caballero}, {Pereira-Santaella}, {Levenson}, {Packham},
  {Perlman}, {Ichikawa}, {Esquej}, \& {D{\'\i}az-Santos}}]{GB2016}
{Garc{\'\i}a-Bernete}, I., {Ramos Almeida}, C., {Acosta-Pulido}, J.~A.,
  {et~al.} 2016, \mnras, 463, 3531, \dodoi{10.1093/mnras/stw2125}

\bibitem[{{Garc{\'\i}a-Bernete}
  {et~al.}(2022{\natexlab{a}}){Garc{\'\i}a-Bernete},
  {Gonz{\'a}lez-Mart{\'\i}n}, {Ramos Almeida}, {Alonso-Herrero},
  {Mart{\'\i}nez-Paredes}, {Ward}, {Roche}, {Acosta-Pulido},
  {L{\'o}pez-Rodr{\'\i}guez}, {Rigopoulou}, \& {Esparza-Arredondo}}]{GB2022}
{Garc{\'\i}a-Bernete}, I., {Gonz{\'a}lez-Mart{\'\i}n}, O., {Ramos Almeida}, C.,
  {et~al.} 2022{\natexlab{a}}, \aap, 667, A140,
  \dodoi{10.1051/0004-6361/202244230}

\bibitem[{{Garc{\'\i}a-Bernete}
  {et~al.}(2022{\natexlab{b}}){Garc{\'\i}a-Bernete}, {Rigopoulou},
  {Alonso-Herrero}, {Donnan}, {Roche}, {Pereira-Santaella}, {Labiano}, {Peralta
  de Arriba}, {Izumi}, {Ramos Almeida}, {Shimizu}, {H{\"o}nig},
  {Garc{\'\i}a-Burillo}, {Rosario}, {Ward}, {Bellocchi}, {Hicks}, {Fuller}, \&
  {Packham}}]{GB2022b}
{Garc{\'\i}a-Bernete}, I., {Rigopoulou}, D., {Alonso-Herrero}, A., {et~al.}
  2022{\natexlab{b}}, \aap, 666, L5, \dodoi{10.1051/0004-6361/202244806}

\bibitem[{{Garc{\'\i}a-Bernete} {et~al.}(2024){Garc{\'\i}a-Bernete},
  {Alonso-Herrero}, {Rigopoulou}, {Pereira-Santaella}, {Shimizu}, {Davies},
  {Donnan}, {Roche}, {Gonz{\'a}lez-Mart{\'\i}n}, {Ramos Almeida}, {Bellocchi},
  {Boorman}, {Combes}, {Efstathiou}, {Esparza-Arredondo},
  {Garc{\'\i}a-Burillo}, {Gonz{\'a}lez-Alfonso}, {Hicks}, {H{\"o}nig},
  {Labiano}, {Levenson}, {L{\'o}pez-Rodr{\'\i}guez}, {Ricci}, {Packham},
  {Rouan}, {Stalevski}, \& {Ward}}]{GATOSIII}
{Garc{\'\i}a-Bernete}, I., {Alonso-Herrero}, A., {Rigopoulou}, D., {et~al.}
  2024, \aap, 681, L7, \dodoi{10.1051/0004-6361/202348266}

\bibitem[{{Garc{\'\i}a-Burillo} {et~al.}(2016){Garc{\'\i}a-Burillo}, {Combes},
  {Ramos Almeida}, {Usero}, {Krips}, {Alonso-Herrero}, {Aalto}, {Casasola},
  {Hunt}, {Mart{\'\i}n}, {Viti}, {Colina}, {Costagliola}, {Eckart}, {Fuente},
  {Henkel}, {M{\'a}rquez}, {Neri}, {Schinnerer}, {Tacconi}, \& {van der
  Werf}}]{GB2016b}
{Garc{\'\i}a-Burillo}, S., {Combes}, F., {Ramos Almeida}, C., {et~al.} 2016,
  \apjl, 823, L12, \dodoi{10.3847/2041-8205/823/1/L12}

\bibitem[{{Garc{\'\i}a-Burillo} {et~al.}(2019){Garc{\'\i}a-Burillo}, {Combes},
  {Ramos Almeida}, {Usero}, {Alonso-Herrero}, {Hunt}, {Rouan}, {Aalto},
  {Querejeta}, {Viti}, {van der Werf}, {Vives-Arias}, {Fuente}, {Colina},
  {Mart{\'\i}n-Pintado}, {Henkel}, {Mart{\'\i}n}, {Krips}, {Gratadour}, {Neri},
  \& {Tacconi}}]{GB2019}
---. 2019, \aap, 632, A61, \dodoi{10.1051/0004-6361/201936606}

\bibitem[{{Garc{\'\i}a-Burillo} {et~al.}(2021){Garc{\'\i}a-Burillo},
  {Alonso-Herrero}, {Ramos Almeida}, {Gonz{\'a}lez-Mart{\'\i}n}, {Combes},
  {Usero}, {H{\"o}nig}, {Querejeta}, {Hicks}, {Hunt}, {Rosario}, {Davies},
  {Boorman}, {Bunker}, {Burtscher}, {Colina}, {D{\'\i}az-Santos}, {Gandhi},
  {Garc{\'\i}a-Bernete}, {Garc{\'\i}a-Lorenzo}, {Ichikawa}, {Imanishi},
  {Izumi}, {Labiano}, {Levenson}, {L{\'o}pez-Rodr{\'\i}guez}, {Packham},
  {Pereira-Santaella}, {Ricci}, {Rigopoulou}, {Rouan}, {Shimizu}, {Stalevski},
  {Wada}, \& {Williamson}}]{GATOSI}
{Garc{\'\i}a-Burillo}, S., {Alonso-Herrero}, A., {Ramos Almeida}, C., {et~al.}
  2021, \aap, 652, A98, \dodoi{10.1051/0004-6361/202141075}

\bibitem[{{Garc{\'\i}a-Burillo} {et~al.}(2024){Garc{\'\i}a-Burillo}, {Hicks},
  {Alonso-Herrero}, {Pereira-Santaella}, {Usero}, {Querejeta},
  {Gonz{\'a}lez-Mart{\'\i}n}, {Delaney}, {Ramos Almeida}, {Combes},
  {Angl{\'e}s-Alc{\'a}zar}, {Audibert}, {Bellocchi}, {Davies}, {Davis},
  {Elford}, {Garc{\'\i}a-Bernete}, {H{\"o}nig}, {Labiano}, {Leist}, {Levenson},
  {L{\'o}pez-Rodr{\'\i}guez}, {Mercedes-Feliz}, {Packham}, {Ricci}, {Rosario},
  {Shimizu}, {Stalevski}, \& {Zhang}}]{GB2024}
{Garc{\'\i}a-Burillo}, S., {Hicks}, E.~K.~S., {Alonso-Herrero}, A., {et~al.}
  2024, \aap, 689, A347, \dodoi{10.1051/0004-6361/202450268}

\bibitem[{{Gonz{\'a}lez-Mart{\'\i}n} {et~al.}(2023){Gonz{\'a}lez-Mart{\'\i}n},
  {Ramos Almeida}, {Fritz}, {Alonso-Herrero}, {H{\"o}nig}, {Roche},
  {Esparza-Arredondo}, {Garc{\'\i}a-Bernete}, {Garc{\'\i}a-Burillo},
  {Osorio-Clavijo}, {Reyes-Amador}, {Stalevski}, \&
  {Victoria-Ceballos}}]{GM2023}
{Gonz{\'a}lez-Mart{\'\i}n}, O., {Ramos Almeida}, C., {Fritz}, J., {et~al.}
  2023, \aap, 676, A73, \dodoi{10.1051/0004-6361/202345858}

\bibitem[{{Groves} {et~al.}(2006){Groves}, {Dopita}, \&
  {Sutherland}}]{Groves2006}
{Groves}, B., {Dopita}, M., \& {Sutherland}, R. 2006, \aap, 458, 405,
  \dodoi{10.1051/0004-6361:20065097}

\bibitem[{Harris {et~al.}(2020)Harris, Millman, van~der Walt, Gommers,
  Virtanen, Cournapeau, Wieser, Taylor, Berg, Smith, Kern, Picus, Hoyer, van
  Kerkwijk, Brett, Haldane, del R{\'{i}}o, Wiebe, Peterson,
  G{\'{e}}rard-Marchant, Sheppard, Reddy, Weckesser, Abbasi, Gohlke, \&
  Oliphant}]{numpy2020}
Harris, C.~R., Millman, K.~J., van~der Walt, S.~J., {et~al.} 2020, Nature, 585,
  357, \dodoi{10.1038/s41586-020-2649-2}

\bibitem[{{Harrison} \& {Ramos Almeida}(2024)}]{HRA2024}
{Harrison}, C.~M., \& {Ramos Almeida}, C. 2024, Galaxies, 12, 17,
  \dodoi{10.3390/galaxies12020017}

\bibitem[{{Heckman} {et~al.}(1981){Heckman}, {Miley}, {van Breugel}, \&
  {Butcher}}]{Heckman1981}
{Heckman}, T.~M., {Miley}, G.~K., {van Breugel}, W.~J.~M., \& {Butcher}, H.~R.
  1981, \apj, 247, 403, \dodoi{10.1086/159050}

\bibitem[{{Hermosa Mu{\~n}oz} {et~al.}(2024){Hermosa Mu{\~n}oz},
  {Alonso-Herrero}, {Pereira-Santaella}, {Garc{\'\i}a-Bernete},
  {Garc{\'\i}a-Burillo}, {Garc{\'\i}a-Lorenzo}, {Davies}, {Shimizu},
  {Esparza-Arredondo}, {Hicks}, {Haidar}, {Leist}, {L{\'o}pez-Rodr{\'\i}guez},
  {Ramos Almeida}, {Rosario}, {Zhang}, {Audibert}, {Bellocchi}, {Boorman},
  {Bunker}, {Combes}, {Campbell}, {D{\'\i}az-Santos}, {Fuller}, {Gandhi},
  {Gonz{\'a}lez-Mart{\'\i}n}, {H{\"o}nig}, {Imanishi}, {Izumi}, {Labiano},
  {Levenson}, {Packham}, {Ricci}, {Rigopoulou}, {Rouan}, {Stalevski},
  {Villar-Mart{\'\i}n}, \& {Ward}}]{HM2024}
{Hermosa Mu{\~n}oz}, L., {Alonso-Herrero}, A., {Pereira-Santaella}, M.,
  {et~al.} 2024, \aap, 690, A350, \dodoi{10.1051/0004-6361/202450262}

\bibitem[{{H{\"o}nig}(2019)}]{Hoenig2019}
{H{\"o}nig}, S.~F. 2019, \apj, 884, 171, \dodoi{10.3847/1538-4357/ab4591}

\bibitem[{{H{\"o}nig} \& {Kishimoto}(2017)}]{Hoenig2017}
{H{\"o}nig}, S.~F., \& {Kishimoto}, M. 2017, \apjl, 838, L20,
  \dodoi{10.3847/2041-8213/aa6838}

\bibitem[{{H{\"o}nig} {et~al.}(2012){H{\"o}nig}, {Kishimoto}, {Antonucci},
  {Marconi}, {Prieto}, {Tristram}, \& {Weigelt}}]{Hoenig2012}
{H{\"o}nig}, S.~F., {Kishimoto}, M., {Antonucci}, R., {et~al.} 2012, \apj, 755,
  149, \dodoi{10.1088/0004-637X/755/2/149}

\bibitem[{{H{\"o}nig} {et~al.}(2013){H{\"o}nig}, {Kishimoto}, {Tristram},
  {Prieto}, {Gandhi}, {Asmus}, {Antonucci}, {Burtscher}, {Duschl}, \&
  {Weigelt}}]{Hoenig2013}
{H{\"o}nig}, S.~F., {Kishimoto}, M., {Tristram}, K.~R.~W., {et~al.} 2013, \apj,
  771, 87, \dodoi{10.1088/0004-637X/771/2/87}

\bibitem[{{Hull} {et~al.}(2022){Hull}, {Yang}, {Cort{\'e}s}, {Dent}, {Kral},
  {Li}, {Le Gouellec}, {Hughes}, {Milli}, {Teague}, \& {Wyatt}}]{Hull2022}
{Hull}, C. L.~H., {Yang}, H., {Cort{\'e}s}, P.~C., {et~al.} 2022, \apj, 930,
  49, \dodoi{10.3847/1538-4357/ac6023}

\bibitem[{Hunter(2007)}]{matplotlib2007}
Hunter, J.~D. 2007, Computing in Science \& Engineering, 9, 90,
  \dodoi{10.1109/MCSE.2007.55}

\bibitem[{{Imanishi} {et~al.}(2018){Imanishi}, {Nakanishi}, {Izumi}, \&
  {Wada}}]{Imanishi2018}
{Imanishi}, M., {Nakanishi}, K., {Izumi}, T., \& {Wada}, K. 2018, \apjl, 853,
  L25, \dodoi{10.3847/2041-8213/aaa8df}

\bibitem[{{Isbell} {et~al.}(2022){Isbell}, {Meisenheimer}, {Pott}, {Stalevski},
  {Tristram}, {Sanchez-Bermudez}, {Hofmann}, {G{\'a}mez Rosas}, {Jaffe},
  {Burtscher}, {Leftley}, {Petrov}, {Lopez}, {Henning}, {Weigelt}, {Allouche},
  {Berio}, {Bettonvil}, {Cruzalebes}, {Dominik}, {Heininger}, {Hogerheijde},
  {Lagarde}, {Lehmitz}, {Matter}, {Meilland}, {Millour}, {Robbe-Dubois},
  {Schertl}, {van Boekel}, {Varga}, \& {Woillez}}]{Isbell2022}
{Isbell}, J.~W., {Meisenheimer}, K., {Pott}, J.~U., {et~al.} 2022, \aap, 663,
  A35, \dodoi{10.1051/0004-6361/202243271}

\bibitem[{{Isbell} {et~al.}(2023){Isbell}, {Pott}, {Meisenheimer}, {Stalevski},
  {Tristram}, {Leftley}, {Asmus}, {Weigelt}, {G{\'a}mez Rosas}, {Petrov},
  {Jaffe}, {Hofmann}, {Henning}, \& {Lopez}}]{Isbell2023}
{Isbell}, J.~W., {Pott}, J.~U., {Meisenheimer}, K., {et~al.} 2023, \aap, 678,
  A136, \dodoi{10.1051/0004-6361/202347307}

\bibitem[{{Izumi} {et~al.}(2023){Izumi}, {Wada}, {Imanishi}, {Nakanishi},
  {Kohno}, {Kudoh}, {Kawamuro}, {Baba}, {Matsumoto}, {Fujita}, \&
  {Tristram}}]{Izumi2023}
{Izumi}, T., {Wada}, K., {Imanishi}, M., {et~al.} 2023, Science, 382, 554,
  \dodoi{10.1126/science.adf0569}

\bibitem[{{Jaffe} {et~al.}(2004){Jaffe}, {Meisenheimer}, {R{\"o}ttgering},
  {Leinert}, {Richichi}, {Chesneau}, {Fraix-Burnet}, {Glazenborg-Kluttig},
  {Granato}, {Graser}, {Heijligers}, {K{\"o}hler}, {Malbet}, {Miley},
  {Paresce}, {Pel}, {Perrin}, {Przygodda}, {Schoeller}, {Sol}, {Waters},
  {Weigelt}, {Woillez}, \& {de Zeeuw}}]{Jaffe2004}
{Jaffe}, W., {Meisenheimer}, K., {R{\"o}ttgering}, H.~J.~A., {et~al.} 2004,
  \nat, 429, 47, \dodoi{10.1038/nature02531}

\bibitem[{{Jones}(2004)}]{Jones2004}
{Jones}, A.~P. 2004, in Astronomical Society of the Pacific Conference Series,
  Vol. 309, Astrophysics of Dust, ed. A.~N. {Witt}, G.~C. {Clayton}, \& B.~T.
  {Draine}, 347

\bibitem[{{Kinney} {et~al.}(2000){Kinney}, {Schmitt}, {Clarke}, {Pringle},
  {Ulvestad}, \& {Antonucci}}]{Kinney2000}
{Kinney}, A.~L., {Schmitt}, H.~R., {Clarke}, C.~J., {et~al.} 2000, \apj, 537,
  152, \dodoi{10.1086/309016}

\bibitem[{{Koss} {et~al.}(2017){Koss}, {Trakhtenbrot}, {Ricci}, {Lamperti},
  {Oh}, {Berney}, {Schawinski}, {Balokovi{\'c}}, {Baronchelli}, {Crenshaw},
  {Fischer}, {Gehrels}, {Harrison}, {Hashimoto}, {Hogg}, {Ichikawa}, {Masetti},
  {Mushotzky}, {Sartori}, {Stern}, {Treister}, {Ueda}, {Veilleux}, \&
  {Winter}}]{Koss2017}
{Koss}, M., {Trakhtenbrot}, B., {Ricci}, C., {et~al.} 2017, \apj, 850, 74,
  \dodoi{10.3847/1538-4357/aa8ec9}

\bibitem[{{Labiano} {et~al.}(2021){Labiano}, {Argyriou},
  {{\'A}lvarez-M{\'a}rquez}, {Glasse}, {Glauser}, {Patapis}, {Law}, {Brandl},
  {Justtanont}, {Lahuis}, {Mart{\'\i}nez-Galarza}, {Mueller}, {Noriega-Crespo},
  {Royer}, {Shaughnessy}, \& {Vandenbussche}}]{Labiano2021}
{Labiano}, A., {Argyriou}, I., {{\'A}lvarez-M{\'a}rquez}, J., {et~al.} 2021,
  \aap, 656, A57, \dodoi{10.1051/0004-6361/202140614}

\bibitem[{{Lazarian} \& {Hoang}(2007)}]{LH2007}
{Lazarian}, A., \& {Hoang}, T. 2007, \mnras, 378, 910,
  \dodoi{10.1111/j.1365-2966.2007.11817.x}

\bibitem[{{Le Gouellec} {et~al.}(2023){Le Gouellec}, {Andersson}, {Soam},
  {Schirmer}, {Michail}, {Lopez-Rodriguez}, {Flores}, {Chuss}, {Vaillancourt},
  {Hoang}, \& {Lazarian}}]{LG2023}
{Le Gouellec}, V. J.~M., {Andersson}, B.~G., {Soam}, A., {et~al.} 2023, \apj,
  951, 97, \dodoi{10.3847/1538-4357/accff7}

\bibitem[{{Leftley} {et~al.}(2018){Leftley}, {Tristram}, {H{\"o}nig},
  {Kishimoto}, {Asmus}, \& {Gandhi}}]{Leftley2018}
{Leftley}, J.~H., {Tristram}, K. R.~W., {H{\"o}nig}, S.~F., {et~al.} 2018,
  \apj, 862, 17, \dodoi{10.3847/1538-4357/aac8e5}

\bibitem[{{Levenson} {et~al.}(2009){Levenson}, {Radomski}, {Packham}, {Mason},
  {Schaefer}, \& {Telesco}}]{Levenson2009}
{Levenson}, N.~A., {Radomski}, J.~T., {Packham}, C., {et~al.} 2009, \apj, 703,
  390, \dodoi{10.1088/0004-637X/703/1/390}

\bibitem[{{Libralato} {et~al.}(2023){Libralato}, {Argyriou}, {Dicken},
  {Garc{\'\i}a Mar{\'\i}n}, {Guillard}, {Hines}, {Kavanagh}, {Kendrew}, {Law},
  {Noriega-Crespo}, \& {{\'A}lvarez-M{\'a}rquez}}]{Libralato2023}
{Libralato}, M., {Argyriou}, I., {Dicken}, D., {et~al.} 2023, arXiv e-prints,
  arXiv:2311.12145, \dodoi{10.48550/arXiv.2311.12145}

\bibitem[{{L{\'o}pez-Gonzaga} {et~al.}(2016){L{\'o}pez-Gonzaga}, {Burtscher},
  {Tristram}, {Meisenheimer}, \& {Schartmann}}]{LG2016}
{L{\'o}pez-Gonzaga}, N., {Burtscher}, L., {Tristram}, K.~R.~W., {Meisenheimer},
  K., \& {Schartmann}, M. 2016, \aap, 591, A47,
  \dodoi{10.1051/0004-6361/201527590}

\bibitem[{{L{\'o}pez-Gonzaga} \& {Jaffe}(2016)}]{LG2016b}
{L{\'o}pez-Gonzaga}, N., \& {Jaffe}, W. 2016, \aap, 591, A128,
  \dodoi{10.1051/0004-6361/201527149}

\bibitem[{{L{\'o}pez-Gonzaga} {et~al.}(2014){L{\'o}pez-Gonzaga}, {Jaffe},
  {Burtscher}, {Tristram}, \& {Meisenheimer}}]{LG2014}
{L{\'o}pez-Gonzaga}, N., {Jaffe}, W., {Burtscher}, L., {Tristram}, K.~R.~W., \&
  {Meisenheimer}, K. 2014, \aap, 565, A71, \dodoi{10.1051/0004-6361/201323002}

\bibitem[{{Lopez-Rodriguez} {et~al.}(2015){Lopez-Rodriguez}, {Packham},
  {Jones}, {Nikutta}, {McMaster}, {Mason}, {Elvis}, {Shenoy}, {Alonso-Herrero},
  {Ram{\'\i}rez}, {Gonz{\'a}lez Mart{\'\i}n}, {H{\"o}nig}, {Levenson}, {Ramos
  Almeida}, \& {Perlman}}]{ELR2015}
{Lopez-Rodriguez}, E., {Packham}, C., {Jones}, T.~J., {et~al.} 2015, \mnras,
  452, 1902, \dodoi{10.1093/mnras/stv1410}

\bibitem[{{Lopez-Rodriguez} {et~al.}(2016){Lopez-Rodriguez}, {Packham},
  {Roche}, {Alonso-Herrero}, {D{\'\i}az-Santos}, {Nikutta},
  {Gonz{\'a}lez-Mart{\'\i}n}, {{\'A}lvarez}, {Esquej}, {Espinosa}, {Perlman},
  {Ramos Almeida}, \& {Telesco}}]{ELR2016}
{Lopez-Rodriguez}, E., {Packham}, C., {Roche}, P.~F., {et~al.} 2016, \mnras,
  458, 3851, \dodoi{10.1093/mnras/stw541}

\bibitem[{{Lopez-Rodriguez} {et~al.}(2017){Lopez-Rodriguez}, {Packham},
  {Jones}, {Siebenmorgen}, {Roche}, {Levenson}, {Alonso-Herrero}, {Perlman},
  {Ichikawa}, {Ramos Almeida}, {Gonz{\'a}lez-Mart{\'\i}n}, {Nikutta},
  {Martinez-Paredez}, {Shenoy}, {Gordon}, \& {Telesco}}]{ELR2017}
{Lopez-Rodriguez}, E., {Packham}, C., {Jones}, T.~J., {et~al.} 2017, \mnras,
  464, 1762, \dodoi{10.1093/mnras/stw2491}

\bibitem[{{Lopez-Rodriguez} {et~al.}(2018{\natexlab{a}}){Lopez-Rodriguez},
  {Alonso-Herrero}, {Diaz-Santos}, {Gonzalez-Martin}, {Ichikawa}, {Levenson},
  {Martinez-Paredes}, {Nikutta}, {Packham}, {Perlman}, {Ramos Almeida},
  {Rodriguez-Espinosa}, \& {Telesco}}]{ELR2018}
{Lopez-Rodriguez}, E., {Alonso-Herrero}, A., {Diaz-Santos}, T., {et~al.}
  2018{\natexlab{a}}, \mnras, 478, 2350, \dodoi{10.1093/mnras/sty1197}

\bibitem[{{Lopez-Rodriguez} {et~al.}(2018{\natexlab{b}}){Lopez-Rodriguez},
  {Fuller}, {Alonso-Herrero}, {Efstathiou}, {Ichikawa}, {Levenson}, {Packham},
  {Radomski}, {Ramos Almeida}, {Benford}, {Berthoud}, {Hamilton}, {Harper},
  {Kov{\'a}vcs}, {Santos}, {Staguhn}, \& {Herter}}]{ELR2018b}
{Lopez-Rodriguez}, E., {Fuller}, L., {Alonso-Herrero}, A., {et~al.}
  2018{\natexlab{b}}, \apj, 859, 99, \dodoi{10.3847/1538-4357/aabd7b}

\bibitem[{{Lopez-Rodriguez} {et~al.}(2020){Lopez-Rodriguez}, {Alonso-Herrero},
  {Garc{\'\i}a-Burillo}, {Gordon}, {Ichikawa}, {Imanishi}, {Kameno},
  {Levenson}, {Nikutta}, \& {Packham}}]{ELR2020}
{Lopez-Rodriguez}, E., {Alonso-Herrero}, A., {Garc{\'\i}a-Burillo}, S.,
  {et~al.} 2020, \apj, 893, 33, \dodoi{10.3847/1538-4357/ab8013}

\bibitem[{{Lopez-Rodriguez} {et~al.}(2025){Lopez-Rodriguez},
  {Sanchez-Bermudez}, {Gonzalez-Martin}, {Nikutta}, {Lau}, {Thatte},
  {Garcia-Bernete}, {Girard}, \& {Hankins}}]{LR2025}
{Lopez-Rodriguez}, E., {Sanchez-Bermudez}, J., {Gonzalez-Martin}, O., {et~al.}
  2025, arXiv e-prints, arXiv:2506.08077, \dodoi{10.48550/arXiv.2506.08077}

\bibitem[{{MacAlpine}(1985)}]{MacAlpine1985}
{MacAlpine}, G.~M. 1985, in Astrophysics of Active Galaxies and Quasi-Stellar
  Objects, ed. J.~S. {Miller}, 259--288

\bibitem[{{Maiolino} {et~al.}(2001{\natexlab{a}}){Maiolino}, {Marconi}, \&
  {Oliva}}]{Maiolino2001b}
{Maiolino}, R., {Marconi}, A., \& {Oliva}, E. 2001{\natexlab{a}}, \aap, 365,
  37, \dodoi{10.1051/0004-6361:20000012}

\bibitem[{{Maiolino} {et~al.}(2001{\natexlab{b}}){Maiolino}, {Marconi},
  {Salvati}, {Risaliti}, {Severgnini}, {Oliva}, {La Franca}, \&
  {Vanzi}}]{Maiolino2001a}
{Maiolino}, R., {Marconi}, A., {Salvati}, M., {et~al.} 2001{\natexlab{b}},
  \aap, 365, 28, \dodoi{10.1051/0004-6361:20000177}

\bibitem[{{Malkan} {et~al.}(1998){Malkan}, {Gorjian}, \& {Tam}}]{Malkan1998}
{Malkan}, M.~A., {Gorjian}, V., \& {Tam}, R. 1998, \apjs, 117, 25,
  \dodoi{10.1086/313110}

\bibitem[{{Mason} {et~al.}(2006){Mason}, {Geballe}, {Packham}, {Levenson},
  {Elitzur}, {Fisher}, \& {Perlman}}]{Mason2006}
{Mason}, R.~E., {Geballe}, T.~R., {Packham}, C., {et~al.} 2006, \apj, 640, 612,
  \dodoi{10.1086/500299}

\bibitem[{{Mediavilla} \& {Arribas}(1995)}]{MA1995}
{Mediavilla}, E., \& {Arribas}, S. 1995, \mnras, 276, 579,
  \dodoi{10.1093/mnras/276.2.579}

\bibitem[{{Morganti} {et~al.}(1999){Morganti}, {Tsvetanov}, {Gallimore}, \&
  {Allen}}]{Morganti1999}
{Morganti}, R., {Tsvetanov}, Z.~I., {Gallimore}, J., \& {Allen}, M.~G. 1999,
  \aaps, 137, 457, \dodoi{10.1051/aas:1999258}

\bibitem[{{Mundell} {et~al.}(2009){Mundell}, {Ferruit}, {Nagar}, \&
  {Wilson}}]{Mundell2009}
{Mundell}, C.~G., {Ferruit}, P., {Nagar}, N., \& {Wilson}, A.~S. 2009, \apj,
  703, 802, \dodoi{10.1088/0004-637X/703/1/802}

\bibitem[{{Nagar} {et~al.}(1999){Nagar}, {Wilson}, {Mulchaey}, \&
  {Gallimore}}]{Nagar1999}
{Nagar}, N.~M., {Wilson}, A.~S., {Mulchaey}, J.~S., \& {Gallimore}, J.~F. 1999,
  \apjs, 120, 209, \dodoi{10.1086/313183}

\bibitem[{{Nenkova} {et~al.}(2008{\natexlab{a}}){Nenkova}, {Sirocky},
  {Ivezi{\'c}}, \& {Elitzur}}]{Nenkova2008a}
{Nenkova}, M., {Sirocky}, M.~M., {Ivezi{\'c}}, {\v{Z}}., \& {Elitzur}, M.
  2008{\natexlab{a}}, \apj, 685, 147, \dodoi{10.1086/590482}

\bibitem[{{Nenkova} {et~al.}(2008{\natexlab{b}}){Nenkova}, {Sirocky},
  {Nikutta}, {Ivezi{\'c}}, \& {Elitzur}}]{Nenkova2008b}
{Nenkova}, M., {Sirocky}, M.~M., {Nikutta}, R., {Ivezi{\'c}}, {\v{Z}}., \&
  {Elitzur}, M. 2008{\natexlab{b}}, \apj, 685, 160, \dodoi{10.1086/590483}

\bibitem[{{Netzer} \& {Laor}(1993)}]{Netzer1993}
{Netzer}, H., \& {Laor}, A. 1993, \apjl, 404, L51, \dodoi{10.1086/186741}

\bibitem[{{Nikutta} {et~al.}(2021{\natexlab{a}}){Nikutta}, {Lopez-Rodriguez},
  {Ichikawa}, {Levenson}, {Packham}, {H{\"o}nig}, \&
  {Alonso-Herrero}}]{Nikutta2021a}
{Nikutta}, R., {Lopez-Rodriguez}, E., {Ichikawa}, K., {et~al.}
  2021{\natexlab{a}}, \apj, 919, 136, \dodoi{10.3847/1538-4357/ac06a6}

\bibitem[{{Nikutta} {et~al.}(2021{\natexlab{b}}){Nikutta}, {Lopez-Rodriguez},
  {Ichikawa}, {Levenson}, {Packham}, {H{\"o}nig}, \&
  {Alonso-Herrero}}]{Nikutta2021b}
---. 2021{\natexlab{b}}, \apj, 923, 127, \dodoi{10.3847/1538-4357/ac2949}

\bibitem[{{Osterbrock}(1989)}]{Osterbrock1989}
{Osterbrock}, D.~E. 1989, {Astrophysics of gaseous nebulae and active galactic
  nuclei}

\bibitem[{{Packham} {et~al.}(1997){Packham}, {Young}, {Hough}, {Axon}, \&
  {Bailey}}]{Packham1997}
{Packham}, C., {Young}, S., {Hough}, J.~H., {Axon}, D.~J., \& {Bailey}, J.~A.
  1997, \mnras, 288, 375, \dodoi{10.1093/mnras/288.2.375}

\bibitem[{pandas~development team(2020)}]{reback2020pandas}
pandas~development team, T. 2020, pandas-dev/pandas: Pandas, latest,  Zenodo,
  \dodoi{10.5281/zenodo.3509134}

\bibitem[{{Pereira-Santaella} {et~al.}(2022){Pereira-Santaella},
  {{\'A}lvarez-M{\'a}rquez}, {Garc{\'\i}a-Bernete}, {Labiano}, {Colina},
  {Alonso-Herrero}, {Bellocchi}, {Garc{\'\i}a-Burillo}, {H{\"o}nig}, {Ramos
  Almeida}, \& {Rosario}}]{PereiraSantaella2022}
{Pereira-Santaella}, M., {{\'A}lvarez-M{\'a}rquez}, J., {Garc{\'\i}a-Bernete},
  I., {et~al.} 2022, \aap, 665, L11, \dodoi{10.1051/0004-6361/202244725}

\bibitem[{{Prada} \& {Guti{\'e}rrez}(1999)}]{Prada1999}
{Prada}, F., \& {Guti{\'e}rrez}, C.~M. 1999, \apj, 517, 123,
  \dodoi{10.1086/307199}

\bibitem[{{Prieto} {et~al.}(2014){Prieto}, {Mezcua}, {Fern{\'a}ndez-Ontiveros},
  \& {Schartmann}}]{Prieto2014}
{Prieto}, M.~A., {Mezcua}, M., {Fern{\'a}ndez-Ontiveros}, J.~A., \&
  {Schartmann}, M. 2014, \mnras, 442, 2145, \dodoi{10.1093/mnras/stu1006}

\bibitem[{{Proga} {et~al.}(1998){Proga}, {Stone}, \& {Drew}}]{Proga1998}
{Proga}, D., {Stone}, J.~M., \& {Drew}, J.~E. 1998, \mnras, 295, 595,
  \dodoi{10.1046/j.1365-8711.1998.01337.x}

\bibitem[{{Quillen} {et~al.}(1999){Quillen}, {Alonso-Herrero}, {Rieke},
  {McDonald}, {Falcke}, \& {Rieke}}]{Quillen1999}
{Quillen}, A.~C., {Alonso-Herrero}, A., {Rieke}, M.~J., {et~al.} 1999, \apj,
  525, 685, \dodoi{10.1086/307933}

\bibitem[{{Raban} {et~al.}(2008){Raban}, {Heijligers}, {R{\"o}ttgering},
  {Meisenheimer}, {Jaffe}, {K{\"a}ufl}, \& {Henning}}]{Raban2008}
{Raban}, D., {Heijligers}, B., {R{\"o}ttgering}, H., {et~al.} 2008, \aap, 484,
  341, \dodoi{10.1051/0004-6361:20077444}

\bibitem[{{Raban} {et~al.}(2009){Raban}, {Jaffe}, {R{\"o}ttgering},
  {Meisenheimer}, \& {Tristram}}]{Raban2009}
{Raban}, D., {Jaffe}, W., {R{\"o}ttgering}, H., {Meisenheimer}, K., \&
  {Tristram}, K. R.~W. 2009, \mnras, 394, 1325,
  \dodoi{10.1111/j.1365-2966.2009.14439.x}

\bibitem[{{Radomski} {et~al.}(2003){Radomski}, {Pi{\~n}a}, {Packham},
  {Telesco}, {De Buizer}, {Fisher}, \& {Robinson}}]{Radomski2003}
{Radomski}, J.~T., {Pi{\~n}a}, R.~K., {Packham}, C., {et~al.} 2003, \apj, 587,
  117, \dodoi{10.1086/367612}

\bibitem[{{Ramakrishnan} {et~al.}(2019){Ramakrishnan}, {Nagar}, {Finlez},
  {Storchi-Bergmann}, {Slater}, {Schnorr-M{\"u}ller}, {Riffel}, {Mundell}, \&
  {Robinson}}]{Ramakrishnan2019}
{Ramakrishnan}, V., {Nagar}, N.~M., {Finlez}, C., {et~al.} 2019, \mnras, 487,
  444, \dodoi{10.1093/mnras/stz1244}

\bibitem[{{Ramos Almeida} \& {Ricci}(2017)}]{Ramos2017}
{Ramos Almeida}, C., \& {Ricci}, C. 2017, Nature Astronomy, 1, 679,
  \dodoi{10.1038/s41550-017-0232-z}

\bibitem[{{Ricci} {et~al.}(2017){Ricci}, {Trakhtenbrot}, {Koss}, {Ueda},
  {Schawinski}, {Oh}, {Lamperti}, {Mushotzky}, {Treister}, {Ho}, {Weigel},
  {Bauer}, {Paltani}, {Fabian}, {Xie}, \& {Gehrels}}]{Ricci2017}
{Ricci}, C., {Trakhtenbrot}, B., {Koss}, M.~J., {et~al.} 2017, \nat, 549, 488,
  \dodoi{10.1038/nature23906}

\bibitem[{{Ricci} {et~al.}(2023){Ricci}, {Ichikawa}, {Stalevski}, {Kawamuro},
  {Yamada}, {Ueda}, {Mushotzky}, {Privon}, {Koss}, {Trakhtenbrot}, {Fabian},
  {Ho}, {Asmus}, {Bauer}, {Chang}, {Gupta}, {Oh}, {Powell}, {Pfeifle}, {Rojas},
  {Ricci}, {Temple}, {Toba}, {Tortosa}, {Treister}, {Harrison}, {Stern}, \&
  {Urry}}]{Ricci23}
{Ricci}, C., {Ichikawa}, K., {Stalevski}, M., {et~al.} 2023, \apj, 959, 27,
  \dodoi{10.3847/1538-4357/ad0733}

\bibitem[{{Rieke} {et~al.}(2015){Rieke}, {Ressler}, {Morrison}, {Bergeron},
  {Bouchet}, {Garc{\'\i}a-Mar{\'\i}n}, {Greene}, {Regan}, {Sukhatme}, \&
  {Walker}}]{Rieke2015}
{Rieke}, G.~H., {Ressler}, M.~E., {Morrison}, J.~E., {et~al.} 2015, \pasp, 127,
  665, \dodoi{10.1086/682257}

\bibitem[{{Rigby} {et~al.}(2023){Rigby}, {Perrin}, {McElwain}, {Kimble},
  {Friedman}, {Lallo}, {Doyon}, {Feinberg}, {Ferruit}, {Glasse}, {Rieke},
  {Rieke}, {Wright}, {Willott}, {Colon}, {Milam}, {Neff}, {Stark}, {Valenti},
  {Abell}, {Abney}, {Abul-Huda}, {Acton}, {Adams}, {Adler}, {Aguilar}, {Ahmed},
  {Albert}, {Alberts}, {Aldridge}, {Allen}, {Altenburg},
  {{\'A}lvarez-M{\'a}rquez}, {Alves de Oliveira}, {Andersen}, {Anderson},
  {Anderson}, {Argyriou}, {Armstrong}, {Arribas}, {Artigau}, {Arvai},
  {Atkinson}, {Bacon}, {Bair}, {Banks}, {Barrientes}, {Barringer}, {Bartosik},
  {Bast}, {Baudoz}, {Beatty}, {Bechtold}, {Beck}, {Bergeron}, {Bergkoetter},
  {Bhatawdekar}, {Birkmann}, {Blazek}, {Blome}, {Boccaletti}, {B{\"o}ker},
  {Boia}, {Bonaventura}, {Bond}, {Bosley}, {Boucarut}, {Bourque}, {Bouwman},
  {Bower}, {Bowers}, {Boyer}, {Bradley}, {Brady}, {Braun}, {Breda},
  {Bresnahan}, {Bright}, {Britt}, {Bromenschenkel}, {Brooks}, {Brooks},
  {Brown}, {Brown}, {Brown}, {Bunker}, {Burger}, {Bushouse}, {Cale}, {Cameron},
  {Cameron}, {Canipe}, {Caplinger}, {Caputo}, {Cara}, {Carey}, {Carniani},
  {Carrasquilla}, {Carruthers}, {Case}, {Catherine}, {Chance}, {Chapman},
  {Charlot}, {Charlow}, {Chayer}, {Chen}, {Cherinka}, {Chichester}, {Chilton},
  {Chonis}, {Clampin}, {Clark}, {Clark}, {Coe}, {Coleman}, {Comber}, {Comeau},
  {Connolly}, {Cooper}, {Cooper}, {Coppock}, {Correnti}, {Cossou}, {Coulais},
  {Coyle}, {Cracraft}, {Curti}, {Cuturic}, {Davis}, {Davis}, {Dean}, {DeLisa},
  {deMeester}, {Dencheva}, {Dencheva}, {DePasquale}, {Deschenes}, {Hunor
  Detre}, {Diaz}, {Dicken}, {DiFelice}, {Dillman}, {Dixon}, {Doggett},
  {Donaldson}, {Douglas}, {DuPrie}, {Dupuis}, {Durning}, {Easmin}, {Eck},
  {Edeani}, {Egami}, {Ehrenwinkler}, {Eisenhamer}, {Eisenhower}, {Elie},
  {Elliott}, {Elliott}, {Ellis}, {Engesser}, {Espinoza}, {Etienne}, {Etxaluze},
  {Falini}, {Feeney}, {Ferry}, {Filippazzo}, {Fincham}, {Fix}, {Flagey},
  {Florian}, {Flynn}, {Fontanella}, {Ford}, {Forshay}, {Fox}, {Franz}, {Fu},
  {Fullerton}, {Galkin}, {Galyer}, {Garc{\'\i}a Mar{\'\i}n}, {Gardner},
  {Gardner}, {Garland}, {Garrett}, {Gasman}, {Gaspar}, {Gaudreau}, {Gauthier},
  {Geers}, {Geithner}, {Gennaro}, {Giardino}, {Girard}, {Giuliano},
  {Glassmire}, {Glauser}, {Glazer}, {Godfrey}, {Golimowski}, {Gollnitz},
  {Gong}, {Gonzaga}, {Gordon}, {Gordon}, {Goudfrooij}, {Greene}, {Greenhouse},
  {Grimaldi}, {Groebner}, {Grundy}, {Guillard}, {Gutman}, {Ha}, {Haderlein},
  {Hagedorn}, {Hainline}, {Haley}, {Hami}, {Hamilton}, {Hammel}, {Hansen},
  {Harkins}, {Harr}, {Hart}, {Hart}, {Hartig}, {Hashimoto}, {Haskins},
  {Hathaway}, {Havey}, {Hayden}, {Hecht}, {Heller-Boyer}, {Henriques}, {Henry},
  {Hermann}, {Hernandez}, {Hesman}, {Hicks}, {Hilbert}, {Hines}, {Hoffman},
  {Holfeltz}, {Holler}, {Hoppa}, {Hott}, {Howard}, {Howard}, {Hunter},
  {Hunter}, {Hurst}, {Husemann}, {Hustak}, {Ilinca Ignat}, {Illingworth},
  {Irish}, {Jackson}, {Jahromi}, {Jakobsen}, {James}, {James}, {Januszewski},
  {Jenkins}, {Jirdeh}, {Johnson}, {Johnson}, {Jones}, {Jones}, {Jones},
  {Jones}, {Jordan}, {Jordan}, {Jurczyk}, {Jurling}, {Kaleida}, {Kalmanson},
  {Kammerer}, {Kang}, {Kao}, {Karakla}, {Kavanagh}, {Kelly}, {Kendrew},
  {Kennedy}, {Kenny}, {Keski-kuha}, {Keyes}, {Kidwell}, {Kinzel}, {Kirk},
  {Kirkpatrick}, {Kirshenblat}, {Klaassen}, {Knapp}, {Knight}, {Knollenberg},
  {Koehler}, {Koekemoer}, {Kovacs}, {Kulp}, {Kumari}, {Kyprianou}, {La Massa},
  {Labador}, {Labiano}, {Lagage}, {Lajoie}, {Lallo}, {Lam}, {Lamb}, {Lambros},
  {Lampenfield}, {Langston}, {Larson}, {Law}, {Lawrence}, {Lee}, {Leisenring},
  {Lepo}, {Leveille}, {Levenson}, {Levine}, {Levy}, {Lewis}, {Lewis},
  {Libralato}, {Lightsey}, {Link}, {Liu}, {Lo}, {Lockwood}, {Logue}, {Long},
  {Long}, {Loomis}, {Lopez-Caniego}, {Lorenzo Alvarez}, {Love-Pruitt}, {Lucy},
  {Luetzgendorf}, {Maghami}, {Maiolino}, {Major}, {Malla}, {Malumuth},
  {Manjavacas}, {Mannfolk}, {Marrione}, {Marston}, {Martel}, {Maschmann},
  {Masci}, {Masciarelli}, {Maszkiewicz}, {Mather}, {McKenzie}, {McLean},
  {McMaster}, {Melbourne}, {Mel{\'e}ndez}, {Menzel}, {Merz}, {Meyett}, {Meza},
  {Miskey}, {Misselt}, {Moller}, {Morrison}, {Morse}, {Moseley}, {Mosier},
  {Mountain}, {Mueckay}, {Mueller}, {Mullally}, {Murphy}, {Murray}, {Murray},
  {Mustelier}, {Muzerolle}, {Mycroft}, {Myers}, {Myrick}, {Nanavati}, {Nance},
  {Nayak}, {Naylor}, {Nelan}, {Nickson}, {Nielson}, {Nieto-Santisteban},
  {Nikolov}, {Noriega-Crespo}, {O'Shaughnessy}, {O'Sullivan}, {Ochs}, {Ogle},
  {Oleszczuk}, {Olmsted}, {Osborne}, {Ottens}, {Owens}, {Pacifici}, {Pagan},
  {Page}, {Park}, {Parrish}, {Patapis}, {Paul}, {Pauly}, {Pavlovsky}, {Pedder},
  {Peek}, {Pena-Guerrero}, {Penanen}, {Perez}, {Perna}, {Perriello},
  {Phillips}, {Pietraszkiewicz}, {Pinaud}, {Pirzkal}, {Pitman}, {Piwowar},
  {Platais}, {Player}, {Plesha}, {Pollizi}, {Polster}, {Pontoppidan},
  {Porterfield}, {Proffitt}, {Pueyo}, {Pulliam}, {Quirt}, {Quispe Neira},
  {Ramos Alarcon}, {Ramsay}, {Rapp}, {Rapp}, {Rauscher}, {Ravindranath},
  {Rawle}, {Regan}, {Reichard}, {Reis}, {Ressler}, {Rest}, {Reynolds}, {Rhue},
  {Richon}, {Rickman}, {Ridgaway}, {Ritchie}, {Rix}, {Robberto}, {Robinson},
  {Robinson}, {Robinson}, {Rock}, {Rodriguez}, {Rodriguez Del Pino}, {Roellig},
  {Rohrbach}, {Roman}, {Romelfanger}, {Rose}, {Roteliuk}, {Roth}, {Rothwell},
  {Rowlands}, {Roy}, {Royer}, {Royle}, {Rui}, {Rumler}, {Runnels}, {Russ},
  {Rustamkulov}, {Ryden}, {Ryer}, {Sabata}, {Sabatke}, {Sabbi}, {Samuelson},
  {Sapp}, {Sappington}, {Sargent}, {Sauer}, {Scheithauer}, {Schlawin},
  {Schlitz}, {Schmitz}, {Schneider}, {Schreiber}, {Schulze}, {Schwab}, {Scott},
  {Sembach}, {Shanahan}, {Shaughnessy}, {Shaw}, {Shawger}, {Shay}, {Sheehan},
  {Shen}, {Sherman}, {Shiao}, {Shih}, {Shivaei}, {Sienkiewicz}, {Sing},
  {Sirianni}, {Sivaramakrishnan}, {Skipper}, {Sloan}, {Slocum}, {Slowinski},
  {Smith}, {Smith}, {Smith}, {Smith}, {Snyder}, {Soh}, {Sohn}, {Soto},
  {Spencer}, {Stallcup}, {Stansberry}, {Starr}, {Starr}, {Stewart},
  {Stiavelli}, {Straughn}, {Strickland}, {Stys}, {Summers}, {Sun}, {Sunnquist},
  {Swade}, {Swam}, {Swaters}, {Swoish}, {Taylor}, {Taylor}, {Te Plate}, {Tea},
  {Teague}, {Telfer}, {Temim}, {Thatte}, {Thompson}, {Thompson}, {Thomson},
  {Tikkanen}, {Tippet}, {Todd}, {Toolan}, {Tran}, {Trejo}, {Truong},
  {Tsukamoto}, {Tustain}, {Tyra}, {Ubeda}, {Underwood}, {Uzzo}, {Van Campen},
  {Vandal}, {Vandenbussche}, {Vila}, {Volk}, {Wahlgren}, {Waldman}, {Walker},
  {Wander}, {Warfield}, {Warner}, {Wasiak}, {Watkins}, {Weaver}, {Weilert},
  {Weiser}, {Weiss}, {Weissman}, {Welty}, {West}, {Wheate}, {Wheatley},
  {Wheeler}, {White}, {Whiteaker}, {Whitehouse}, {Whiteleather}, {Whitman},
  {Williams}, {Willmer}, {Willoughby}, {Wilson}, {Wirth}, {Wislowski}, {Wolf},
  {Wolfe}, {Wolff}, {Workman}, {Wright}, {Wu}, {Wu}, {Wymer}, {Yates},
  {Yeager}, {Yeates}, {Yerger}, {Yoon}, {Young}, {Yu}, {Zak}, {Zeidler},
  {Zhou}, {Zielinski}, {Zincke}, \& {Zonak}}]{Rigby2023}
{Rigby}, J., {Perrin}, M., {McElwain}, M., {et~al.} 2023, \pasp, 135, 048001,
  \dodoi{10.1088/1538-3873/acb293}

\bibitem[{Robitaille(2019)}]{aplpy2019}
Robitaille, T. 2019, {APLpy v2.0: The Astronomical Plotting Library in Python},
  \dodoi{10.5281/zenodo.2567476}

\bibitem[{{Robitaille} \& {Bressert}(2012)}]{aplpy2012}
{Robitaille}, T., \& {Bressert}, E. 2012, {APLpy: Astronomical Plotting Library
  in Python}, Astrophysics Source Code Library.
\newblock \doeprint{1208.017}

\bibitem[{{Rodrigo} \& {Solano}(2020)}]{SVO2020}
{Rodrigo}, C., \& {Solano}, E. 2020, in XIV.0 Scientific Meeting (virtual) of
  the Spanish Astronomical Society, 182

\bibitem[{{Rodrigo} {et~al.}(2012){Rodrigo}, {Solano}, \& {Bayo}}]{SVO2012}
{Rodrigo}, C., {Solano}, E., \& {Bayo}, A. 2012, {SVO Filter Profile Service
  Version 1.0}, IVOA Working Draft 15 October 2012,
  \dodoi{10.5479/ADS/bib/2012ivoa.rept.1015R}

\bibitem[{{Rosario} {et~al.}(2018){Rosario}, {Burtscher}, {Davies}, {Koss},
  {Ricci}, {Lutz}, {Riffel}, {Alexander}, {Genzel}, {Hicks}, {Lin},
  {Maciejewski}, {M{\"u}ller-S{\'a}nchez}, {Orban de Xivry}, {Riffel},
  {Schartmann}, {Schawinski}, {Schnorr-M{\"u}ller}, {Saintonge}, {Shimizu},
  {Sternberg}, {Storchi-Bergmann}, {Sturm}, {Tacconi}, {Treister}, \&
  {Veilleux}}]{Rosario2018}
{Rosario}, D.~J., {Burtscher}, L., {Davies}, R.~I., {et~al.} 2018, \mnras, 473,
  5658, \dodoi{10.1093/mnras/stx2670}

\bibitem[{{Ruiz} {et~al.}(2005){Ruiz}, {Crenshaw}, {Kraemer}, {Bower}, {Gull},
  {Hutchings}, {Kaiser}, \& {Weistrop}}]{Ruiz2005}
{Ruiz}, J.~R., {Crenshaw}, D.~M., {Kraemer}, S.~B., {et~al.} 2005, \aj, 129,
  73, \dodoi{10.1086/426372}

\bibitem[{{Satyapal} {et~al.}(2021){Satyapal}, {Kamal}, {Cann}, {Secrest}, \&
  {Abel}}]{Satyapal2021}
{Satyapal}, S., {Kamal}, L., {Cann}, J.~M., {Secrest}, N.~J., \& {Abel}, N.~P.
  2021, \apj, 906, 35, \dodoi{10.3847/1538-4357/abbfaf}

\bibitem[{{Schnorr-M{\"u}ller} {et~al.}(2016){Schnorr-M{\"u}ller},
  {Storchi-Bergmann}, {Robinson}, {Lena}, \& {Nagar}}]{SM2016}
{Schnorr-M{\"u}ller}, A., {Storchi-Bergmann}, T., {Robinson}, A., {Lena}, D.,
  \& {Nagar}, N.~M. 2016, \mnras, 457, 972, \dodoi{10.1093/mnras/stw037}

\bibitem[{{Schommer} {et~al.}(1988){Schommer}, {Caldwell}, {Wilson}, {Baldwin},
  {Phillips}, {Williams}, \& {Turtle}}]{Schommer1988}
{Schommer}, R.~A., {Caldwell}, N., {Wilson}, A.~S., {et~al.} 1988, \apj, 324,
  154, \dodoi{10.1086/165887}

\bibitem[{{Sebastian} {et~al.}(2020){Sebastian}, {Kharb}, {O'Dea}, {Gallimore},
  \& {Baum}}]{Sebastian2020}
{Sebastian}, B., {Kharb}, P., {O'Dea}, C.~P., {Gallimore}, J.~F., \& {Baum},
  S.~A. 2020, \mnras, 499, 334, \dodoi{10.1093/mnras/staa2473}

\bibitem[{{Shimizu} {et~al.}(2019){Shimizu}, {Davies}, {Lutz}, {Burtscher},
  {Lin}, {Baron}, {Davies}, {Genzel}, {Hicks}, {Koss}, {Maciejewski},
  {M{\"u}ller-S{\'a}nchez}, {Orban de Xivry}, {Price}, {Ricci}, {Riffel},
  {Riffel}, {Rosario}, {Schartmann}, {Schnorr-M{\"u}ller}, {Sternberg},
  {Sturm}, {Storchi-Bergmann}, {Tacconi}, \& {Veilleux}}]{Shimizu2019}
{Shimizu}, T.~T., {Davies}, R.~I., {Lutz}, D., {et~al.} 2019, \mnras, 490,
  5860, \dodoi{10.1093/mnras/stz2802}

\bibitem[{{Simpson} {et~al.}(2002){Simpson}, {Colgan}, {Erickson}, {Hines},
  {Schultz}, \& {Trammell}}]{Simpson2002}
{Simpson}, J.~P., {Colgan}, S. W.~J., {Erickson}, E.~F., {et~al.} 2002, \apj,
  574, 95, \dodoi{10.1086/340946}

\bibitem[{{Smajic{\'c}} {et~al.}(2012){Smajic{\'c}}, {Fischer}, {Zuther}, \&
  {Eckart}}]{Smajic2012}
{Smajic{\'c}}, S., {Fischer}, S., {Zuther}, J., \& {Eckart}, A. 2012, \aap,
  544, A105, \dodoi{10.1051/0004-6361/201118256}

\bibitem[{{Tazaki} {et~al.}(2017){Tazaki}, {Lazarian}, \&
  {Nomura}}]{Tazaki2017}
{Tazaki}, R., {Lazarian}, A., \& {Nomura}, H. 2017, \apj, 839, 56,
  \dodoi{10.3847/1538-4357/839/1/56}

\bibitem[{{Theios} {et~al.}(2016){Theios}, {Malkan}, \& {Ross}}]{Theios2016}
{Theios}, R.~L., {Malkan}, M.~A., \& {Ross}, N.~R. 2016, \apj, 822, 45,
  \dodoi{10.3847/0004-637X/822/1/45}

\bibitem[{{Tristram} {et~al.}(2014){Tristram}, {Burtscher}, {Jaffe},
  {Meisenheimer}, {H{\"o}nig}, {Kishimoto}, {Schartmann}, \&
  {Weigelt}}]{Tristram2014}
{Tristram}, K. R.~W., {Burtscher}, L., {Jaffe}, W., {et~al.} 2014, \aap, 563,
  A82, \dodoi{10.1051/0004-6361/201322698}

\bibitem[{{Tristram} {et~al.}(2009){Tristram}, {Raban}, {Meisenheimer},
  {Jaffe}, {R{\"o}ttgering}, {Burtscher}, {Cotton}, {Graser}, {Henning},
  {Leinert}, {Lopez}, {Morel}, {Perrin}, \& {Wittkowski}}]{Tristram2009}
{Tristram}, K.~R.~W., {Raban}, D., {Meisenheimer}, K., {et~al.} 2009, \aap,
  502, 67, \dodoi{10.1051/0004-6361/200811607}

\bibitem[{{Venanzi} {et~al.}(2020){Venanzi}, {H{\"o}nig}, \&
  {Williamson}}]{Venanzi2020}
{Venanzi}, M., {H{\"o}nig}, S., \& {Williamson}, D. 2020, \apj, 900, 174,
  \dodoi{10.3847/1538-4357/aba89f}

\bibitem[{Virtanen {et~al.}(2020)Virtanen, Gommers, Oliphant, Haberland, Reddy,
  Cournapeau, Burovski, Peterson, Weckesser, Bright, {van der Walt}, Brett,
  Wilson, Millman, Mayorov, Nelson, Jones, Kern, Larson, Carey, Polat, Feng,
  Moore, {VanderPlas}, Laxalde, Perktold, Cimrman, Henriksen, Quintero, Harris,
  Archibald, Ribeiro, Pedregosa, {van Mulbregt}, \& {SciPy 1.0
  Contributors}}]{2020SciPy-NMeth}
Virtanen, P., Gommers, R., Oliphant, T.~E., {et~al.} 2020, Nature Methods, 17,
  261, \dodoi{10.1038/s41592-019-0686-2}

\bibitem[{{Watanabe} {et~al.}(2003){Watanabe}, {Nagata}, {Sato}, {Nakaya}, \&
  {Hough}}]{Watanabe2003}
{Watanabe}, M., {Nagata}, T., {Sato}, S., {Nakaya}, H., \& {Hough}, J.~H. 2003,
  \apj, 591, 714, \dodoi{10.1086/375514}

\bibitem[{{Wehrle} \& {Morris}(1987)}]{Wehrle1987}
{Wehrle}, A.~E., \& {Morris}, M. 1987, \apjl, 313, L43, \dodoi{10.1086/184828}

\bibitem[{{Weingartner} \& {Draine}(2001)}]{WD2001}
{Weingartner}, J.~C., \& {Draine}, B.~T. 2001, \apj, 548, 296,
  \dodoi{10.1086/318651}

\bibitem[{{Wells} {et~al.}(2015){Wells}, {Pel}, {Glasse}, {Wright},
  {Aitink-Kroes}, {Azzollini}, {Beard}, {Brandl}, {Gallie}, {Geers}, {Glauser},
  {Hastings}, {Henning}, {Jager}, {Justtanont}, {Kruizinga}, {Lahuis}, {Lee},
  {Martinez-Delgado}, {Mart{\'\i}nez-Galarza}, {Meijers}, {Morrison},
  {M{\"u}ller}, {Nakos}, {O'Sullivan}, {Oudenhuysen}, {Parr-Burman}, {Pauwels},
  {Rohloff}, {Schmalzl}, {Sykes}, {Thelen}, {van Dishoeck}, {Vandenbussche},
  {Venema}, {Visser}, {Waters}, \& {Wright}}]{Wells2015}
{Wells}, M., {Pel}, J.~W., {Glasse}, A., {et~al.} 2015, \pasp, 127, 646,
  \dodoi{10.1086/682281}

\bibitem[{{W}es {M}c{K}inney(2010)}]{mckinney-proc-scipy-2010}
{W}es {M}c{K}inney. 2010, in {P}roceedings of the 9th {P}ython in {S}cience
  {C}onference, ed. {S}t\'efan van~der {W}alt \& {J}arrod {M}illman, 56 -- 61,
  \dodoi{10.25080/Majora-92bf1922-00a}

\bibitem[{{Whittet}(1992)}]{Whittet1992}
{Whittet}, D.~C.~B. 1992, {Dust in the galactic environment}

\bibitem[{{Whittet}(2022)}]{Whittet2022}
---. 2022, {Dust in the Galactic Environment (Third Edition)},
  \dodoi{10.1088/2514-3433/ac7204}

\bibitem[{{Whittle}(1985)}]{Whittle1985}
{Whittle}, M. 1985, \mnras, 213, 1, \dodoi{10.1093/mnras/213.1.1}

\bibitem[{{Wiebe} {et~al.}(2009){Wiebe}, {Ade}, {Bock}, {Chapin}, {Devlin},
  {Dicker}, {Griffin}, {Gundersen}, {Halpern}, {Hargrave}, {Hughes}, {Klein},
  {Marsden}, {Martin}, {Mauskopf}, {Netterfield}, {Olmi}, {Pascale},
  {Patanchon}, {Rex}, {Scott}, {Semisch}, {Thomas}, {Truch}, {Tucker},
  {Tucker}, \& {Viero}}]{Wiebe2009}
{Wiebe}, D.~V., {Ade}, P. A.~R., {Bock}, J.~J., {et~al.} 2009, \apj, 707, 1809,
  \dodoi{10.1088/0004-637X/707/2/1809}

\bibitem[{{Williamson} {et~al.}(2020){Williamson}, {H{\"o}nig}, \&
  {Venanzi}}]{Williamson2020}
{Williamson}, D., {H{\"o}nig}, S., \& {Venanzi}, M. 2020, \apj, 897, 26,
  \dodoi{10.3847/1538-4357/ab989e}

\bibitem[{{Wilson} {et~al.}(1985){Wilson}, {Baldwin}, \&
  {Ulvestad}}]{Wilson1985}
{Wilson}, A.~S., {Baldwin}, J.~A., \& {Ulvestad}, J.~S. 1985, \apj, 291, 627,
  \dodoi{10.1086/163103}

\bibitem[{{Wright} {et~al.}(2015){Wright}, {Wright}, {Goodson}, {Rieke},
  {Aitink-Kroes}, {Amiaux}, {Aricha-Yanguas}, {Azzollini}, {Banks},
  {Barrado-Navascues}, {Belenguer-Davila}, {Blommaert}, {Bouchet}, {Brandl},
  {Colina}, {Detre}, {Diaz-Catala}, {Eccleston}, {Friedman},
  {Garc{\'\i}a-Mar{\'\i}n}, {G{\"u}del}, {Glasse}, {Glauser}, {Greene},
  {Groezinger}, {Grundy}, {Hastings}, {Henning}, {Hofferbert}, {Hunter},
  {Jessen}, {Justtanont}, {Karnik}, {Khorrami}, {Krause}, {Labiano}, {Lagage},
  {Langer}, {Lemke}, {Lim}, {Lorenzo-Alvarez}, {Mazy}, {McGowan}, {Meixner},
  {Morris}, {Morrison}, {M{\"u}ller}, {rgaard-Nielson}, {Olofsson},
  {O'Sullivan}, {Pel}, {Penanen}, {Petach}, {Pye}, {Ray}, {Renotte}, {Renouf},
  {Ressler}, {Samara-Ratna}, {Scheithauer}, {Schneider}, {Shaughnessy},
  {Stevenson}, {Sukhatme}, {Swinyard}, {Sykes}, {Thatcher}, {Tikkanen}, {van
  Dishoeck}, {Waelkens}, {Walker}, {Wells}, \& {Zhender}}]{Wright2015}
{Wright}, G.~S., {Wright}, D., {Goodson}, G.~B., {et~al.} 2015, \pasp, 127,
  595, \dodoi{10.1086/682253}

\bibitem[{{Wright} {et~al.}(2023){Wright}, {Rieke}, {Glasse}, {Ressler},
  {Garc{\'\i}a Mar{\'\i}n}, {Aguilar}, {Alberts}, {{\'A}lvarez-M{\'a}rquez},
  {Argyriou}, {Banks}, {Baudoz}, {Boccaletti}, {Bouchet}, {Bouwman}, {Brandl},
  {Breda}, {Bright}, {Cale}, {Colina}, {Cossou}, {Coulais}, {Cracraft}, {De
  Meester}, {Dicken}, {Engesser}, {Etxaluze}, {Fox}, {Friedman}, {Fu},
  {Gasman}, {G{\'a}sp{\'a}r}, {Gastaud}, {Geers}, {Glauser}, {Gordon},
  {Greene}, {Greve}, {Grundy}, {G{\"u}del}, {Guillard}, {Haderlein},
  {Hashimoto}, {Henning}, {Hines}, {Holler}, {Detre}, {Jahromi}, {James},
  {Jones}, {Justtanont}, {Kavanagh}, {Kendrew}, {Klaassen}, {Krause},
  {Labiano}, {Lagage}, {Lambros}, {Larson}, {Law}, {Lee}, {Libralato}, {Lorenzo
  Alverez}, {Meixner}, {Morrison}, {Mueller}, {Murray}, {Mycroft}, {Myers},
  {Nayak}, {Naylor}, {Nickson}, {Noriega-Crespo}, {{\"O}stlin}, {O'Sullivan},
  {Ottens}, {Patapis}, {Penanen}, {Pietraszkiewicz}, {Ray}, {Regan},
  {Roteliuk}, {Royer}, {Samara-Ratna}, {Samuelson}, {Sargent}, {Scheithauer},
  {Schneider}, {Schreiber}, {Shaughnessy}, {Sheehan}, {Shivaei}, {Sloan},
  {Tamas}, {Teague}, {Temim}, {Tikkanen}, {Tustain}, {van Dishoeck},
  {Vandenbussche}, {Weilert}, {Whitehouse}, \& {Wolff}}]{Wright2023}
{Wright}, G.~S., {Rieke}, G.~H., {Glasse}, A., {et~al.} 2023, \pasp, 135,
  048003, \dodoi{10.1088/1538-3873/acbe66}

\bibitem[{{Zhang} {et~al.}(2024){Zhang}, {Packham}, {Hicks}, {Davies},
  {Shimizu}, {Alonso-Herrero}, {Hermosa Mu{\~n}oz}, {Garc{\'\i}a-Bernete},
  {Pereira-Santaella}, {Audibert}, {L{\'o}pez-Rodr{\'\i}guez}, {Bellocchi},
  {Bunker}, {Combes}, {D{\'\i}az-Santos}, {Gandhi}, {Garc{\'\i}a-Burillo},
  {Garc{\'\i}a-Lorenzo}, {Gonz{\'a}lez-Mart{\'\i}n}, {Imanishi}, {Labiano},
  {Leist}, {Levenson}, {Ramos Almeida}, {Ricci}, {Rigopoulou}, {Rosario},
  {Stalevski}, {Ward}, {Esparza-Arredondo}, {Delaney}, {Fuller}, {Haidar},
  {H{\"o}nig}, {Izumi}, \& {Rouan}}]{GATOSIV}
{Zhang}, L., {Packham}, C., {Hicks}, E. K.~S., {et~al.} 2024, \apj, 974, 195,
  \dodoi{10.3847/1538-4357/ad6a4b}

\bibitem[{{Zubko} {et~al.}(2004){Zubko}, {Dwek}, \& {Arendt}}]{Zubko2004}
{Zubko}, V., {Dwek}, E., \& {Arendt}, R.~G. 2004, \apjs, 152, 211,
  \dodoi{10.1086/382351}

\end{thebibliography}
\bibliographystyle{aasjournal}



\end{document}